\documentclass{bmcart}

\usepackage{amsthm,amsmath}
\theoremstyle{definition}

\RequirePackage{natbib}
\usepackage{url}
\RequirePackage{hyperref}
\usepackage[utf8]{inputenc} 

\usepackage[modulo]{lineno}
\usepackage{lmodern}
\usepackage{microtype}
\usepackage{gensymb} 
\usepackage{textcomp}
\usepackage[onelanguage,tworuled,noline,noend]{algorithm2e} 
\usepackage{tabularx}
\usepackage{enumitem}

\usepackage{xcolor} 
\SetAlCapHSkip{0pt}
\SetCommentSty{normalfont}
\SetKwComment{Comment}{$\triangleright$}{}
\SetKwComment{tcp}{}{}
\SetNlSty{normalfont}{}{}

\usepackage{graphicx}
\usepackage{tikz}
\usetikzlibrary{arrows.meta}
\usepackage{pgfplots}
\usepackage{pgfplotstable}

\startlocaldefs
\newcommand{\Sup}[1]{\textsuperscript{#1}}
\newcommand{\Sub}[1]{\textsubscript{#1}}
\DeclareMathOperator{\mass}{mass}
\DeclareMathOperator{\calibration}{calibration}
\DeclareMathOperator{\measurement}{measurement}
\DeclareMathOperator{\incorrectfragment}{incorrect~fragment}
\DeclareMathOperator{\correctfragment}{correct~fragment}
\DeclareMathOperator{\correct}{correct}
\DeclareMathOperator{\total}{total}
\DeclareMathOperator{\FWHM}{FWHM}
\DeclareMathOperator{\ToF}{ToF}
\DeclareMathOperator{\DBE}{DBE}

\newcommand{\refsupDAG}[1]{5.1} %
\newcommand{\refsuptabisotopocules}[1]{8} 
\newcommand{\refsupenumisotopocule}[1]{5.3} %
\newcommand{\refsupeqrelativeproba}[1]{7} 
\newcommand{\refsupfullexample}[1]{6} 
\newcommand{\refsupsubsubsecknapsacktwonaivesets}[1]{6.1}
\newcommand{\refsupsubsubsecknapsackDBEopt}[1]{6.2}

\endlocaldefs

\begin{document}

\begin{frontmatter}

\begin{fmbox}
\dochead{Preprint}


\title{Automated fragment formula annotation for electron ionisation, high resolution mass spectrometry: application to atmospheric measurements of halocarbons}


\author[
   addressref={aff1},                   
   email={myriam.guillevic@empa.ch}   
]{\inits{MG}\fnm{Myriam} \snm{Guillevic}}
\author[
   addressref={aff2},
   email={aurore.guillevic@inria.fr}
]{\inits{AG}\fnm{Aurore} \snm{Guillevic}}
\author[
   addressref={aff1},                   
   email={martin.vollmer@empa.ch}   
]{\inits{MKV}\fnm{Martin K.} \snm{Vollmer}}
\author[
   addressref={aff1},                   
   email={paul.schlauri@empa.ch}   
]{\inits{PS}\fnm{Paul} \snm{Schlauri}}
\author[
   addressref={aff1},                   
   email={matthias.hill@empa.ch}   
]{\inits{MH}\fnm{Matthias} \snm{Hill}}
\author[
   addressref={aff1},                   
   email={lukas.emmenegger@empa.ch}   
]{\inits{LE}\fnm{Lukas} \snm{Emmenegger}}
\author[
   addressref={aff1},                   
   email={stefan.reimann@empa.ch}   
]{\inits{SR}\fnm{Stefan} \snm{Reimann}}

\address[id=aff1]{
  \orgname{Laboratory for Air Pollution /Environmental Technology, Empa, Swiss Federal Laboratories for Materials Science and Technology}, 
  \street{Ueberlandstrasse 129},                     %
  \postcode{8600}                                
  \city{D\"ubendorf},                              
  \cny{Switzerland}                                    
}
\address[id=aff2]{%
  \orgname{Université de Lorraine, CNRS, Inria, LORIA},
  \street{}
  \postcode{54000}
  \city{Nancy},
  \cny{France}
}


\begin{artnotes}
\end{artnotes}

\end{fmbox}


\begin{abstractbox}

\begin{abstract} 
\parttitle{Background} 

Non-target screening consists in searching a sample for all present substances, suspected or unknown, with very little prior knowledge about the sample. This approach has been introduced  more than a decade ago in the field of water analysis, together with dedicated compound identification tools, but is still very scarce for indoor and atmospheric trace gas measurements, despite the clear need for a better understanding of the atmospheric trace gas composition. 

For a systematic detection of emerging trace gases in the atmosphere, a new and powerful analytical method is gas chromatography (GC) of preconcentrated samples, followed by electron ionisation, high resolution mass spectrometry (EI-HRMS).
In this work, we present  data analysis tools to enable automated fragment formula annotation for unknown compounds measured by GC-EI-HRMS.

\parttitle{Results} 
 
Based on co-eluting mass/charge fragments, we developed an innovative data analysis method to reliably reconstruct the chemical formulae of the fragments, using efficient combinatorics and graph theory. The method does not require the presence of the molecular ion, which is absent in $\sim$40\% of EI spectra.
Our method has been trained and validated on \textgreater50 halocarbons and hydrocarbons, with 3 to 20 atoms and molar masses of 30~--~330~g~mol\Sup{-1}, measured with a mass resolution of approx.~3500.
For \textgreater90\% of the compounds, more than 90\% of the annotated fragment formulae are correct. Cases of wrong identification can be attributed to the scarcity of detected fragments per compound or the lack of isotopic constraint (no minor isotopocule detected).

\parttitle{Conclusions}
Our method enables to reconstruct most probable chemical formulae independently from spectral databases.
Therefore, it demonstrates the suitability of EI-HRMS data for non-target analysis and paves the way for the identification of substances for which no EI mass spectrum is registered in databases. We illustrate the performances of our method for atmospheric trace gases and suggest that it may be well suited for many other types of samples. The L-GPL licenced Python code is released under the name ALPINAC for ALgorithmic Process for Identification of Non-targeted Atmospheric Compounds.

\end{abstract}




\end{abstractbox}
%

\end{frontmatter}



\section*{Background}

Non-target screening (NTS) is an emerging approach for analysing environmental samples, with potentially revolutionary outcomes.  NTS aims to detect, identify and quantify substances that are unknown in a sample, with no or very little a priori knowledge. This approach contrasts with the more established target or suspect approaches, where a sample is screened only for compounds already known or suspected to be present.

So far, NTS has been developed mostly in the fields of drinking water monitoring, food and soil analysis, forensics and metabolomics \citep[e.g.,][]{Acierno2016, Hollender2017, Alygizakis2018, Kruve2018, Nason2021}, with human health or economic interests as the major underlying motivation. Yet, for the analysis of trace compounds in ambient or indoor air, only very limited NTS-related research has been done (e.g., the discovery of the greenhouse gas SF\Sub{5}CF\Sub{3} \citep{Sturges2000, Veenaas2020}), despite the need for a better understanding of the composition of the air. To look for emerging gases relevant for climate or air quality, suspect approaches are still nearly exclusively used  \citep{Vollmer2015a, Laube2014, Ivy2012}.

NTS requires to measure properties that are specific for one given compound. 
In practice this is usually achieved by chromatographic time separation of the compounds. 
Further, the type of molecule ionisation and the mass range and mass accuracy are particularly relevant for NTS.

Originally, NTS was developed for medium to large molecules, therefore using
soft ionisation such as chemical ionisation (CI) or electrospray ionisation (ESI), 
producing only a few relatively large fragments.
As the molecular ion (entire molecule without one electron) is normally present and detected with soft ionisation,
it is possible to reconstruct the chemical formula (i.e., the atomic
assemblage, without any structural information) of the compound. To elucidate its structure, additional fragmentation and detection is required. Most
(semi-)automated identification software packages were  developed for 
CI or ESI so far~\citep{Loos2015, Jaeger2016, Loos2017, Ruttkies2016, Leon2019} or tandem MS \citep{Duerkhop2019, Koelmel2020}. 

In contrast,
atmospheric trace gases consist of relatively small molecules which are best ionised by
the hard electron ionisation (EI) technique. This causes a fragmentation
cascade, producing many relatively small fragments; the resulting mass spectra contain valuable
structural information but often lack the molecular ion
\citep[e.g.,][Chap.~6]{Gross2017}. Consequently, the identification of the
original molecule becomes highly challenging. 
To circumvent this, specific instrumental source tuning may enhance the detectability of the molecular ion \citep{Abate2010}.  Alternatively, measurements
could be repeated using soft ionisation, such as chemical ionisation, field desorption or field ionisation, but such a combined analytical approach is
expensive and time consuming.

A well-established approach to identify a compound based on its assemblage of masses measured by EI-MS or EI-HRMS, under the absence of the molecular ion, is to perform a mass spectrum library search. Indeed, EI ionisation has been standardised already before the 90's and produces reproducible mass spectra \citep[e.g.~Chap.~5]{Gross2017}.  
One of the best known EI libraries is the NIST/EPA/NIH Mass Spectral Library, with more than 250'000 experimental spectra, including approx. 140 spectra for C1 molecules \citep{NISTwebsite}. 
However, only known and analysed compounds are present in these libraries, and identification results are therefore biased towards these compounds. Unknown emerging pollutant cannot be found by such library search.

For unknown compounds absent from spectral libraries, the identification challenge remains twofold: to identify the chemical formula of the molecular ion (also known as molecular formula annotation) and, in a second step, to identify its structure. Methods exist to identify the formula of the molecular ion in case it is present (e.g., \citep{Hufsky2012}). Previous attempts have been made to predict the mass of the absent molecular ion \citep{Scott1993} and thereby its molecular formula \citep{molgenms_manual}. However, these last methods do not make use of high resolution mass data now available. Alternatively, classifiers have been develop to predict to which class(es) the unknown compound may belong \citep[e.g.,][]{Stein1995, Varmuza1996, Hummel2010, Tsugawa2011}.

Once candidate molecular ions have been generated, and potential classes identified, structure-generation programs (e.g., commercial software MOLGEN-MS \citep{Wieland1996, Gugisch2015}, open-source software OMG \citep{Peironcely2012}) followed by fragmentation programs (e.g., QCEIMS \citep{Grimme2013, Bauer2016}, CFM-ID \citep{Allen2016}, MetFrag \citep{Ruttkies2016}, MOLGEN-MS \citep[][and references therein]{Schymanski2011, Schymanski2012}) can be used to produce candidate mass spectra otherwise absent from libraries. 
In addition, when chromatographic information is available, it can be compared with retention indices if available or with a retention prediction for candidate compounds \citep[e.g.][]{Ruttkies2016, Schymanski2012}.



While high resolution mass spectrometers have been used for at least 30 years, and may provide sufficient information for broad, non-target screening approaches, recent developments have made this technology accessible to a larger number of laboratories.
In the early 2000s, fast response, large coverage and high accuracy mass analyser, such as Orbitrap and time-of-flight (ToF) mass spectrometers, were introduced for water analysis, but only  recently, first approaches have been made to use these powerful detectors also for organic atmospheric trace gases \citep[]{Obersteiner2016, Roehler2020}. 
Due to the challenge of identifying compounds, EI-HRMS are currently used as large mass-range coverage target or suspect screening instruments \citep[e.g.,][]{Schuck2018, Lefrancois2021} but only rarely as NTS instruments.
For state-of-the art EI-HRMS, there is currently a huge divide between what it can deliver in terms of sample coverage, throughput and mass resolution, compared to what identification tools can provide.

In this article, we present a workflow to reconstruct the chemical formula of fragments produced by the fragmentation of a precursor molecule in GC-EI-ToFMS. In addition, we develop a ranking method to identify most probable solutions and the reconstructed fragments that are most similar to the molecular ion. We evaluate our method by quantification of the correct results, on a training set of molecules and on an additional validation set. The entire method is written in Python and publicly available under the name ALPINAC (for ALgorithmic Process for Identification of Non-targeted Atmospheric Compounds, see Section Data availability). While Python may not be the fastest programming language (compared to e.g. C++), it is now widely used in teaching computer science, including to students in environmental sciences and chemistry, and we hope this work will therefore be accessible to a large public.

\newpage

\section{Experimental data}

\subsection{Training data set}

\label{sec:train-set}
To develop our methodology, we use known compounds routinely measured within
the Advanced Global Atmospheric Gases Experiment (AGAGE) network \citep{Prinn2018},
reported in Table~\ref{tab:training-set}. Most of the substances are halocarbons, i.e.~molecules made of a carbon chain, with halogen atoms, and are present in the atmosphere as trace gases. Structures include saturated and unsaturated chains and the presence of rings.

Within AGAGE, the chromatographic and mass spectrometric properties  are obtained by measuring diluted aliquots of a
pure compound \citep{Vollmer2018, Prinn2000, Arnold2012, Guillevic2018} (identification at Level 1 according to the classification for non-target analysis introduced by Schymanski et al. \citep{Schymanski2014}). Subsequently, an unbroken chain of calibration from the prepared synthetic
primary standards to measurements on our instrument ensures that the correct compounds are measured, with the correct quantification \citep{Miller2008b}.  

\subsection{Validation data set}
\label{sec:val-set}

To validate the model after its training phase, we use a set of potentially emerging compounds, listed in Table~\ref{tab:validation-set}.
We  prepared a qualitative standard containing 18 new hydrofluorocarbons (HFCs), listed under the Kigali Amendment to the Montreal Protocol \citep{MP_Kigali2016}. The use of these substances will be progressively restricted in the coming years. Developing the capacity to check for their presence in the air, and their future molar fraction decrease, is part of supporting the application of the Kigali Amendment. The preparation of the qualitative standard is described in the Supplement.
In addition, we use three hydrofloroolefins (HFOs) newly detected in air \citep{Vollmer2015a}, which are replacing  the HFCs in applications such as foam blowing and refrigeration \citep[e.g., ][]{Brown2013}: HFO-1234yf, HFO-1234ze(E) and HCFO-1233zd(E). We use already available standards prepared for these HFOs \citep{Vollmer2015a, Guillevic2018}. Finally, we use two halogenated substances of high boiling point, which are potential emerging contaminants, which were identified and measured at Empa during a specific campaign \citep{Reimann2020}.

For both the training and the validation set, the correct identification of compounds containing the monoisotopic elements fluorine and iodine may be challenging. Indeed for low abundance peaks, the absence of isotopocule may be due to a mono-isotopic element or to a non detected isotopocule, containing e.g. nitrogen or oxygen isotopes.

\subsection{Measurement by GC-EI-TOFMS}

Since the 80s, specific instrumentation has been developed to tackle the challenges of measuring atmospheric halogenated trace gases: pre-concentration of the gases of interest, often present only at picomole per mole levels, chromatographic separation of substances of boiling points as low as $-128$\degree{}C, and precise measurements to allow detection of annual trends \citep[][and references therein]{Prinn2018}.

Our measurement system is very similar to earlier setups \citep{Miller2008b, Arnold2012}. In brief, it starts with a preconcentration trap, refrigerated at approx.~$-150$\degree{}C using a Stirling engine, able to concentrate trace gases from up to six litres of gas (atmospheric air or reference gas mixture). 
Stepwise thawing of the trap eliminates the most abundant air constituents, carbon dioxide and methane, and any remaining oxygen or nitrogen, that would otherwise saturate the detector. Remaining compounds are separated by a gas chromatograph (GC), equipped with a Gaspro pre- and main column (5~m and 60~m, respectively, 0.32~mm inner diameter, Agilent), ionised and detected by a time-of-flight detector (H-TOF, Tofwerk AG, Thun, Switzerland). The detector is set to measure fragments with masses from 24~m/z (mass to charge ratio) to 300~m/z. Masses below 24~m/z are prevented from hitting the detector, to avoid saturation by potential water contamination.  The mass resolution is approximately 3000 below m/z of 50 and up to 4000 above m/z of 100. The raw intensity data at each time-of-flight and each time bin are saved in a file of format \texttt{hdf5} \citep{HDF5website}, which constitutes the used raw data. The total analysis time for one sample is 70~min, with 40~min of preconcentration and stepwise thawing, followed by 30~min of gas chromatography and detection by TOFMS. 

Intensity data, defined as the number of ions that hit the detector at a certain time, are recorded along two axes, the time-of-flight axis (later converted to a mass axis) and the retention time (RT) axis. While the signal along the RT axis reflects the separation of molecules by the GC, the signal along the mass axis reflects the fragmentation pattern of the molecules measured by EI-MS. One fragmented, detected molecule can be visualised as a mountain ridge, producing a variety of mass peaks with various intensities, all aligned perpendicular to the time axis. First, peaks are detected and fitted along the mass axis, and afterwards along the RT axis.

Along the time-of-flight axis, at each time bin, peaks are fitted using a pseudo-Voigt function (Supplement). The obtained time-of-flight centres of the peaks are then converted to masses, using the mass calibration function (Supplement). We assume that all masses have been ionised just once.
This produces a set of 20 to 30 centres of mass values with associated RT and
intensity, from which the mean and standard deviation are computed,
weighted by intensity.
For each detected peak, the mass uncertainty ($u_{\mass}$) is the Euclidean distance of the calibration uncertainty ($u_{\calibration}$, see Supplement) and the measured standard deviation ($u_{\measurement}$):
\begin{linenomath}
\begin{equation}\label{eq:u-mass}
u_{\mass} = 2.5    \sqrt{ u_{\calibration}^2 + u_{\measurement}^2}
\end{equation}
\end{linenomath}
This $u_{\mass}$ is computed using a coverage factor of 2.5 to constrain the range of possible masses for the knapsack algorithm described below. This corresponds approximately to a 98.5\% confidence interval. 
The resulting, expanded mass uncertainty is on average $\approx$~6 mDa or $\approx$~70 ppm.

Along the RT axis, data are saved with a frequency of six points per second (6~Hz). Usually, chromatography peaks last for a minimum of four seconds, producing 20 to 30 points per peak in the RT domain. The observed peak shape along the time axis is typical for gas chromatography and is fitted using the equation proposed by Pap et al. \citep{Pap2001}, that in our case fits well the observed tailing. When computing ratio of intensities, the obtained isotopic pattern accuracy ranges from 1~\% to 5~\%, depending on peak intensities, and is on average 2.0~\%. This value mostly reflects the precision of the entire measurement system. Finally, co-eluting mass peaks are grouped together. 

Routine quality control of instrumental performance includes measuring blanks to check for potential contaminants coming from the measurement setup itself, drifts in retention time due to column ageing or water contamination, and stability of intensity ratios of mass fragments belonging to the same compound.

\section{Method for automated fragment formula identification}

\subsection{Method overview}

The output after peak fitting and mass calibration is a dataset of mass/charge ratio ($m/z$), each with
intensity (in $V$) and uncertainty (in ppm), at a precise retention time (in seconds). Co-eluting peaks may correspond to chemical fragments of a unique molecule, or a small number of distinct molecules. They are therefore grouped into one time slot of approx. 2~s width to be treated together by the identification algorithm.
We consider each time slot separately.

The overview of our method is depicted in Fig.~\ref{fig:method-diagram}. The general approach is to
consider separately each group of co-eluting fragments, and to reconstruct the chemical formula of each fragment based on two types of information:
\begin{itemize}
	\item from the experimental data produced by GC-EI-ToF analysis, we use the measured mass and measured signal intensity of each peak. For the measured masses, the uncertainty (Eq.~\eqref{eq:u-mass}) is computed following metrology principles (Fig.~\ref{fig:method-diagram}, yellow box \textbf{Input}). It is known that using mass information only is not enough to correctly reconstruct chemical formulae, even at \textless1~ppm accuracy \citep{Kind2006};
	\item therefore, we combine these experimental data with chemical information that is universal, i.e.~true for any given molecule: exact mass and valence of chemical elements, known environmental stable isotopic abundances (Fig.~\ref{fig:method-diagram}, two mauve boxes).
\end{itemize}

In practice, the identification method combines algorithms for two purposes: (i) algorithms that enumerate solutions in an exhaustive way,
according to given constrains (Fig.~\ref{fig:method-diagram}, steps 1 and 3); 
(ii) algorithms that eliminate unlikely solutions, based on other constraints (Fig.~\ref{fig:method-diagram}, steps 2, 7, 8 and 9).

The developed workflow is as follows~(steps are numbered as in Fig.~\ref{fig:method-diagram}):

\begin{enumerate}[label={Step~\arabic*:}, ref={Step~\arabic*}]
	\item \label{step1} To start, possible atom assemblages matching the measured masses, within uncertainty, are exhaustively generated. This step usually produces a large number of possible chemical formulae;
	\item \label{step2} all generated formulae are organised in a pseudo-fragmentation
          graph. This step relies on the specificity of EI-MS, producing many
          various fragments from the same precursor ion molecule. The irrelevant
          or unlikely formulae are discarded;
	\item \label{step3} isotopocules (i.e.~molecules having the same type and number of atoms, but where at least one atom is a different isotope) are generated;
	\item \label{step4} for each set of isotopocules, the  isotopic pattern of fragments (theoretical intensity profile) along the mass axis is generated. The optimum contribution to the measured profile, of each set separately, is computed;
	\item \label{step5} using a specifically designed likelihood estimator, all candidate chemical formulae are set out in order of preference, number one being the most likely, and the last one being the least likely;
	\item \label{step6} This ordering enables the selection of the most likely candidate(s);
	\item \label{step7} the intensity of each isotopocule set is optimised by comparison to the measured mass profile, using a machine learning technique. This is by far the  computationally most costly step. Isotopocule series characterised by a total intensity below a given threshold, usually, the instrumental limit of detection (LOD), are eliminated.
	\item \label{step8} The pseudo-fragmentation tree is updated, and the optimisation procedure resumes at \ref{step5}, until a predefined fraction of the measured signal is reproduced.
	\item \label{step9} All remaining, not optimised candidate chemical formulae are deleted. The remaining candidates constitute the final list of most likely correct chemical formulae. Each measured mass may have zero, one or several assigned chemical formulae.
	\item \label{step10} The last step is a tentative reconstruction of the molecular ion(s). Most likely molecular ion(s) are generated based on the largest fragments from the graph and they are set out in order according to their computed likelihood value.

\end{enumerate}

Each step of the method is explained in details hereafter.
In the Supplement, Sect.~\refsupfullexample{}, we give a numerical example with the
mass spectrum that will turn out to correspond to CCl\Sub{4}.

\subsection{\ref{step1}: Generating fragment formulae: the knapsack algorithm}

The aim of the \emph{knapsack step} is to recover all chemical formulae that could correspond to each
detected fragment, given its mass and uncertainty, and excluding all other chemical formulae that would not fulfil the criteria of measured mass and uncertainty. This knapsack step produces the correct chemical formulae, usually along with many other incorrect formulae. The aim of subsequent steps will be to eliminate the incorrect formulae using additional constrains.

 We use combinatorics to generate the chemical formulae of candidate fragments, for each
mass detected in a spectrum. In particular, we develop a variant of the knapsack algorithm
\cite{Schroeppel1981,Odlyzko1991}, dedicated to our setting, which will be described below.

\subsubsection{The knapsack algorithm in combinatorics.}
\label{subsubsect:knapsack-combinatorics}
Combinatorics are mathematical algorithms of fast and exhaustive enumeration,
and the knapsack problem is a well-known topic in this area (see e.g. the handbook on algorithms \cite{Cormen2009}).
The problem is usually stated as follows: given a knapsack of maximum available
capacity (e.g.~mass), and a set of items each of a specific capacity, find
subset(s) of items that can fit into the knapsack, while optimising some other
quantity (usually maximizing the total price of the items). In our setting, the knapsack
is a fragment of given mass (within the uncertainty range), and the items are
atoms of exact given mass. We are interested in enumerating all possible
combinations of atoms so that the sum of their masses fits the measured fragment mass
within the uncertainty margin. An unbounded number of each atom is available i.e.~each atom type can be used multiple times, this is
also known as the \emph{unbounded} knapsack problem. Contrary to the classical
problem in combinatorics, we do not optimise another parameter. Instead, we are
interested in listing all possibilities. In this work, we design an algorithm for
a fast and exhaustive enumeration of all the solutions to the knapsack problem.

The inputs of our dedicated knapsack algorithm are the measured masses of
the fragments with uncertainties, and a list of masses of
elements that are expected to form the fragments. Because the exhaustive
list of solutions  grows exponentially with the number of elements, we will introduce
different techniques to avoid as early as possible enumerating wrong chemical
formulae, while still being exhaustive.

\subsubsection{Targeted mass with uncertainty.}
\label{subsubsect:knapsack-input-output}
This section describes the basic algorithm to enumerate all the possibilities.
The target mass, which is one measured mass, is denoted by $m$; the set of item masses, which are the exact masses of chemical elements (IUPAC: \citep{Meija2016}), is made of $I$
distinct positive values $m_i$, labelled $m_0$ to $m_{I-1}$, sorted in
increasing order, that is, $m_i < m_{i+1}$ for all $i$. 
We do not consider uncertainties of the atomic masses, which are negligible
compared to the ToF analytical mass uncertainties.
A solution of a knapsack problem is encoded as a vector of positive integers
$[a_0,a_1,\ldots,a_{I-1}]$ where $a_i$ is the number of items of mass $m_i$; it
is $0$ if the item $i$ is not in the solution.
Algorithms~\ref{alg:knapsack-basic-rec}, \ref{alg:knapsack-basic-rec-aux}
(in pseudo-code) describe the basic recursive enumeration.
An iterative (non-recursive) function is also possible but we implemented a
recursive function.

\begin{algorithm}[htb]
  \caption{\texttt{unbounded\_knapsack\_rec} header function,
    Exhaustive enumeration of solutions to an unbounded knapsack
    problem (see Sect.~\ref{subsubsect:knapsack-input-output} for description)}
  \label{alg:knapsack-basic-rec}
  \DontPrintSemicolon
  \KwIn{positive target mass bounds $m_{\min}, m_{\max}$,
    ordered item mass list $[m_0,m_1, \ldots, m_{I-1}]$ where $0 < m_i < m_{i+1}$}
  \KwOut{set $S$ of solutions $[a_0,a_1, \ldots, a_{I-1}]$ such that
    $m_{\min} \leq \sum_{i=0}^{I-1}a_i m_i \leq m_{\max}$, $a_i \geq 0$}
  $S \gets \{\}$ (empty set) \;
  $\boldsymbol{a} = [0, 0, \ldots, 0]$ (zero vector of length $I$) \;
  $S \gets $ \texttt{unbounded\_knapsack\_rec\_aux}
  $(m_{\min}, m_{\max}, [m_i]_{0\leq i < I}, \boldsymbol{a}, I-1, S)$ \;
  \Return $S$
\end{algorithm}

\begin{algorithm}[htb]
  \DontPrintSemicolon
  \caption{\texttt{unbounded\_knapsack\_rec\_aux} Recursive function (see Sect.~\ref{subsubsect:knapsack-combinatorics} for description)}
  \label{alg:knapsack-basic-rec-aux}
  \KwIn{positive target mass bounds $m_{\min}, m_{\max}$,
    ordered item mass list $[m_0,m_1, \ldots, m_{I-1}]$ where $0 < m_i < m_{i+1}$,
    vector $\boldsymbol{a}=[a_0,\ldots,a_{I-1}]$, index $i$ ($0 \leq i < I$),
    partial solution set $S$}
  \KwOut{updated set $S$ of solutions $[a_0,a_1, \ldots, a_{I-1}]$ such that
    $\sum_{i=0}^{I-1}a_i m_i = m \in [m_{\min},m_{\max}]$, $a_i \geq 0$}
  \If{$i < 0$ or $m_{\max} < 0$ or ($m_{\min} > 0$ and $m_{\max} < m_0$)}{
    \Return $S$
  }
  \If{$i$ = 0}{
    $a_{0}^{\max} \gets \lfloor m_{\max} / m_0 \rfloor$ \;
    $a_{0}^{\min} \gets \max(0, \lceil m_{\min} / m_0\rceil)$ \;
    \For{$a_0$ from $a_{0}^{\max}$ downto $a_{0}^{\min}$ by $-1$}{
      $\boldsymbol{a}[0] \gets a_0$ \tcp{(we have $m_{\min} \leq a_0 m_0 \leq m_{\max})$}
      Append $\boldsymbol{a}$ to $S$ \;
    }
    \Return $S$ \;
  }
  \If{$i > 0$}{
    $a_{i}^{\max} \gets \lfloor m_{\max} / m_i \rfloor$ \;
    \For{$a_i$ from $a_{i}^{\max}$ downto $0$ by $-1$}{
      $\boldsymbol{a}[i] \gets a_i$ \;
      $S \gets $ \texttt{unbounded\_knapsack\_rec\_aux}
      $(m_{\min}-a_im_i, m_{\max}-a_im_i, [m_i]_{0\leq i<I}, \boldsymbol{a}, i-1, S)$ \;
    }
    \Return $S$
  }
\end{algorithm}

\subsubsection{Only the most abundant isotope of each element used as input.}
We consider 9 elements (H, C, N, O, F, S, Cl, Br, I) with their stable isotopes (if any), making a list of 19 different exact masses
\footnote{H, \Sup{2}H, C, \Sup{13}C, N, \Sup{15}N, O, \Sup{17}O,
  \Sup{18}O, F, S, \Sup{33}S, \Sup{34}S, Cl,
  \Sup{36}S, \Sup{37}Cl, Br, \Sup{81}Br, I. }
that can be combined to form a molecule (we omit the  elements that are rarely found in volatile atmospheric trace gases, such as Si, P and metals \footnote{B, Si, P and noble gases are also supported by ALPINAC and can be added to the list of chemical elements to use if needed.}).
The electron ionisation fragmentation produces isotopocule fragments. For example for the molecule CCl\Sub{4}, we may observe CCl\Sub{4} made of only abundant isotopes (\Sup{12}C or C in short notation, \Sup{35}Cl or Cl in short notation),
and isotopocules containing \Sup{13}C and \Sup{37}Cl (see the complete list of isotopocules provided in Fig.~\ref{fig:isotopocules-CCl4} and in the Supplement, Table~\refsuptabisotopocules{}).

To reduce the enumeration of the knapsack, the input is limited to the mass of
the most abundant isotope of each element
(e.g.~ C of mass $12.000000$~g~mol\Sup{-1} for carbon, Cl of mass $34.96885271$~g~mol\Sup{-1} for
chlorine), making a list of 9 exact masses to be used for the enumeration, instead of 19. 
For relatively small molecules, the fragment made of only abundant isotopes has usually the
highest intensity (or second highest, Fig.~\ref{fig:isotopocules-CCl4}), hence producing a much
smaller mass uncertainty than for the other isotopocules. The target mass range is
thinner, reducing the knapsack enumeration. By doing so, we obtain possible
solutions with the knapsack only for some of the most abundant fragments.
Once a fragment made of abundant isotopes is generated by the knapsack, we later enumerate
all its isotopocules containing minor isotopes, in \ref{step3} (Fig.~\ref{fig:method-diagram}). This is further explained in Section~\ref{sec:iso-dist}.

\subsubsection{Optimisation of the knapsack enumeration: double bond equivalent (DBE) criterion with meet-in-the-middle optimisation algorithm.}

Not all sets of atoms form a valid chemical formula. Indeed, each atom allows a maximum number of chemical bonds with other atoms, according to its valence.  
This can be expressed using the \emph{double bond equivalent} (DBE), or sum of number of rings and double bonds in a given chemical formula. 
The DBE is computed with~\citep[\S6.4.4 Eq.~(6.9)][]{Gross2017}
\begin{linenomath}
\begin{equation}\label{eq:DBE}
  \DBE = 1 + 0.5 \times \sum_{i}^{i_{\max}}N_i(V_i-2)
\end{equation}
\end{linenomath}
where $V_i$ is the valence of element $i$ and $N_i$ is the number of atoms of element $i$ in the chemical formula. 
For EI-MS, since we expect no cluster formation in the ionisation source, the minimum valence for a chemical formula is 0. We do not set any maximum valence value. For the sulphur element, where multiple valences are possible, we chose the maximum value (6), according to one of the golden rules for identification \citep{Kind2007}. 

Of all chemical formulae matching the considered mass domain, only a fraction are chemically valid. This means that to reduce the enumeration time of the knapsack, one strategy is to avoid enumerating chemical formulae with a negative DBE value. We explain hereafter how we implement this.

In the early 80's, cryptologists\footnote{A variant of the knapsack problem was
  used to build cryptosystems to securely hide secrets in the 70's. It was later
  broken with the LLL algorithm. We leave to future work the application of the LLL
  algorithm to our setting.}
formulated a meet-in-the-middle strategy to speed-up the enumeration of all
solutions of a knapsack problem \cite{Schroeppel1981}.
The key-ingredient is to partition the candidate items in two sets.
Applying this strategy to our topic, one enumerates all possibilities made of items of the first set and whose mass
is smaller or equal to the target mass. The possibilities are ordered by
increasing mass. Meanwhile, one does the same with the items of the second
set. 
The two sets are chosen so that the respective running-time of the two enumerations are balanced,
in order to minimize the total running-time.
We end up with two lists of masses in increasing order, of value between 0 and the target mass.
Then reading onward the first list and downwards the second list, one looks for
pairs of partial solutions, one from each list, so that the paired mass matches
the target mass.
A numerical example is given in the Supplement in Section~\refsupsubsubsecknapsacktwonaivesets{}.

We adapt this strategy to speed up the solution of our problem.
We partition the input atoms in two sets: the set of
multivalent atoms (C, N, O, S) and the set of monovalent atoms (H, F, Cl, Br, I). 
First, all solutions made of multivalent atoms, and smaller or equal to the target mass, are generated.
To a generated multivalent-atom-solution, a maximum number of monovalent atoms, twice the DBE value (cf. Eq.~\eqref{eq:DBE}), can be added and still form a valid chemical formula. 
In this way, the partial solutions made of multivalent atoms give us an upper bound
on the number of monovalent atoms that can complete the fragment, reducing
considerably the enumeration of partial possibilities with monovalent atoms.
In particular, it gives an upper bound on the number of hydrogen H.
A numerical example is given in the Supplement in Section~\refsupsubsubsecknapsackDBEopt{}.

The list made of multivalent atoms is precomputed for the heaviest mass first,
and can be re-used for the smaller masses. We also implemented a way to re-use
the list of monovalent atoms precomputed for the heaviest target mass, when
processing the lighter target masses.
Our strategy turned out to be fast enough for the considered mass ranges (see runtime in Sect.~\ref{sec:runtime})
and we did not investigate further optimisations.
Dührkop et al.~\cite{Duhrkop13} have a very different approach well-suited for molecular masses of around
1000~Da, implemented in the SIRIUS software suite for tandem MS \citep{SIRIUS_website}.

After \ref{step1} (Fig.~\ref{fig:method-diagram}), for each measured mass, all chemical formulae that are in agreement with the measured mass within its uncertainty, made of abundant isotopes, and having a positive DBE value, are generated. At this stage, the fragment formulae are not uniquely identified by the knapsack: for each measured mass there are
either too many possibilities, or none (usually because the fragment may contain
non-abundant isotopes).

\subsection{\ref{step2}: Organisation of the results in a pseudo-fragmentation graph}

The aim of \ref{step2} is to organise all chemical formulae generated in \ref{step1} according to a specific order, to help identify and delete unlikely chemical formulae.

With EI-MS, a fragmentation cascade happens due to the high ionisation energy,
i.e.~several fragmentation steps one after the other
\citep[e.g.,][]{Gross2017}. A fragmentation step produces an ionised fragment (detected) and a neutral (not detected). Each detected fragment may result from one or several fragmentation steps. 
As the EI fragmentation takes place under vacuum with pressure usually below $10^{-5}$ bar,
we do not expect to see clusters originating from agglomeration of (fragments of) 
the molecular ion with other chemical species. On the contrary, all detected fragments are pieces of the original molecule.
If all fragmentation steps are known,
one can organise the fragments in an acyclic directed graph (Fig.~\ref{fig:dag-CCl4}). The
nodes are the fragments.
One edge is one fragmentation step.
This forms a \emph{fragmentation graph}.
Potentially, several fragmentation pathways in the EI source may produce the
same end fragment(s). But thanks to directions, the graph is acyclic.

We organise all the candidate formulae from the knapsack in a directed graph
(with the class \texttt{DiGraph} provided in the Python package \texttt{Networkx}, \citep{Hagberg2008}). Each node on the graph is a candidate fragment, with associated attributes, such as its chemical formula, its exact mass, the associated measured mass(es), and the list of minor isotopocules that will be generated at the next step.
An edge is set from a node
to another if the chemical formula of the first fragment admits the chemical formula of the second one as subformula
(e.g.~CCl\Sub{3} has subfragment CCl, see Fig.~\ref{fig:dag-CCl4} presenting the directed graph obtained with all knapsack solutions of CCl\Sub{4}).
This mimics the possible fragmentation pathways.
In other words, we define a \emph{partial} ordering of the fragments (it is not
\emph{total} because, for example, there is no relation between candidate fragments CCl\Sub{3} and CSBr, cf.~Fig.~\ref{fig:dag-CCl4}).
The \emph{maximal fragments}\footnote{in the usual mathematical meaning,
e.g.~\url{https://en.wikipedia.org/wiki/Partially_ordered_set\#Extrema}} have no
ancestor but may have children (Fig.~\ref{fig:dag-CCl4}, nodes in orange and
yellow), they are the maximal elements of the ensemble of fragments.
If the molecular ion is present, it is one of the maximal fragments. As with EI, the molecular ion is often absent (as for 14 compounds of the training set, see Table~\ref{tab:training-set-mol-ion}), several maximal fragments are allowed.
Also, to
account for potential co-elution of different molecules, several connected
components are allowed. 
This algorithm does not use any structural information, only the
candidate chemical formulae, producing only a pseudo-fragmentation graph, not a chemically realistic one, contrary to previous algorithms \citep{Hufsky2012}. We do not use a list of expected neutral losses (as in e.g., \citep{Jaeger2016}) due to the high heterogeneity of our chosen molecules. Therefore, it is likely that some edges are actually not structurally
possible, but this is not relevant at this stage. Optimisation strategies for an efficient construction of this directed graph are reported in the Supplement (Sect.~\refsupDAG{}).

From this pseudo-fragmentation graph, we would like to eliminate the
isolated nodes, or nodes not being connected to any other node (singletons), with neither ancestors nor children (Fig.~\ref{fig:dag-CCl4}, nodes in yellow).  They may correspond to (i) impurities produced for example by
GC bleed or (ii) solutions from the knapsack that are unlikely, in particular
with an atom absent from all other nodes.
But we need to account for very small molecules such as CFC-11 (CFCl\Sub{3}) or
CFC-13 (CF\Sub{3}Cl) that produce a very limited number of different fragments,
without the molecular ion: 
if one measured mass has only singletons as candidate fragments, we do not eliminate them.

At the end of \ref{step2}, usually a few unlikely knapsack solutions being singletons have been eliminated.

\subsection{\ref{step3}: Enumeration of chemical formulae containing minor isotopes}
\label{sec:rare-iso-enum}

The knapsack algorithm produces candidate chemical formulae made of abundant isotopes
only. But all isotopocules of a fragment are expected to be present in a given time
slot (the very rare ones which are below the detection limit of the
mass spectrometer may not be detected). Therefore, for each candidate chemical formula, we now generate a set of all
isotopocules including their relative
intensities based on their natural isotopic abundances \citep{Meija2016}).

Hereafter, we name \emph{knapsack formula} a chemical formula from the knapsack, and
\emph{minor-isotope formula} a chemical formula with at least one minor isotope,
even if this isotopocule is expected to be more
abundant than the \emph{abundant chemical formula}. For example, 
CCl\Sub{4} is called knapsack formula, while
CCl\Sub{3}[\Sup{37}Cl] is called minor-isotope formula.

For each possible knapsack formula, we generate the list of all possible isotopocules.
Again, this is an enumeration process using combinatorics.
If the chemical formula is made of atoms that are monoisotopic, the list contains the
knapsack formula only, whose intensity is one, that is, 100\%.
Otherwise, for each minor-isotope formula, we compute its exact mass and expected intensity
relative to the knapsack formula \citep{Yergey1983, Meija2016}. 
The isotopocules of
the knapsack formula CCl\Sub{4} with their relative intensities are shown in Fig.~\ref{fig:isotopocules-CCl4} and the corresponding numerical values can be found in Table~\refsuptabisotopocules{}
   of the Supplement.

Optimisation strategies to speed up the enumeration are reported in the Supplement (Sect.~\refsupenumisotopocule{}). In particular, isotopocules with intensity expected below the instrumental limit of detection are not enumerated.

At the end of \ref{step3}, knapsack-generated chemical formulae, containing only abundant isotopes, are organised as nodes in a pseudo-fragmentation graph. Each knapsack formula of a node is complemented by its minor-isotope formula(e) if the latter is above the instrumental limit of detection. 

\subsection{\ref{step4}: Computing the optimum contribution for each isotopocule set individually}
\label{sec:iso-dist}

We now consider the measured mass intensity profile. Potentially, any candidate chemical formula may contribute to the measured intensity profile along the mass axis.
First, one generates the theoretical mass profile for each node, i.e. for each knapsack
formula together with its minor-isotope formulae.
Then, one optimises a certain contribution for each node taken individually, to match the measured mass profile.

\subsubsection{Computing a profile of intensity vs mass for a given set of isotopocules}
For each set of isotopocules belonging to the same node, we generate
an expected mass profile.
A measured intensity profile is not continuous, it is a discrete set of
coordinates $(m_a, I_{m_a})$ where $m_a$ is a mass abscissa, and
$I_{m_a}$ is the intensity for this mass.
We consider that
a knapsack fragment and its associated minor-isotope fragments have an expected
intensity profile made of the sum of contributions of all isotopocules along
the mass range. Each mass peak is generated as a pseudo-Voigt
profile, with a prescribed peak width as obtained from the mass calibration
(Supplement) and a mass resolution of about 8~ppm, which is
sufficient given our instrumental mass resolution.
We obtain an expected discrete mass profile (a set of coordinates) for the whole
isotopocule set of the node, of the form
$\{(m_a,\tilde{I}_{m_a})\colon m_a\in \text{mass abscissa}\}$.

\subsubsection{Computing the contribution of a given set of isotopocules}

At this point, over the mass domain covered by a given candidate set of isotopocules $i$, we look for a non-negative scaling factor $k_i$, such that the measured signal
$s_{\text{measured}} = \{(m_a, I_{m_a})\colon m_a \in \text{mass abscissa}\}$
best fits the theoretical profile of the set.
This can be seen 
as an optimisation problem, where the difference between measured and generated profile is minimised:
\begin{linenomath}
\begin{equation}\label{eq:opt-one-ki}
  s_{\text{measured}}  -  k_i \cdot \tilde{I}_i =
  \sum_{m_a \in \text{mass abscissa}} \left( I_{m_a} -  k_i \cdot {\tilde{I}}_{i,m_a}\right) = 0
\end{equation}
\end{linenomath}
where $s_{\text{measured}}$ and $\tilde{I}_i$ can be seen as vectors. With only one $k_i$ value to optimise, Eq.~\eqref{eq:opt-one-ki} can be simplified as the average value of the measured profile divided by the computed isotopocule profile:
\begin{linenomath}
\begin{equation}
  k_i  = \frac{1}{\#\{m_a \in \text{mass abscissa}\}}
  \sum_{m_a \in \text{mass abscissa}} \frac{I_{m_a}}{{\tilde{I}}_{i,m_a}}
\end{equation}
\end{linenomath}

After optimisation of the factor $k_i$, if the entire profile falls below the LOD, the candidate node is removed from the graph of solutions. 

This step can be seen as computing the maximum contribution for a given node (as if no other nodes were present to contribute to the signal). The computed maximum contribution, that we denote $k_{i}^{\max}$, will be used as starting value for the first computation of likelihood in \ref{step5}.

\subsection{\ref{step5}: Ranking of candidates according to a likelihood estimator}

To help us select the most probably correct formulae among all candidate formulae, 
we define an artificial likelihood estimator based on two quantifications, as explained
hereafter.
The estimator takes as input a knapsack fragment, together with its set of minor-isotope fragments.

We decide to capture in the likelihood estimator the other knapsack fragments
that are sub-fragments, and their isotopocules.
For example, for CCl\Sub{2} as chosen knapsack formula, we
would consider its five minor-isotope formulae (e.g. [\Sup{13}C]Cl\Sub{2}), the subformulae CCl and
Cl and their own set of isotopocules (C of m/z~12 is filtered out in our ToF MS).
At \ref{step4}, we estimated $k_{i}^{\max}$, the maximum value that $k_i$ could take.
We now estimate the maximum proportion $g$ of signal that
a candidate fragment $n$, its isotopocules, and all its sub-fragments $i$
could explain.
These considerations lead to a first possible estimator:
\begin{linenomath}
\begin{equation}\label{eq:bad-estimator}
g(n) = \frac{\sum\limits_{i~\text{sub-fragment of}~n} k_i^{\max}
\sum\limits_{j~\text{isotopocule of}~i}p_{i,j}}
{\sum\limits_{m_{\text{spec}} \in \text{mass spectrum}} I_{m_{\text{spec}}}}
\end{equation}
\end{linenomath}
where $p_{i,j}$ is the theoretical abundance of the given isotopocule $j$ for the given
fragment $i$ computed with Eq.~\refsupeqrelativeproba{}
in the Supplement,
$m_{\text{spec}}$ ranges over the mass spectrum,
$I_{m_{\text{spec}}}$ is the intensity at that mass ($I_{m_{\text{spec}}}$ is
computed as a discrete integral, this is the area of the peak of the measured
signal around $m_{\text{spec}}$).
Numerical values of $g(n)$ for knapsack fragments of CCl\Sub{4} are given in Table~\ref{tab:CFC11-results}, fourth column.
In practice, we have observed a misbehaviour of the estimator $g(n)$ from
Eq.~\eqref{eq:bad-estimator}: knapsack formulae that
contain many atomic elements have many more sub-fragments, of various
masses, hence a higher probability to match a larger fraction of the
signal. Eq.~\eqref{eq:bad-estimator} therefore gives advantage to ``complicated''
formulae. This effect can be seen for knapsack fragments of CFC-11 (Table~\ref{tab:CFC11-results}, col.~4): 
fragment HCFCl (wrong) gets a higher score than CFCl (correct) using Eq.~\eqref{eq:bad-estimator}.
Misbehaviour of similar estimators, also capturing the total matched signal, has been reported previously \citep{Schymanski2009}.

To correct for this effect, we multiply Eq.~\eqref{eq:bad-estimator} by
the number of found sub-formulae, divided by
the expected maximum number of sub-formulae. In practice, this maximum number
of sub-formulae is computed with the knapsack algorithm, using the
minimum detectable mass as lower bound (in our case, $m/z = 23$, as all smaller
masses are filtered out in our ToF detector). As the number of knapsack
solutions increases with i) the total number of atoms and
ii) the number of elements present in the fragment, using this number as
denominator will favour solutions constituted of a limited number of elements. The advantage of this technique is to favour ``simple'' solutions,
without setting any parameter to limit the number of elements to
use. Equation~\eqref{eq:bad-estimator}
is therefore completed as:
\begin{linenomath}
\begin{equation}
  \label{eq:likelihood-estimator-2}
g(n) = \frac{\sum\limits_{i~\text{sub-fragment of}~n} k_i^{\max}
\sum\limits_{j~\text{isotopocule of}~i}p_{i,j}}
{\sum\limits_{m_{\text{spec}} \in \text{mass spectrum}} I_{m_{\text{spec}}}}
\frac{\# \{\text{sub-fragments}~i~\text{of}~n\}}
{\# {\genfrac{\{}{\}}{0pt}{1}{\text{all theoretical sub-fragments}}{\text{of }n\text{, mass }\geq 23m/z}} }
\end{equation}
\end{linenomath}
where $\# \{\text{sub-fragments}~i~\text{of}~n\}$ is the number of existing subfragments in the directed graph, for fragment $n$.

All knapsack fragments are set out in order by decreasing likelihood value. 
Looking again at the same example for CFC-11 in Table~\ref{tab:CFC11-results}, using Eq.~\eqref{eq:likelihood-estimator-2}
now fragment CFCl (correct) gets a higher likelihood score than HCFCl (col.~5), and is therefore ranked better (col.~6). We underline that this defined likelihood estimator has no chemical signification. Its aim is only to highlight the simplest knapsack solutions that explain the maximum proportion of the measured signal. It is therefore only a practical tool to speed up the identification process.

\subsection{\ref{step5} to \ref{step8}: Iterating loop to compute the optimum contribution of multiple sets of candidates together}

Overall, the measured signal profile
$s_{\text{measured}} = \{(m_a, I_{m_a})\colon m_a \in \text{mass abscissa}\}$
should match a linear combination of the expected
profiles of all correct candidate sets, an approach already found in
e.g.~\citep{Roussis2003}. Formally, this approach allows several fragments to form together the signal of one measured peak, which is realistic given our mass resolution.  Equation \eqref{eq:opt-one-ki} is therefore completed as:
\begin{linenomath}
\begin{equation}
  s_{\text{measured}}  - \sum k_i \cdot \tilde{I}_i =
  \sum_{m_a \in \text{mass abscissa}} \left( I_{m_a} - \sum_{\text{profile}~i} k_i \cdot {\tilde{I}}_{i,m_a}\right) = 0
\end{equation}
\end{linenomath}
This equation cannot be simplified. Instead, the optimisation of multiple $k_i$ is handled by a machine learning technique, using the Python package \texttt{lmfit}, based on the Levenberg-Marquardt algorithm \citep[for details, see][]{Newville2014}.
This optimisation of multiple $k_i$ together is by far the most expensive computation step in
the entire method. The \texttt{lmfit} package is most efficient when only a
limited number of parameters $k_i$ are optimised. In particular, it is outside
the computing power of a regular desktop machine to optimize all the
profiles of the candidate formulae together.
We implement three techniques to lower the running time of \texttt{lmfit}.

First, the measured profile is itself not perfect but affected by measurement noise. This noise also affects the expected isotopic pattern accuracy. To account for this, 
we do not aim at reconstructing 100\% of the signal, but only a significant portion of it.
Since our experimental precision is in the order of 1 \% to 5\%, as well as the isotopic pattern accuracy, we set the
threshold at 95\% of the signal: when 95\% of the ``area'' below the signal is reconstructed, the
optimisation is stopped.
The fraction of explained signal $f$ is calculated as follow, iterating over
the $i$ selected candidate fragments, their respective set of isotopocules, and the
mass spectrum:
\begin{linenomath}
\begin{equation}
\label{eq:likelihood-estimator-1}
f = \frac{ \sum\limits_{\text{selected candidate}~i}~ k_i
\sum\limits_{\text{isotopocule}~j~\text{of}~i}p_{i,j}}
{\sum\limits_{m_{\text{spec}} \in \text{mass spectrum}}I_{m_{\text{spec}}}}
\end{equation}
\end{linenomath}
where $k_i$ are the linear factors optimised by the lmfit package.
We aim at reaching $f \geq 0.95$.

Second, a further reduction of computation time is achieved by splitting the mass domain:
all candidates are grouped into smaller, non-overlapping mass domains, where the
optimisation is run separately. Indeed, optimising a small number of $k_i$ multiple times is more efficient than optimising a large number of $k_i$ just at once.

Third, we observe that at this stage, many wrong solutions generated by the knapsack are still present, while usually, only a limited number of chemical formulae are really present (see later discussion in Sect.~\ref{sec:runtime} and also Fig.~\ref{fig:method-behaviour}). We therefore adopt the following 
greedy-like strategy%
\footnote{\url{https://en.wikipedia.org/wiki/Greedy_algorithm}}, 
in order to reduce the number of fitted isotopic
profiles.
The most likely solutions
are processed first, until the reconstructed signal
reaches 95\% of the measured signal. This way, unlikely solutions left after 95\%
of the signal has been reconstructed are not considered at all. This approach
requires all candidate fragments to be ordered according to a well chosen likelihood
estimator, as done in \ref{step5}. 

In practice, from the list of ranked knapsack formulae (or nodes on the graph), we take the one ranked first, together with all its sub-fragments (or children of the node) and all associated isotopocules, optimise the contributions of these selected nodes (\ref{step7}), and then eliminate any node below the LOD, as well as any node becoming a singleton (\ref{step8}). Since the $k_i$ have updated values, the ranking is updated (back to \ref{step5}). If less than 95\% of the signal is reconstructed, the next most likely node is added to the selected nodes (\ref{step6}). This iterating procedure is depicted in Fig.~\ref{fig:method-diagram}, green boxes.

At the end of this iterating procedure, nodes that have not been selected to be optimised are deleted from the pseudo-fragmentation graph, and any new singletons are deleted as well (\ref{step9}). For each remaining node, its factor $k_i$ is used to compute the final contribution of this node to each measured mass peak.

\subsection{\ref{step10}: Tentative identification of the molecular ion}

At the end of \ref{step9}, the majority of detected fragments have been assigned a chemical formula. Using this information, we develop a simple algorithm to identify or reconstruct likely molecular ions. The three conditions of the SENIOR theorem have to be fulfilled, as listed by Kind and Fiehn \citep{Kind2007}:
\begin{enumerate}
	\item The sum of valences or the total number of atoms having odd valences is even.
	\item The sum of valences is greater than or equal to twice the maximum valence. This rule prevents fragments such as CFCl to be considered as valid molecular ion.
	\item The sum of valences is greater than or equal to twice the number of atoms minus 1. In our settings, by construction all fragment formulae have a non-negative DBE value, therefore this rule is fulfilled.
\end{enumerate}

We start from the list of maximal fragments still part of the graph at the end of \ref{step9}, and separate them into two groups, with odd or even sum valence.

All maximal fragments with even valence fulfil the first condition of the SENIOR theorem. We then test for the second condition. All maximal fragments fulfilling these two criteria are added to the list of potential molecular ions.

Using all the maximal fragments with odd valence, we enumerate all possibilities of adding one monovalent atom to each of them, using all monovalent atoms present in all fragments on the graph. Each newly constructed maximal fragment, if fulfilling the second SENIOR condition, is considered a potential molecular ion. It is added to the graph with all other fragments, and its likelihood value is computed.

This algorithm implicitly makes the assumption that all multivalent atoms present in the true molecule have been detected in at least one fragment and correctly identified.

\section{Results and discussion}

\subsection{Validation data from standard measurements}

To evaluate the reconstructed fragment formulae, we adopt the following pragmatic approach, already used in e.g. \cite{Kwiecien2015}: a fragment formula is considered to be correct if it is a sub-formula of the true molecular formula. Fragments resulting from re-arrangements, which do happen with EI ionisation, are thus considered as correct solutions.

\subsection{Estimators of method performance}

A first quantitative measurement of method performance is how much of the measured signal $I_{m_{\text{spec}}}$ is reconstructed, written $f_{I\!,\total}$, irrespective of whether the signal corresponds to correct or incorrect fragments:
\begin{linenomath}
\begin{equation}\label{eq:frac-rec-signal-total-signal}
f_{I\!,\total} = \frac{ \sum I_{\correctfragment} + I_{\incorrectfragment}}
{\sum\limits_{m_{\text{spec}} \in \text{mass spectrum}} I_{m_{\text{spec}}}}
\end{equation}
\end{linenomath}
Reconstruction of fragment formulae produces a qualitative information. To quantify method performance,
we construct two metrics based on recommendations for qualitative measurements \citep[][Annex D]{Eurachem}.

We define the ratio of correctly identified fragment formulae $f_n$ as the number of correctly identified fragment formulae divided by the total number of reconstructed fragment formulae, per compound:
\begin{linenomath}
\begin{equation}\label{eq:frac-correct-frag-all-frag}
f_{n,\correct} = \frac{n_{\correctfragment}}{n_{\correctfragment} + n_{\incorrectfragment}}
\end{equation}
\end{linenomath}
Then, we define the ratio of correctly assigned signal $f_{I\!,\correct}$ as the sum of intensities of correctly assigned fragments divided by the total intensity of all reconstructed fragments:
\begin{linenomath}
\begin{equation}\label{eq:frac-correct-I-all-I}
f_{I\!,\correct} = \frac{\sum I_{\correctfragment}}{ \sum I_{\correctfragment} + I_{\incorrectfragment}}
\end{equation}
\end{linenomath}
For comparison, we also compute per training compound the fraction of correct fragments and signal on the top-10 results, which are the list of maximum 10 fragments with maximum likelihood.

\subsection{Performances of the method on the training set}

An example of the output produced by the identification method is plotted in Fig.~\ref{fig:CCl4-mass-spectrum}. All numerical values can be found in the supplementary information (see Section Additional Information). What is delivered to the user is a list
of the chemical formulae of the generated and non-eliminated fragments, their
exact mass, assigned signal, likelihood value according to
Eq.~\eqref{eq:likelihood-estimator-2} and ranking. The method also informs if a
fragment is a maximal fragment and if it is a potential molecular ion fulfilling the SENIOR theorem. 
For example, for CFC-11 the maximal
fragment CFCl\Sub{3} with ranking of 1 (maximum likelihood) is the molecular ion
(Table~\ref{tab:CFC11-results}). This example demonstrates the usefulness of our likelihood estimation to identify the fragments closest to, or being,
the molecular ion. 

\subsubsection{Fraction of reconstructed signal}
\label{subsubsect-frac-signal}

For 34 compounds in the training set (all but two), the fraction of reconstructed signal is above 0.95. For  CF\Sub{4} and C\Sub{2}H\Sub{6}, it is 0.06 and 0.88, respectively.  In these two cases, the measured mass and its uncertainty envelope did not contain the true mass, causing the knapsack to fail in generating the correct chemical formulae.
Given that we multiply the mass uncertainty with a coverage factor of 2.5 (Eq.~\eqref{eq:u-mass}), corresponding to 98.5\% of the expected mass interval, it can be expected that in a few cases, the considered mass domain does not contain the true answer. We observe that no wrong fragment fills this gap, but the signal is rather not reconstituted. It is therefore easier for the user to identify such extreme cases, and e.g.~run the identification process again using a larger coverage factor.

For CF\Sub{4}, correct chemical formulae were reconstructed for the three measured masses when using a coverage factor of 6.0. For C\Sub{2}H\Sub{6}, using $k=2.5$, no solution is produced for
the measured fragment at 29.037805~m/z.  Using a coverage factor $k$ of 3.0 instead allows the knapsack to produce the correct fragment formula, C\Sub{2}H\Sub{5}.

\subsubsection{Compounds with very few measured masses}
\label{subsubsect-few-masses}
As a consequence of their simple molecule structure or a very low abundance in the sample, five compounds had fewer than 6 detectable masses (NF\Sub{3}, CF\Sub{4}, CH\Sub{3}I, SO\Sub{2}F\Sub{2}, SF\Sub{5}CF\Sub{3}). We observed that in such cases, our identification algorithm does not have enough constrains to suggest correct results. 
This confirms previous observations from Hufsky et al. \citep{Hufsky2012}.
We therefore developed the following strategy: when the number of measured masses is less than 6, maximal fragments are treated separately through the iterative \ref{step5} to \ref{step8} (green boxes on Fig.~\ref{fig:method-diagram}), so that chemical formulae belonging to different maximal fragments are not optimised together. The list of kept maximal fragments is then returned as result, ordered by likelihood. The program returns a message warning that multiple maximal fragments are possible, and suggests the user to choose the one considered most likely.

For the sake of including these compounds with all other results in the following figures,  the maximal fragment ranked first is kept, and all others are eliminated (assuming no co-elution). In all cases except for SO\Sub{2}F\Sub{2}, where only two masses were measured, the first ranked maximal fragment was correct.

\subsubsection{Fraction of correct reconstructed fragments and signal}

Fig.~\ref{fig:accuracy-sum-formula} displays the histograms of the fractions of correct fragment formulae and of correct reconstructed signal, for all fragments (left) and for the top-10 fragments for each compound (right). Fig.~\ref{fig:accuracy-sum-formula} shows
better fraction of correct results when taking the fraction of correct signal into account compared to the fraction of correct chemical formulae: wrongly identified fragments tend to have a smaller abundance, mostly due to higher mass uncertainty or lack of companion peak that would provide an isotopic constrain. For only 44\% of the compounds, 90\% or more of the chemical formulae are correct. However, for more than 90\% of the compounds, the signal from correct fragments constitutes at least 90\% of the reconstructed signal. This underlines that the proportion of signal assigned to each chemical formula carries information about how likely this chemical formula is correct.

The ability of the method to produce correct chemical formulae further improves when taking into account the top-10 results only, i.e.~the 10 chemical formulae ranked as most likely according to the likelihood estimation, and their associated signal. For these top-10 fragments, the chemical formulae are at least 90\% correct for 58\% of the compounds; the proportion of compounds for which  at least 90\% of the signal is correct stays unchanged, at 90\%. Indeed, most of the time, wrong fragments are anyway assigned a small portion of the signal.

For the training set, only two compounds were poorly identified, CF\Sub{4} and SO\Sub{2}F\Sub{2}, for reasons discussed previously.

\subsubsection{Information from the likelihood estimation and ranking results}

As we have seen, the identification algorithm does not eliminate all wrong chemical formulae. We now study more in detail the likelihood value and ranking associated to each reconstructed fragment to see if these values can better inform us if a fragment is correct or wrong.

Fig.~\ref{fig:hist-likelihood-frag-att} (left histogram) presents the distribution of likelihood values for correct fragments (in blue) and for incorrect fragments (in red), and the same for maximal fragments only (right histogram). From these distributions, we observe that likelihood values above 20 indicate that the fragment is correct by 95\% ($n$ = 89, 85 correct and 4 incorrect fragments), and the maximal fragment correct by 90\% ($n$ = 35, 32 correct and 3 incorrect maximal fragments).
We could therefore use a likelihood value threshold above which a fragment or maximal fragment could be tagged as most probably correct.
At the other end of the distribution, 90\% of maximal fragments with a likelihood value below 8 are wrong ($n$ = 31, 3 correct and 28 incorrect maximal fragments).
In contrast, small likelihood values do not necessarily indicate that the fragment is false, but rather that the fragment represents a small portion of the signal. Only below a likelihood value of 0.3, 90\% of the fragments are wrong. This implies that using a given likelihood value as cut off to delete fragments  would  be either inefficient and delete very few fragments, or be inaccurate and delete many correct fragments.  

The distribution of ranking values for fragments and maximal fragments are shown in  Fig.~\ref{fig:hist-ranking-frag-att}.
The left histogram illustrates that the ranking of fragments does not help separating between correct and wrong fragments because the corresponding distributions overlap strongly.
However, for the maximal fragments, this overlap is less pronounced (right histogram):
correct maximal fragments have a ranking value usually better than 10, while maximal fragments whose ranking is worse than this value are mostly wrong.

These observations can help the user in the identification process by tagging
the maximal fragments with a likelihood value above 20 or a ranking value better than 10 as probably correct, and the others as likely wrong.

\subsubsection{Reconstructed molecular ions}

For 29 molecules on the training set ($>$80~\%), the reconstructed molecular ion ranked first is the correct one (see detailed results in Tables \ref{tab:training-set-mol-ion}). For CCl\Sub{4} and HCFC-141b, the correct solution is ranked second and the solution ranked first is still quite close to the correct solution (CHCl\Sub{3} instead of CCl\Sub{4} and C\Sub{2}H\Sub{4}FCl instead of C\Sub{2}H\Sub{3}FCl\Sub{2}, respectively). 
For PFC-c318, perfluorohexane and SF\Sub{5}CF\Sub{3}, sub-fragments of the molecular ion are listed.
These last three cases are wrong because our reconstruction method assumes that the correct number of multivalent atoms is detected in at least one fragment, which was not the case. For CF\Sub{4} and SO\Sub{2}F\Sub{2}, no correct solution was suggested, these two cases are discussed in Sect.~\ref{subsubsect-frac-signal} and \ref{subsubsect-few-masses}.

For comparison, we tested two software that suggest the weight of the molecular ion. Both software use unit mass information, not high resolution. We used the same training set, excluding C$_6$F$_{14}$ because its molecular weight is outside our mass detection range.  The NIST library tool \citep{Scott1993} was able to reconstruct the correct molecular weight in 17~cases ($\approx$49~\%). In average, the NIST molecular weight prediction deviates from the correct one by 12~Da, with a median deviation of 2~Da, which is very similar to the performances published initially in 1993 by Scott et al. \citep{Scott1993}. We also tested the commercial software MOLGEN-MS, which contains a module to determine if specific chemical classes are present in a molecule (MSclass module, \citep{Varmuza1996}), and then predicts the weight of the molecular ion (ElCoCo module, \citep{molgenms_manual}). We found better results when using the ElCoCo module alone, without MSclass: the correct weight for the molecular ion was listed in 15 cases ($\approx$43~\%). Technical details are given in the supplement.


\subsection{Performance of the model on the validation set}

The validation set (Table~\ref{tab:validation-set}) is made of 23 compounds. The performance of the reconstructions is similar as with the training set: for 95\% of the compounds, more than 90\% of the reconstructed signal is correct (Fig.~\ref{fig:val-accuracy-formula-signal}). This illustrates that our identification method can be applied to different dataset, while producing similar performances. As an example, the reconstructed mass spectrum for HFO-1234yf is given in Figure~\ref{fig:HFO1234yf-massspectrum}.

The reconstructed molecular ion is correct in 19 cases (83~\%, see detailed results in Table~\ref{tab:validation-set-mol-ion}). For the wrong cases, the suggested molecular ions are mostly sub-fragments of the true molecular ion. Interestingly, when the correct molecular ion is listed, it is always ranked first, suggesting that the likelihood estimator (Eq.~\ref{eq:likelihood-estimator-2}) is quite effective in ranking first the most likely results.

\subsection{How wrong knapsack solutions are rejected and implications for computation runtime}
\label{sec:runtime}

Figure~\ref{fig:method-behaviour} displays, for each compound of the training and validation sets, 
how many knapsack fragments are generated, kept and rejected. 
For compounds where less than $\approx$50 knapsack fragments were generated, 
we observe that most fragments have gone through the iterating steps of the workflow (see Fig.~\ref{fig:method-diagram}), and are therefore validated (Fig.~\ref{fig:method-behaviour}, blue crosses) or deleted, as singletons (grey 'x') or as being below the LOD (mauve stars). 
On the other hand, for compounds with more than $\approx$100 knapsack fragments, the number of fragments gone through the computation-intensive iterating steps do not exceed $\approx$100, even if more than 1000 knapsack fragments were generated. 
In such cases, most fragments are rejected at \ref{step9} of the workflow (Fig.~\ref{fig:method-diagram}).
This behaviour may explain why the computation runtime does not increase linearly with the number of generated knapsack fragments, as discussed hereafter.

Details about the computation runtime are given in Fig.~\ref{fig:method-runtime}. For all compounds, the knapsack step (Step~1) represents only a minor part of the runtime, proving that the optimisation of this computation step is appropriate for the considered compounds.  Above 1000 generated knapsack formulae, the graph construction (Step~2) represents an important portion of the runtime, but remains minor for all compounds with less than 1000 knapsack fragments. Since the runtime for the graph construction is proportional to the number of generated knapsack solutions, if applied to larger molecules, further optimisation will be necessary to limit the computation time. For example, a method to pre-select likely present chemical elements may be useful.

For most compounds, the most computation intensive step is the optimisation of multiple isotopocule profiles (Step~7), which uses the machine learning tool \texttt{lmfit}. However, the runtime of Step~7 does not increase linearly with the number of knapsack solutions, but for most compounds is limited to less than 10 seconds. We attribute this to the fact that only a limited number of most likely knapsack fragments go through this expensive step.

\section{Conclusion}
Adequate information about the presence in the environment and potentially illicit emissions of halogenated and in particular (per)fluorinated compounds is more and more pressing, requiring a broader use of NTS approaches \citep{NYT2016}.
Gas chromatography followed by electron ionisation and high-resolution mass spectrometry is increasingly used
for ambient air quality measurements and is a promising technique for non-target screening of atmospheric trace gases. 
To support automated identification of small unknown (halogenated) substances, especially in cases where the molecular ion is absent from the obtained mass spectrum, specific data analysis tools making use of newly availably high resolution mass spectrometry data are expected to improve current identification performances.

In this work, we have developed a novel algorithm to allow reconstruction of the chemical formula based on the measured mass of fragments likely belonging to the same substance. The developed method specifically addresses the observation that peaks with low signal have a higher mass uncertainty, by using a specific uncertainty for each measured mass peak. This approach allows to use all measured data simultaneously, 
while giving more weight to more accurate mass peaks. 
Our method does not require the molecular ion to be present, and can still reconstruct the chemical formula of all other detected fragments. 
This is important for electron-ionisation spectra, where the molecular ion is absent in approx. 40~\% of the cases. In addition, we developed a simple algorithm to generate and rank possible molecular ions, based on the addition of likely monovalent chemical elements to the largest identified fragments.

Overall, the method performs well on very heterogeneous compounds, comprising nine different chemical elements and many different structures, for molecules from 3~atoms (COS) to 20~atoms (C$_6$F$_{14}$), over a large molar mass domain from 30~g~mol$^{-1}$ to over 300~g~mol$^{-1}$, and measured with an average mass uncertainty of 70~ppm: for more than 90\% of the compounds, more than 90\% of the signal has been assigned to the correct chemical formula. The presented method was able to reconstruct and rank first the molecular ion in \textgreater80~\% of the cases. The reconstruction becomes less reliable with decreasing number of detected mass peaks, in case of compounds measured at very low molar fraction.

Finally, we would like to emphasis that in the difficult field of compound annotation and structure elucidation, robust knowledge can certainly be gained from applying different methods in parallel. For example,
when the retention time information is available, additional confidence could be gained from comparison with predictions (e.g., using Quantitative Structure Property Relationships (QSPR) models or correlations with the boiling point, as previously done in e.g. \citep{Schymanski2012}).
For the compounds where the molecular ion was proven present in the spectrum, the fragmentation tree computation method from Hufsky et al. \citep{Hufsky2012} could then be applied. For further structure elucidation, the identified or generated most likely molecular ion could be fed to subsequent molecular structure generators and EI fragmentation programs, which already exists \citep{Kerber2004, Gugisch2015, Allen2016, Ruttkies2015, Ruttkies2016, Grimme2013, Bauer2016}.

\clearpage


\begin{backmatter}

\section{Data Availability}

The Python code is released with the acronym ALPINAC, for ALgorithmic Process
for Identification of Non-targeted Atmospheric Compounds, under an MIT
open-source license at \url{https://gitlab.inria.fr/guillevi/alpinac/}.
All measured mass spectra are provided in a zip supplement, in a .txt format
compatible with ALPINAC. The same data are also provided as .jdx (NIST
compatible) and .tra (MOLGEN-MS compatible) formats, in zip files.

\section*{Additional information}
Supplementary information is available in the additional file \texttt{sup\_ALPINAC\_20210805.pdf}.

All input mass spectral data for ALPINAC (training and validation set) can be found in \texttt{alpinac\_input\_mass\_spectra.zip}.

All output formulae produced by ALPINAC, using the default settings, can be found in \texttt{alpinac\_output\_formulae.zip}.

All input mass spectra converted to .TRA files, compatible for MOLGEN-MS, can be found in \texttt{input\_mass\_spectra\_forMOLGENMS.zip}.

All input mass spectra converted to .jdx files, compatible for the NIST library, can be found in \texttt{input\_mass\_spectra\_forNIST.zip}.

\section*{Competing interests}
  The authors declare that they have no competing interests.

\section*{Author's contributions}

MG designed the identification algorithms, with inputs from SR, MKV and PS.
AG suggested implementation choices and code improvements.
MG and AG wrote the Python software.
MKV, PS, MH and MG prepared the standards and collected the ToF data. 
MG, AG wrote the manuscript with contributions from all co-authors.
SR, MKV and LE secured the HALOSEARCH Empa funding.

\section*{Acknowledgements}
 
M.~Guillevic thanks Luke Western, Stuart Grange and Minsu Kim for valuable discussions related to machine learning.
A.~Guillevic thanks Paul Zimmermann for discussions on the knapsack problem, and
Stephane Glondu for his valuable help in installing Python packages for
computational chemistry and setting the development environment on Linux platforms.
We thank Georgios Papadopoulos, Mike Cubison (Tofwerk AG) and Harald Stark (Aerodyne Research Inc.) for their support related to the TOF instrument.
We are grateful to Markus Meringer who provided a 90-days free trial version of the commercial software MOLGEN-MS.
We thank G.~and D.~Jones who provided linguistic help. We thank two anonymous reviewers whose comments helped to improve the manuscript.
MG is funded by the Empa research grant HALOSEARCH. We acknowledge funding by the DAAMAA project from the Swiss Polar Institute.


\bibliographystyle{bmc-mathphys} 
\bibliography{ALPINAC_20210319}      




\section*{Figures}

\definecolor{yellow}{HTML}{ffdd55}
\definecolor{green}{HTML}{cdde87}
\definecolor{lavander}{HTML}{afafe9}
\definecolor{rosa}{HTML}{e9afc6}
\definecolor{mauve}{HTML}{ae7181}
\definecolor{lightorange}{HTML}{fdaa48}

\begin{figure}
  \centering
  \begin{sffamily}
\def \boxwidth{0.75\textwidth}
\def \boxheight{3\baselineskip}
\def \smallboxwidth{0.18\textwidth}
\def \hdist{4\baselineskip}
\def \hhdist{5\baselineskip}
\def \ldist{0.48\textwidth}
    \begin{small}
  \begin{tikzpicture}
    [Start/.style = {rectangle,rounded corners,draw=black,thick, minimum width=\boxwidth, minimum height=\boxheight, fill=lightorange, fill opacity=1, inner sep=1pt, text centered},
    LeftInput/.style = {rectangle,rounded corners,draw=black,thick, minimum width=\smallboxwidth, minimum height=\boxheight, fill=mauve, fill opacity=1, inner sep=1pt, text centered},
    StepInit/.style = {rectangle,rounded corners,draw=black,thick, minimum width=\boxwidth, minimum height=\boxheight, fill=yellow, fill opacity=1, inner sep=1pt, text centered},
		  Step/.style = {rectangle,rounded corners,draw=black,thick, minimum width=\boxwidth, minimum height=\boxheight, fill=green, fill opacity=1, inner sep=1pt, text centered},
    Output/.style = {rectangle,rounded corners,draw=black,thick, minimum width=\boxwidth, minimum height=\boxheight, fill=lavander, fill opacity=1, inner sep=1pt, text centered},
]

\node[Start] (A1) {
  \begin{tabular}{@{}c@{}}
    {\bf Input: experimental data}\\
    masses, mass uncertainties, signal intensities, LOD
  \end{tabular}
};
\node[StepInit, below of=A1, node distance=\hdist](B1) {
  \begin{tabular}{@{}c@{}}
    {\bf Step 1: Knapsack} \\
    Generate all fragment formulae matching the measured masses. \\
    Use most abundant isotopes only.
  \end{tabular}
};
\node[StepInit, below of=B1, node distance=\hdist](B2) {
  \begin{tabular}{@{}c@{}}
    {\bf Step 2: Initialise directed graph} \\
    Connect a fragment to closest, larger fragments. \\
    Eliminate singletons.
  \end{tabular}
};
\node[StepInit, below of=B2, node distance=\hdist](B3) {
  \begin{tabular}{@{}c@{}}
    {\bf Step 3: Initialise isotopocule sets} \\
    Generate minor-isotope formulae, compute isotopocule profiles.\\
  \end{tabular}
};
\node[StepInit, below of=B3, node distance=\hdist](B4) {
  \begin{tabular}{@{}c@{}}
    {\bf Step 4: Compute max. contribution of each set} \\
		for each isotopocule set, individually.		 
  \end{tabular}
};
\node[Step, below of=B4, node distance=\hdist](B5) {
  \begin{tabular}{@{}c@{}}
    {\bf Step 5: Rank fragments} \\
    Per fragment including all its children:\\
    rank according to likelihood estimator.
  \end{tabular}
};
\node[Step, below of=B5, node distance=\hdist](B6) {
  \begin{tabular}{@{}c@{}}
    {\bf Step 6: Select most likely fragments} \\
    and all their children.\\
  \end{tabular}
};
\node[Step, below of=B6, node distance=\hdist](B7) {
  \begin{tabular}{@{}c@{}}
    {\bf Step 7: Optimise multiple isotopocule sets} \\
    Optimise contribution of isotopocule sets to fit measured profile.\\
		Use Python package \texttt{lmfit}. Eliminate sets \textless LOD.
  \end{tabular}
};
\node[Step, below of=B7, node distance=\hdist](B8) {
  \begin{tabular}{@{}c@{}}
    {\bf Step 8: Update directed graph} \\
    Eliminate singletons. \\
		Re-connect a fragment to closest, larger fragments if needed.
    
  \end{tabular}
};
\node[StepInit, below of=B8, node distance=\hhdist](B9) {
  \begin{tabular}{@{}c@{}}
    {\bf Step 9: Eliminate not optimised sets} \\
  \end{tabular}
};
\node[StepInit, below of=B9, node distance=\hdist](B10) {
  \begin{tabular}{@{}c@{}}
    {\bf Step 10: Reconstruct molecular ion(s)} \\
  \end{tabular}
};
\node[Output, below of=B10, node distance=\hdist](O1) {
  \begin{tabular}{@{}c@{}}
    {\bf Output:} \\
    chemical formulae of fragments, exact masses, assigned signal intensities\\
  \end{tabular}
};

\node[LeftInput, left of=B1, node distance=\ldist](L1) {
  \begin{tabular}{@{}c@{}}
    Exact mass \\
    and valence \\
    of atoms
  \end{tabular}
};
\node[LeftInput, left of=B3, node distance=\ldist](L2) {
  \begin{tabular}{@{}c@{}}
    environmental \\
    abundance of \\
    isotopes
  \end{tabular}
};

\draw[-Stealth] (L1) -- (B1) ;
\draw[-Stealth] (L2) -- (B3) ;

\draw[-Stealth] (A1) -- (B1) ;
\draw[-Stealth] (B1) -- (B2) ;
\draw[-Stealth] (B2) -- (B3) ;
\draw[-Stealth] (B3) -- (B4) ;
\draw[-Stealth] (B4) -- (B5) ;
\draw[-Stealth] (B5) -- (B6) ;
\draw[-Stealth] (B6) -- (B7) ;
\draw[-Stealth] (B7) -- (B8) ;
\draw[-Stealth] (B8) --node[right]{\begin{tabular}{@{}l@{}}more than 95\% of the measured \\signal is reconstructed\end{tabular}} (B9) ;
\draw[-Stealth] (B9) -- (B10) ;
\draw[-Stealth] (B10) -- (O1) ;

\path (node cs:name=B8,anchor=west) edge [-Stealth,bend left]
node[left]{\parbox{0.11\textwidth}{iterate until 95\% of the measured signal is reconstructed}}
(node cs:name=B5,anchor=west);

\end{tikzpicture}
\end{small}
\end{sffamily}
\caption{Overview of the method for automated identification of fragment formulae. Orange box (top): input measured data. Two mauve boxes (left): input chemical data \citep{Meija2016}. Yellow boxes are steps done just once. Steps 1 to 4: steps of initialisation. Green boxes, steps 5 to 8: steps repeated until a certain fraction of the measured signal has been reconstructed, here 95\%. Steps 9 and 10, yellow box: final steps, done just once. Blue box: output data, list of most likely fragments together with associated likelihood and ranking.}
\label{fig:method-diagram}
\end{figure}

\begin{figure}
  \begin{tikzpicture}
[circn/.style = {circle, minimum width=3em, inner sep=0pt, text centered},
  lb/.style = {fill=white, fill opacity=1, inner sep=0pt}]
\definecolor{sgl}{rgb}{1.0,0.9,0.1}
\definecolor{mxl}{rgb}{1.0,0.6,0.0}
\definecolor{nod}{rgb}{0.1,0.6,0.4}
\definecolor{leaf}{rgb}{0.34,0.78,0.4}
\begin{scriptsize}
\node[circn, fill=sgl] (n0) at (2.85,5.00) {$\mathsf{CSBr}$};
\node[circn, fill=sgl] (n1) at (4.11,5.00) {$\mathsf{C_2S_3}$};
\node[circn, fill=mxl] (n2) at (1.41,1.25) {$\mathsf{FS_2Cl}$};
\node[circn, fill=mxl] (n3) at (0.97,3.75) {$\mathsf{OSCl_2}$};
\node[circn, fill=mxl] (n4) at (5.36,5.00) {$\mathsf{HCCl_3}$};
\node[circn, fill=nod] (n5) at (4.73,3.75) {$\mathsf{CCl_3}$};
\node[circn, fill=mxl] (n6) at (7.24,3.75) {$\mathsf{HS_2Cl}$};
\node[circn, fill=mxl] (n7) at (8.49,3.75) {$\mathsf{HO_2SCl}$};
\node[circn, fill=mxl] (n8) at (2.23,3.75) {$\mathsf{NOCl_2}$};
\node[circn, fill=sgl] (n9) at (6.61,5.00) {$\mathsf{H_2S_3}$};
\node[circn, fill=mxl] (n10) at (3.48,3.75) {$\mathsf{COCl_2}$};
\node[circn, fill=nod] (n11) at (2.23,2.50) {$\mathsf{OCl_2}$};
\node[circn, fill=mxl] (n12) at (0.00,1.25) {$\mathsf{H_2FS_2}$};
\node[circn, fill=mxl] (n13) at (9.75,3.75) {$\mathsf{H_2OSCl}$};
\node[circn, fill=mxl] (n14) at (11.00,3.75) {$\mathsf{HNCl_2}$};
\node[circn, fill=mxl] (n15) at (3.48,2.50) {$\mathsf{CF_2Cl}$};
\node[circn, fill=leaf] (n16) at (0.54,0.00) {$\mathsf{FS_2}$};
\node[circn, fill=nod] (n17) at (5.99,3.75) {$\mathsf{HCCl_2}$};
\node[circn, fill=nod] (n18) at (4.73,2.50) {$\mathsf{CCl_2}$};
\node[circn, fill=sgl] (n19) at (7.87,5.00) {$\mathsf{COS}$};
\node[circn, fill=nod] (n20) at (3.76,1.25) {$\mathsf{CCl}$};
\node[circn, fill=nod] (n21) at (7.87,2.50) {$\mathsf{HCl}$};
\node[circn, fill=leaf] (n22) at (2.96,0.00) {$\mathsf{Cl}$};
\draw (n2) --node[lb]{$\mathsf{Cl}$} (n16);
\draw (n2) --node[lb]{$\mathsf{FS_2}$} (n22);
\draw (n3) --node[lb]{$\mathsf{S}$} (n11);
\draw (n4) --node[lb]{$\mathsf{H}$} (n5);
\draw (n4) --node[lb]{$\mathsf{Cl}$} (n17);
\draw (n5) --node[lb]{$\mathsf{Cl}$} (n18);
\draw (n6) --node[lb]{$\mathsf{S_2}$} (n21);
\draw (n7) --node[lb]{$\mathsf{O_2S}$} (n21);
\draw (n8) --node[lb]{$\mathsf{N}$} (n11);
\draw (n10) --node[lb]{$\mathsf{C}$} (n11);
\draw (n10) --node[lb]{$\mathsf{O}$} (n18);
\draw (n11) --node[lb]{$\mathsf{OCl}$} (n22);
\draw (n12) --node[lb]{$\mathsf{H_2}$} (n16);
\draw (n13) --node[lb]{$\mathsf{HOS}$} (n21);
\draw (n14) --node[lb]{$\mathsf{NCl}$} (n21);
\draw (n15) --node[lb]{$\mathsf{F_2}$} (n20);
\draw (n17) --node[lb]{$\mathsf{H}$} (n18);
\draw (n17) --node[lb]{$\mathsf{CCl}$} (n21);
\draw (n18) --node[lb]{$\mathsf{Cl}$} (n20);
\draw (n20) --node[lb]{$\mathsf{C}$} (n22);
\draw (n21) --node[lb]{$\mathsf{H}$} (n22);
\end{scriptsize}
\end{tikzpicture}

  \caption{Directed acyclic pseudo-fragmentation graph obtained in \ref{step2}, with all the
    candidate fragments (nodes) from the knapsack algorithm for CCl\Sub{4}. One
    observes 23 nodes, with 2 leaves (or smallest possible fragments, in light green), 15 maximal fragments (in orange and yellow), of which 4 have no children and are therefore singletons (in yellow). The latter are eliminated in
    \ref{step2}.}
  \label{fig:dag-CCl4}
\end{figure}

\begin{figure}
  \centering
  \begin{tikzpicture}[scale=1.0]
\pgfkeys{
    /pgf/number format/set thousands separator = {}}
\begin{axis}
[
grid = none,
width = 0.6\textwidth,
height = 0.4\textwidth,
scale only axis,
axis y line=none,
axis x line*=top,
enlargelimits = false,
xmin = 151.1,
xmax = 161.9,
ymin = 0,
ymax = 1.35,
xtick={152,153,154,...,161},
xticklabels={CCl\Sub{4},[\Sup{13}C]Cl\Sub{4},
  CCl\Sub{3}[\Sup{37}Cl],[\Sup{13}C]Cl\Sub{3}[\Sup{37}Cl],
  CCl\Sub{2}[\Sup{37}Cl]\Sub{2},[\Sup{13}C]Cl\Sub{2}[\Sup{37}Cl]\Sub{2},
  CCl[\Sup{37}Cl]\Sub{3},[\Sup{13}C]Cl[\Sup{37}Cl]\Sub{3},
  C[\Sup{37}Cl]\Sub{4},[\Sup{13}C][\Sup{37}Cl]\Sub{4}},
xtick align=outside,
every x tick label/.append style={above,anchor=south west,rotate=45},
]
\end{axis}
\begin{axis}
[
grid = none,
width = 0.6\textwidth,
height = 0.4\textwidth,
scale only axis,
xlabel = {Mass (m/z)},
axis y line*=left,
ylabel = {relative intensity w.r.t.~CCl\Sub{4}},
enlargelimits = false,
xmin = 151.1,
xmax = 161.9,
ymin = 0,
ymax = 1.35,
xtick={152,153,154,...,161},
ybar,
bar width=5pt,
]
\addplot
table[x index=1,y index=3, header=true]{
fragment                                   mass       abundance relative intensity w.r.t. monoisotopic
{CCl\Sub{4}                             } 151.875411 0.325859 1.000000
{CCl\Sub{3}[\Sup{37}Cl]                 } 153.872461 0.416938 1.279504
{CCl\Sub{2}[\Sup{37}Cl]\Sub{2}          } 155.869511 0.200052 0.613923
{CCl[\Sup{37}Cl]\Sub{3}                 } 157.866561 0.042661 0.130920
{C[\Sup{37}Cl]\Sub{4}                   } 159.863610 0.003412 0.010470
{[\Sup{13}C]Cl\Sub{4}                   } 152.878766 0.003650 0.011202
{[\Sup{13}C]Cl\Sub{3}[\Sup{37}Cl]       } 154.875816 0.004671 0.014333
{[\Sup{13}C]Cl\Sub{2}[\Sup{37}Cl]\Sub{2}} 156.872865 0.002241 0.006877
{[\Sup{13}C]Cl[\Sup{37}Cl]\Sub{3}       } 158.869915 0.000478 0.001467
{[\Sup{13}C][\Sup{37}Cl]\Sub{4}         } 160.866965 0.000038 0.000117
};
\node at (151.87,0) {CCl\Sub{4}};
\end{axis}
\end{tikzpicture}
\caption{isotopocules of CCl\Sub{4} with mass and relative intensity
  w.r.t.~CCl\Sub{4}. The abundant formula CCl\Sub{4} has one isotopocule
CCl\Sub{3}[\Sup{37}Cl] of relative intensity greater than one (1.279504). See Table~\refsuptabisotopocules{} in the Supplement for numerical data. }
\label{fig:isotopocules-CCl4}
\end{figure}


\begin{figure}[htb]
  \centering
\includegraphics[width=12cm]{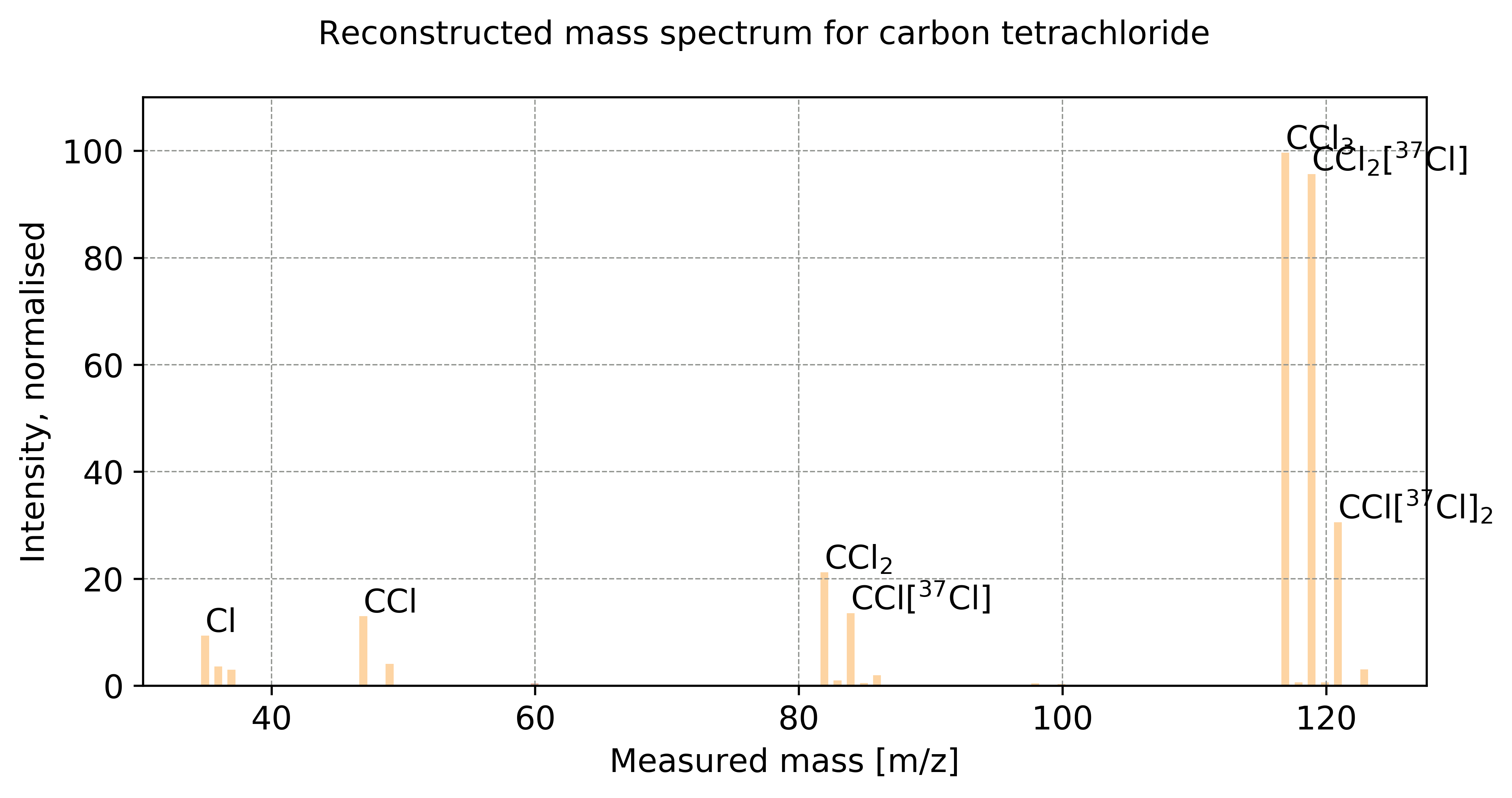}
\caption{Reconstructed mass spectrum for CCl\Sub{4},
when setting as target that 95\% of the signal should be reconstructed. 
Numerical values can be found in the supplementary material.}
\label{fig:CCl4-mass-spectrum}
\end{figure}



\begin{figure}[htb]
  \centering
\includegraphics[width=10cm]{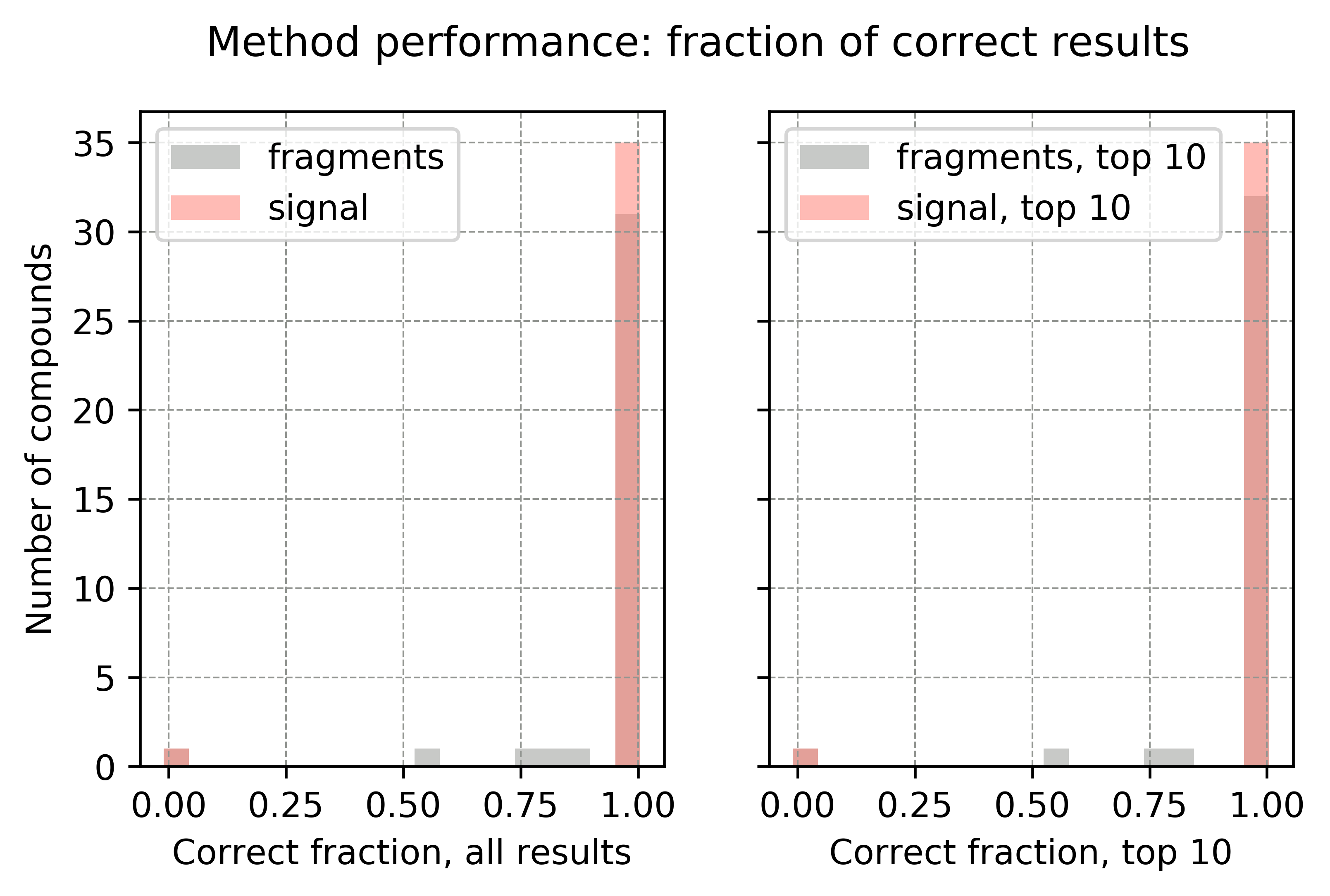}
\caption{Performance of the identification algorithm: fraction of correct reconstructed fragments and signal. A fragment is considered correct if its chemical formula is a subset of the chemical formula of the true molecular ion. The histogram for fragments is shown in grey and for signal in peach. Left: fraction of correct reconstructed fragments compared to all reconstructed fragments (grey); fraction of correct reconstructed signal compared to the sum of reconstructed signal (peach). Right: fraction of correct fragments from the top-10 likelihood list of fragments (grey); fraction of the associated correct signal compared to the signal reconstructed by the top-10 likelihood fragments (peach). If the number of reconstructed fragments is not more than 10, then the top-10 results have same value as the results considering all fragments. Two substances have fragments poorly identified: CF\Sub{4} and SO\Sub{2}F\Sub{2}. See text for discussion.}
\label{fig:accuracy-sum-formula}
\end{figure}

\begin{figure}[htb]
  \centering
\includegraphics[width=10cm]{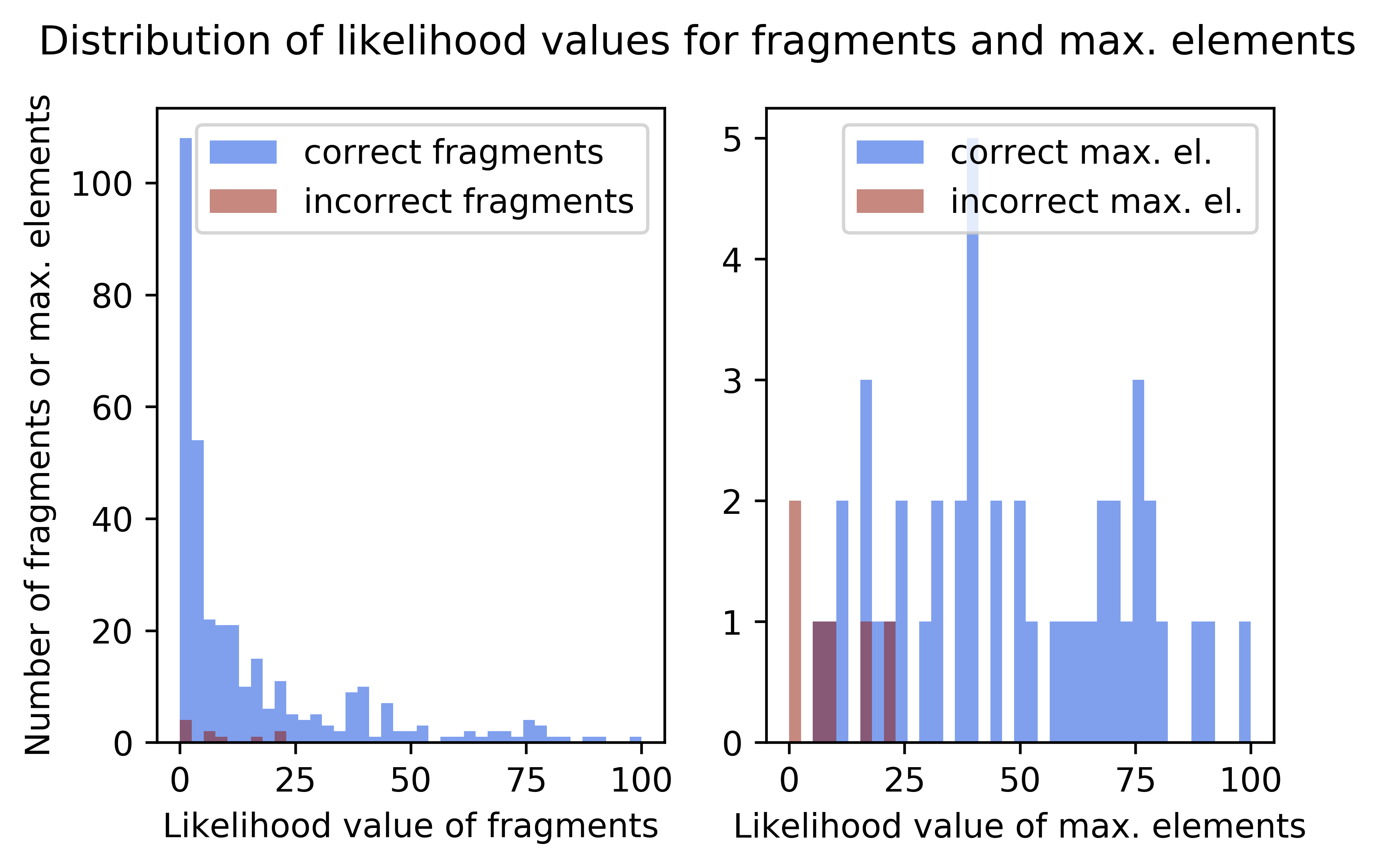}
\caption{Training set: distribution of likelihood values of fragments (left) and maximal fragments (right). A likelihood value of 100 indicates that the chemical formula of the fragment or maximal fragment is highly likely. Blue: distribution for correctly identified fragments/maximal fragments. Red: distribution for wrongly identified fragments/maximal fragments. In total, there were 353 reconstructed fragments, 343 correct and 10 wrong, and 50 maximal fragments, 44 correct and 6 wrong. Above a likelihood value of 20, \textgreater95\% of the fragments are correct, and \textgreater90\% of the maximal fragments.}
\label{fig:hist-likelihood-frag-att}
\end{figure}

\begin{figure}[htb]
  \centering
\includegraphics[width=10cm]{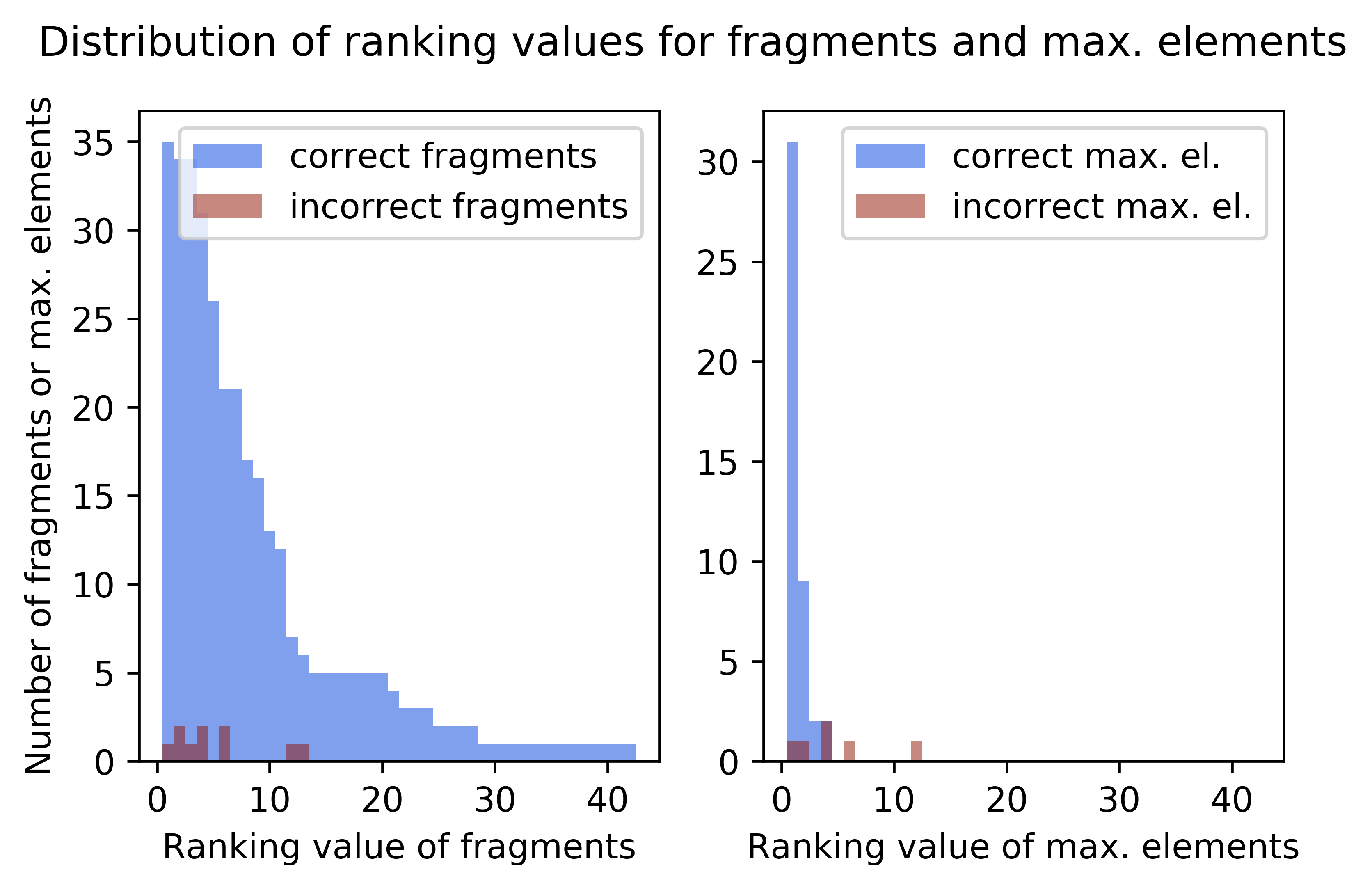}
\caption{Training set: distribution of ranking values for fragments and maximal fragments. A ranking value of 1 means that the fragment/maximal fragment was ranked as most likely (maximum likelihood value within the set of fragments/maximal fragment). Blue: distribution of ranking for correctly identified fragments/maximal fragments. Red: distribution of ranking for wrongly identified fragments/maximal fragments.  }
\label{fig:hist-ranking-frag-att}
\end{figure}

\begin{figure}[htb]
  \centering
\includegraphics[width=8cm]{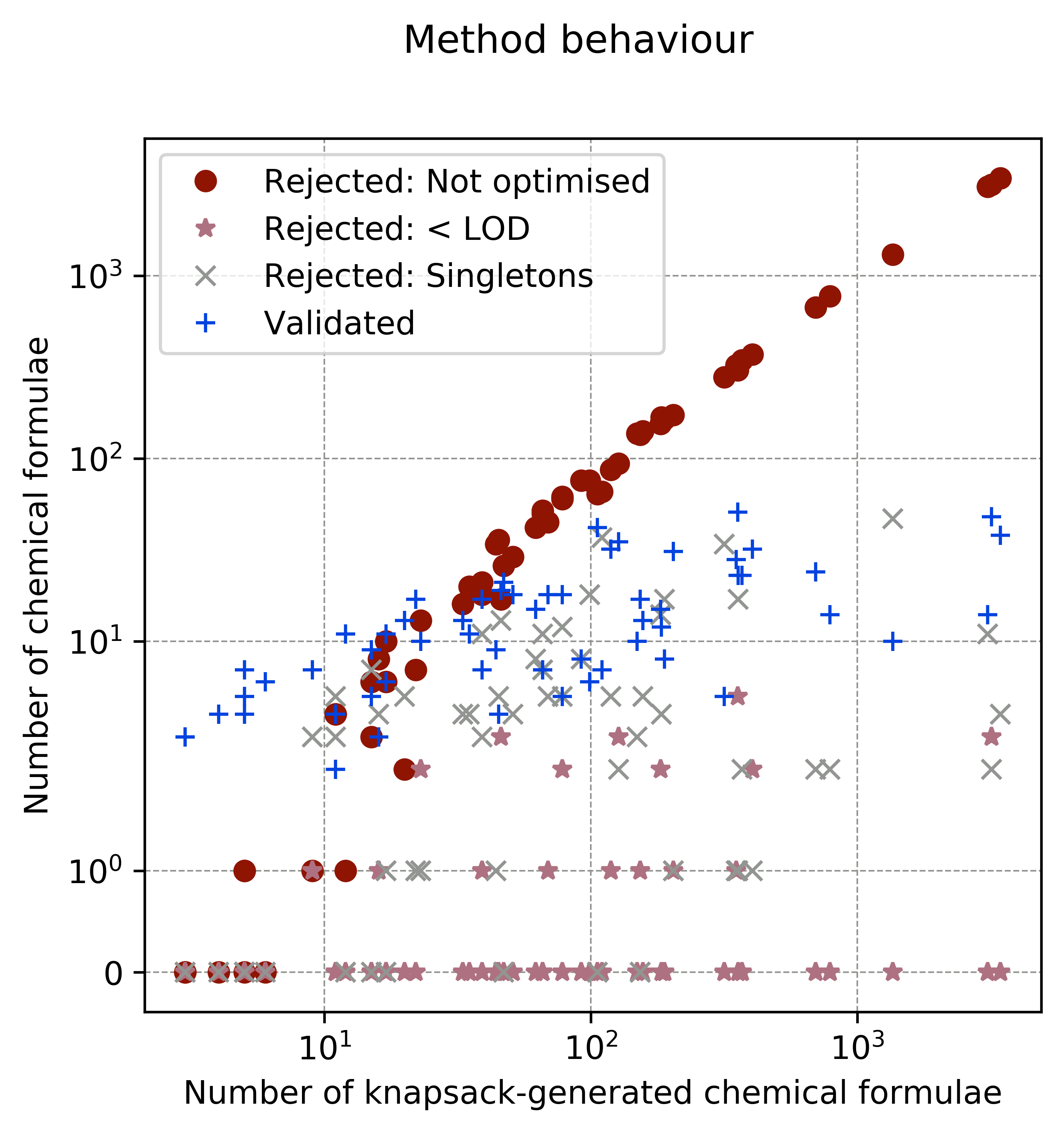}
\caption{Behaviour of the identification algorithm for the training and validation sets: how knapsack solutions are rejected.}
\label{fig:method-behaviour}
\end{figure}

\begin{figure}[htb]
  \centering
\includegraphics[width=8cm]{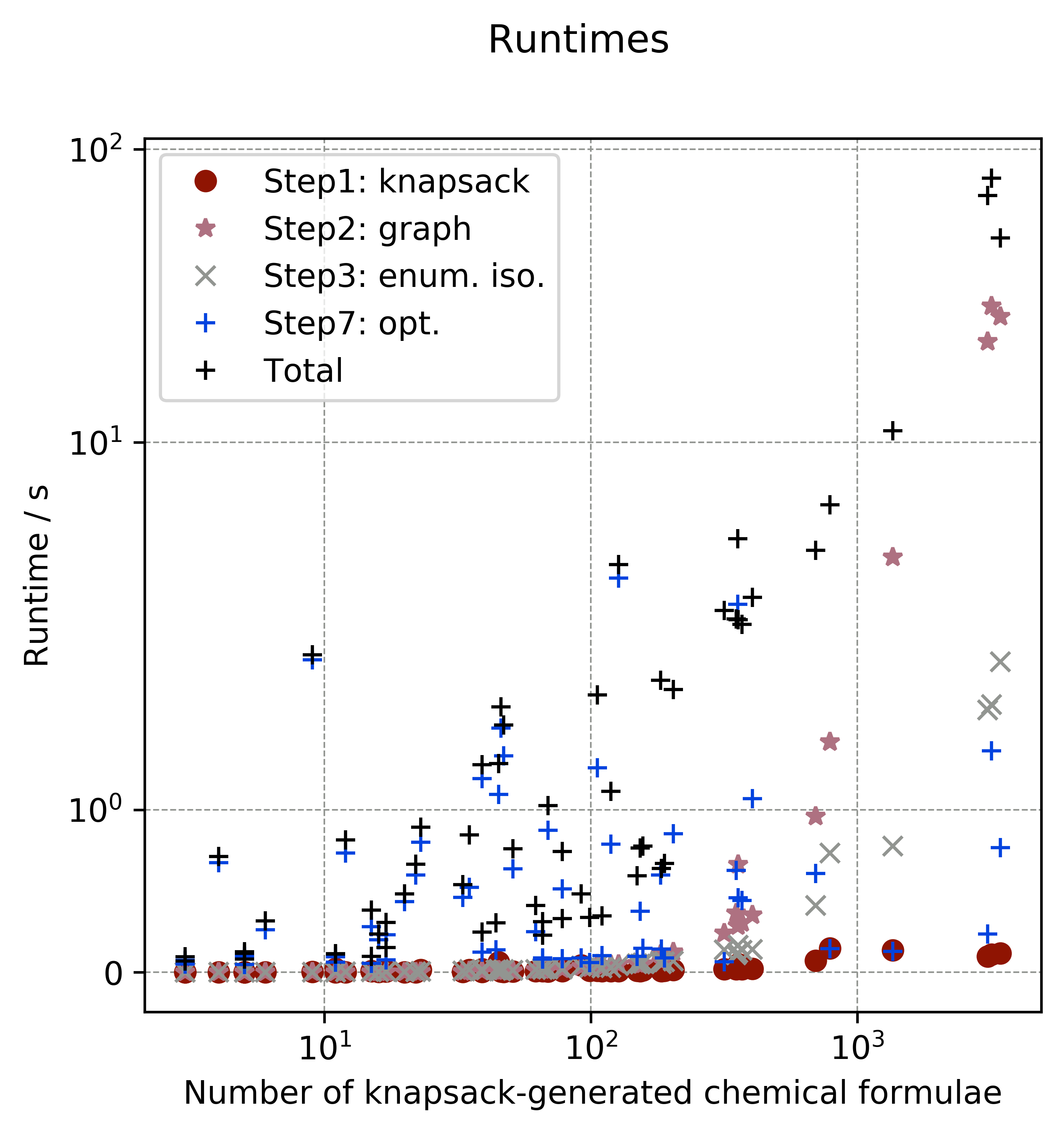}
\caption{Runtimes of the identification algorithm  for the training and validation sets. The total runtime per compound is shown in black. The runtime of specific steps is also depicted: for the knapsack (Step~1), for the graph construction with all knapsack fragments (Step~2), for the enumeration of all minor-isotope chemical formulae above the LOD (Step~3), for the optimisation of contribution of sets of fragments, using a machine learning algorithm (Step~7). For most compounds, Step~7 remains the most time intensive step. The corresponding numerical values are given in Tables~\ref{tab:runtimes} and \ref{tab:runtimes-validation}.}
\label{fig:method-runtime}
\end{figure}

\begin{figure}[htb]
  \centering
\includegraphics[width=10cm]{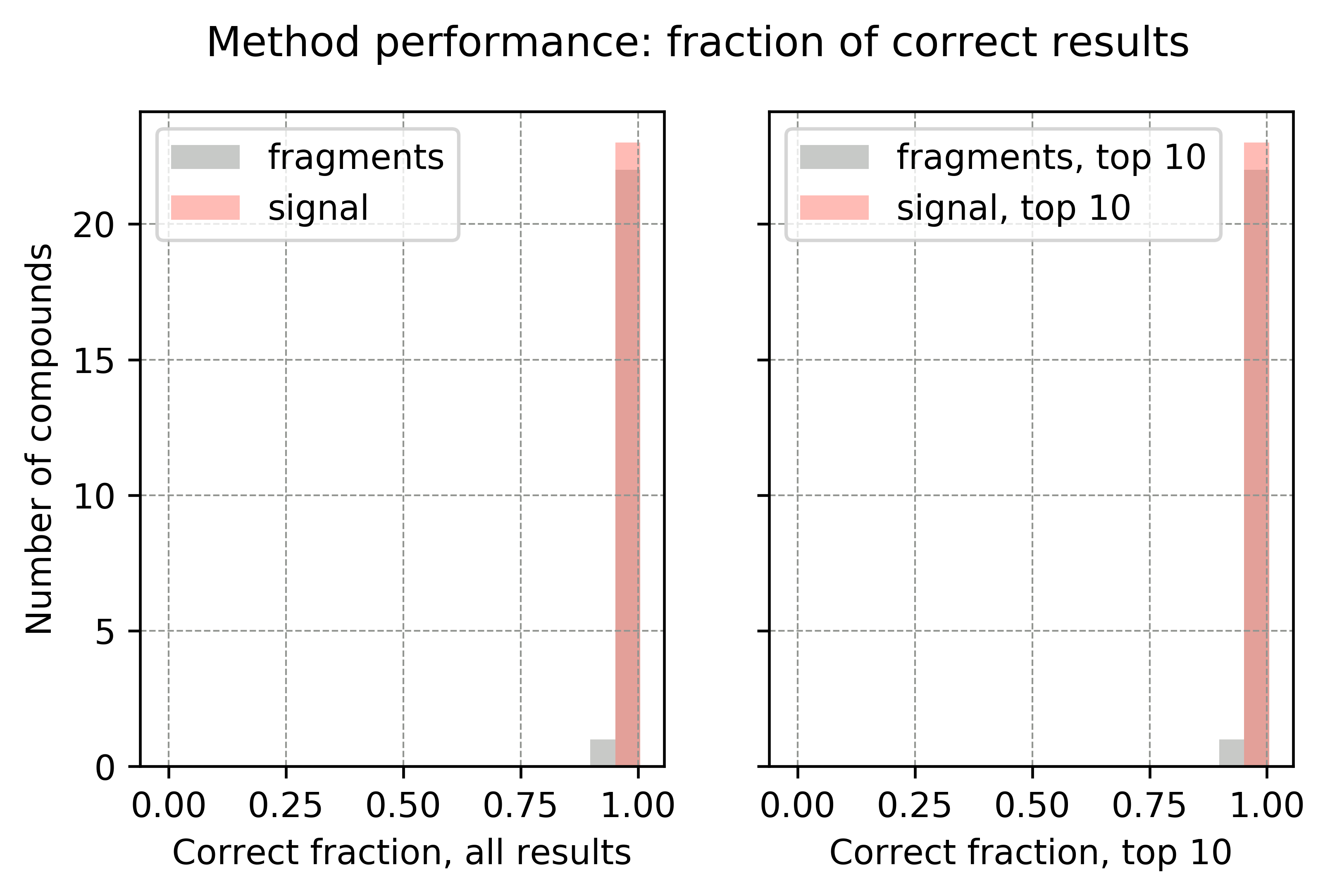}
\caption{Performance of the identification algorithm on the validation set (21 compounds): fraction of correct reconstructed fragments and signal. A fragment is considered correct if its chemical formula is a subset of the chemical formula of the true molecular ion. The histogram for fragments is shown in grey and for signal in peach. Left: fraction of correct reconstructed fragments compared to all reconstructed fragments (grey); fraction of correct reconstructed signal compared to the sum of reconstructed signal (peach). Right: fraction of correct fragments from the top-10 likelihood list of fragments (grey); fraction of the associated correct signal compared to the signal reconstructed by the top-10 likelihood fragments (peach). If the number of reconstructed fragments is not more than 10, then the top-10 results have same value as the results considering all fragments.}
\label{fig:val-accuracy-formula-signal}
\end{figure}

\begin{figure}[htb]
  \centering
\includegraphics[width=10cm]{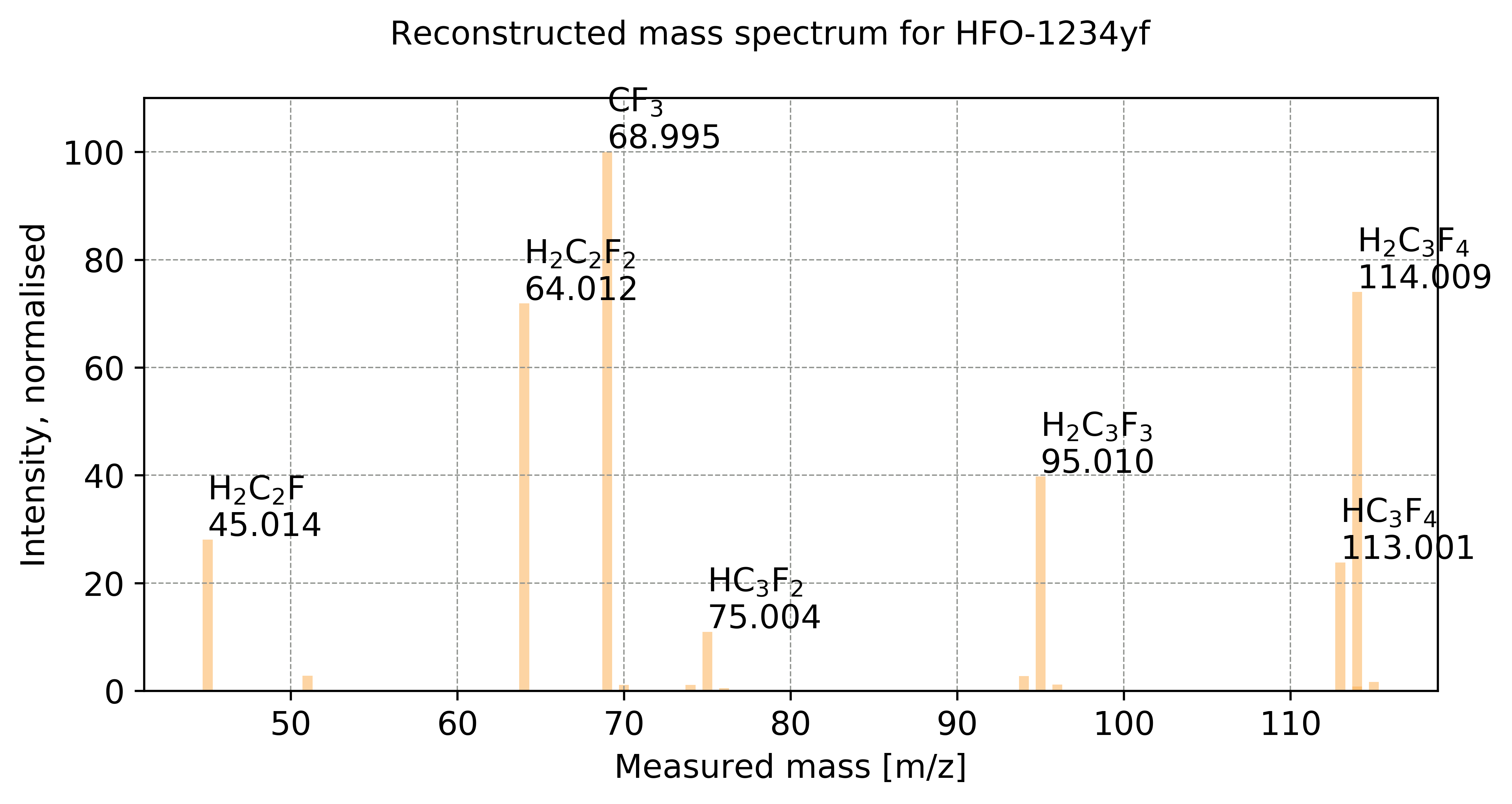}
\caption{Reconstructed mass spectrum for HFO-1234yf,
when setting as target that 95\% of the signal should be reconstructed. 
Numerical values can be found in the supplementary material.}
\label{fig:HFO1234yf-massspectrum}
\end{figure}


\clearpage

\section*{Tables}

\begin{table}[ht]
  \caption{Known compounds used as training set. These 36 substances are routinely measured within the AGAGE network \citep{Prinn2018}. Identification and quantification of these compounds have been done by \citep{Prinn2000, Arnold2012, Vollmer2018, Guillevic2018}. Present chemical elements are: H, C, N, O, F, S, Cl, Br and I. These are the chemical elements used as input for the knapsack algorithm. SMILES codes can be found in the Supplement.}
  \label{tab:training-set}
  \begin{tabular}{l l l r r }
\hline
Compound	&	Chemical name	 &	\begin{tabular}{@{}l@{}}Chemical \\formula\end{tabular}	&	\begin{tabular}{@{}c@{}}Monoisotopic \\molecular mass \\Da \end{tabular}	& \begin{tabular}{@{}c@{}} CAS \\number\end{tabular} \\
\hline

C\Sub{2}H\Sub{6}	&	ethane	&	C\Sub{2}H\Sub{6}	&	30.04695	&	74-84-0	\\
C\Sub{3}H\Sub{8}	&	propane	&	C\Sub{3}H\Sub{8}	&	44.0626	&	74-98-6	\\
CH\Sub{3}Cl	&	chloromethane	&	CH\Sub{3}Cl	&	49.99233	&	74-87-3	\\
COS	&	carbonyl sulphide	&	COS	&	59.96699	&	463-58-1	\\
NF\Sub{3}	&	nitrogen trifluoride	&	NF\Sub{3}	&	70.99828	&	7783-54-2	\\
Benzene	&	benzene	&	C\Sub{6}H\Sub{6}	&	78.04695	&	71-43-2	\\
CH\Sub{2}Cl\Sub{2}	&	dichloromethane	&	CH\Sub{2}Cl\Sub{2}	&	83.95336	&	75-09-2	\\
HCFC-22	&	chlorodifluoromethane	&	HCF\Sub{2}Cl	&	85.97348	&	75-45-6	\\
CF\Sub{4}	&	tetrafluoromethane	&	CF\Sub{4}	&	87.99361	&	75-73-0	\\
Toluene	&	toluene	&	C\Sub{7}H\Sub{8}	&	92.0626	&	108-88-3	\\
CH\Sub{3}Br	&	bromomethane	&	CH\Sub{3}Br	&	93.94181	&	74-83-9	\\
HCFC-142b	&	1-chloro-1,1-difluoroethane	&	H\Sub{3}C\Sub{2}F\Sub{2}Cl	&	99.98913	&	75-68-3	\\
SO\Sub{2}F\Sub{2}	&	sulfuryl difluoride	&	SO\Sub{2}F\Sub{2}	&	101.95871	&	2699-79-8	\\
CFC-13	&	chlorotrifluoromethane	&	CF\Sub{3}Cl	&	103.96406	&	75-72-9	\\
HCFC-141b	&	1,1-dichloro-1-fluoroethane	&	H\Sub{3}C\Sub{2}FCl\Sub{2}	&	115.95958	&	1717-00-6	\\
CHCl\Sub{3}	&	chloroform	&	CHCl\Sub{3}	&	117.91438	&	67-66-3	\\
CFC-12	&	dichlorodifluoromethane	&	CF\Sub{2}Cl\Sub{2}	&	119.93451	&	75-71-8	\\
C\Sub{2}HCl\Sub{3}	&	1,1,2-trichloroethene	&	C\Sub{2}HCl\Sub{3}	&	129.91438	&	79-01-6	\\
CFC-11	&	trichlorofluoromethane	&	CFCl\Sub{3}	&	135.90496	&	75-69-4	\\
HCFC-124	&	2-chloro-1,1,1,2-tetrafluoroethane	&	HC\Sub{2}F\Sub{4}Cl	&	135.97029	&	2837-89-0	\\
PFC-116	&	perfluoroethane	&	C\Sub{2}F\Sub{6}	&	137.99042	&	76-16-4	\\
CH\Sub{3}I	&	iodomethane	&	CH\Sub{3}I	&	141.92795	&	74-88-4	\\
SF\Sub{6}	&	sulfur hexafluoride	&	SF\Sub{6}	&	145.96249	&	2551-62-4	\\
Halon-1301	&	bromo(trifluoro)methane	&	CF\Sub{3}Br	&	147.91355	&	75-63-8	\\
CCl\Sub{4}	&	tetrachloromethane	&	CCl\Sub{4}	&	151.87541	&	56-23-5	\\
CFC-115	&	1-chloro-1,1,2,2,2-pentafluoroethane	&	C\Sub{2}F\Sub{5}Cl	&	153.96087	&	76-15-3	\\
C\Sub{2}Cl\Sub{4}	&	1,1,2,2-tetrachloroethene	&	C\Sub{2}Cl\Sub{4}	&	163.87541	&	127-18-4	\\
Halon-1211	&	bromochlorodifluoromethane	&	CF\Sub{2}ClBr	&	163.884	&	353-59-3	\\
CFC-114	&	1,2-dichloro-1,1,2,2-tetrafluoroethane	&	C\Sub{2}F\Sub{4}Cl\Sub{2}	&	169.93132	&	76-14-2	\\
CH\Sub{2}Br\Sub{2}	&	dibromomethane	&	CH\Sub{2}Br\Sub{2}	&	171.85233	&	74-95-3	\\
CFC-113	&	1,1,2-trichloro-1,2,2-trifluoroethane	&	C\Sub{2}F\Sub{3}Cl\Sub{3}	&	185.90177	&	76-13-1	\\
PFC-218	&	perfluoropropane	&	C\Sub{3}F\Sub{8}	&	187.98723	&	76-19-7	\\
SF\Sub{5}CF\Sub{3}	&	pentafluoro(trifluoromethyl)sulfur	&	SF\Sub{5}CF\Sub{3}	&	195.9593	&	373-80-8	\\
PFC-c318	&	octafluorocyclobutane	&	C\Sub{4}F\Sub{8}	&	199.98723	&	115-25-3	\\
Halon-2402	&	1,2-dibromo-1,1,2,2-tetrafluoroethane	&	C\Sub{2}F\Sub{4}Br\Sub{2}	&	257.83029	&	124-73-2	\\
C\Sub{6}F\Sub{14}	&	perfluorohexane	&	C\Sub{6}F\Sub{14}	&	337.97764	&	355-42-0	\\

\hline
  \end{tabular}
\end{table}

\begin{table}[ht]
  \caption{Known compounds used as validation set. Identification and quantification of these 23 compounds has been done by \citep{Vollmer2015a, Vollmer2015b, Guillevic2018, Reimann2020}.
    Present chemical elements: H, C, N, O, F, Cl.
    Chemical elements used as input for the knapsack algorithm are the same as for the
    training set: H, C, N, O, F, S, Cl, Br and I.
		SMILES codes can be found in the Supplement.
    }
    \label{tab:validation-set}
    \resizebox{\textwidth}{!}{%
  \begin{tabular}{lllrr}
\hline
Compound	&	Chemical name	&	\begin{tabular}{@{}l@{}}Chemical\\ formula\end{tabular}	&	\begin{tabular}{@{}c@{}}Monoisotopic \\molecular mass \\Da \end{tabular}	& \begin{tabular}{@{}c@{}} CAS \\number\end{tabular} \\
\hline
\multicolumn{4}{c}{Kigali Amendment to the Montreal Protocol}\\
\hline

HFC-41	&	fluoromethane	&	CH\Sub{3}F	&	34.021878	&	593-53-3	\\
HFC-32	&	difluoromethane	&	CH\Sub{2}F\Sub{2}	&	52.012456	&	75-10-5	\\
HFC-152	&	1,2-difluoroethane	&	C\Sub{2}H\Sub{4}F\Sub{2}	&	66.028106	&	624-72-6	\\
HFC-152a	&	1,1-difluoroethane	&	C\Sub{2}H\Sub{4}F\Sub{2}	&	66.028106	&	75-37-6	\\
HFC-23	&	fluoroform	&	CHF\Sub{3}	&	70.003035	&	75-46-7	\\
HFC-143	&	1,1,2-trifluoroethane	&	C\Sub{2}H\Sub{3}F\Sub{3}	&	84.018685	&	430-66-0	\\
HFC-143a	&	1,1,1-trifluoroethane	&	C\Sub{2}H\Sub{3}F\Sub{3}	&	84.018685	&	420-46-2	\\
HFC-134	&	1,1,2,2-tetrafluoroethane	&	C\Sub{2}H\Sub{2}F\Sub{4}	&	102.009263	&	359-35-3	\\
HFC-134a	&	1,1,1,2-tetrafluoroethane	&	C\Sub{2}H\Sub{2}F\Sub{4}	&	102.009263	&	811-97-2	\\
HFC-125	&	pentafluoroethane	&	C\Sub{2}HF\Sub{5}	&	119.999841	&	354-33-6	\\
HFC-245ca	&	1,1,2,2,3-pentafluoropropane	&	C\Sub{3}H\Sub{3}F\Sub{5}	&	134.015491	&	679-86-7	\\
HFC-245fa	&	1,1,1,3,3-pentafluoropropane	&	C\Sub{3}H\Sub{3}F\Sub{5}	&	134.015491	&	460-73-1	\\
HFC-365mfc	&	1,1,1,3,3-pentafluorobutane	&	C\Sub{4}H\Sub{5}F\Sub{5}	&	148.031141	&	406-58-6	\\
HFC-236cb	&	1,1,1,2,2,3-hexafluoropropane	&	C\Sub{3}H\Sub{2}F\Sub{6}	&	152.006069	&	677-56-5	\\
HFC-236ea	&	1,1,1,2,3,3-hexafluoropropane	&	C\Sub{3}H\Sub{2}F\Sub{6}	&	152.006069	&	431-63-0	\\
HFC-236fa	&	1,1,1,3,3,3-hexafluoropropane	&	C\Sub{3}H\Sub{2}F\Sub{6}	&	152.006069	&	690-39-1	\\
HFC-227ea	&	1,1,1,2,3,3,3-heptafluoropropane	&	C\Sub{3}HF\Sub{7}	&	169.996647	&	431-89-0	\\
HFC-43-10mee	&	1,1,1,2,2,3,4,5,5,5-decafluoropentane	&	C\Sub{5}H\Sub{2}F\Sub{10}	&	251.999682	&	138495-42-8	\\

\hline
\multicolumn{4}{c}{HFOs}					\\
\hline

HFO-1234yf	&	2,3,3,3-tetrafluoroprop-1-ene	&	H\Sub{2}C\Sub{3}F\Sub{4}	&	114.009263	&	754-12-1	\\
HFO-1234ze(E)	&	(E)-1,3,3,3-tetrafluoroprop-1-ene	&	H\Sub{2}C\Sub{3}F\Sub{4}	&	114.009263	&	29118-24-9	\\
HCFO-1233zd(E)	&	(E)-1-chloro-3,3,3-trifluoro prop-1-ene	&	H\Sub{2}C\Sub{3}F\Sub{3}Cl	&	129.979712	&	102687-65-0	\\

\hline
\multicolumn{4}{c}{Halogenated compounds with high boiling point}							\\
\hline
HCBD	& 1,1,2,3,4,4-hexachlorobuta-1,3-diene & C\Sub{4}Cl\Sub{6} & 257.813116 & 87-68-3 \\
TCHFB	& 1,2,3,4‐Tetrachlorohexafluorobutane & C\Sub{4}Cl\Sub{4}F\Sub{6} & 301.865830  & 375-45-1 \\

\hline
  \end{tabular}%
  }
\end{table}

\begin{table}[ht]
  \caption{Known compounds used as training set: presence of the molecular ion. If the molecular ion is absent, we give the detected maximal fragments instead. Note that several maximal fragments may be detected for one substance. The last column indicates the ranking of the correct molecular ion, if reconstructed by our method, or which molecular ion(s) is (are) reconstructed (if any).}
  \label{tab:training-set-mol-ion}
  \begin{tabular}{l l l l }
\hline
Compound	&	\begin{tabular}{@{}l@{}}Chemical \\formula \end{tabular} &	\begin{tabular}{@{}l@{}}Molecular \\ion present \end{tabular}	& \begin{tabular}{@{}l@{}}Reconstructed \\mol. ion \end{tabular} \\
\hline

C\Sub{2}H\Sub{6}	&	C\Sub{2}H\Sub{6}	&	yes	&	1	\\
C\Sub{3}H\Sub{8}	&	C\Sub{3}H\Sub{8}	&	yes	&	1	\\
CH\Sub{3}Cl	&	CH\Sub{3}Cl	&	yes	&	1	\\
COS	&	COS	&	yes	&	1	\\
NF\Sub{3}	&	NF\Sub{3}	&	yes	&	1	\\
Benzene	&	C\Sub{6}H\Sub{6}	&	yes	&	1	\\
CH\Sub{2}Cl\Sub{2}	&	CH\Sub{2}Cl\Sub{2}	&	yes	&	1	\\
HCFC-22	&	HCF\Sub{2}Cl	&	yes	&	1	\\
CF\Sub{4}	&	CF\Sub{4}	&	CF\Sub{3}	&	none	\\
Toluene	&	C\Sub{7}H\Sub{8}	&	yes	&	1	\\
CH\Sub{3}Br	&	CH\Sub{3}Br	&	yes	&	1	\\
HCFC-142b	&	H\Sub{3}C\Sub{2}F\Sub{2}Cl	&	H\Sub{2}C\Sub{2}F\Sub{2}Cl, H\Sub{3}C\Sub{2}FCl, H\Sub{3}C\Sub{2}F\Sub{2}	&	1	\\
SO\Sub{2}F\Sub{2}	&	SO\Sub{2}F\Sub{2}	&	yes	&	O3FCl	\\
CFC-13	&	CF\Sub{3}Cl	&	CF\Sub{2}Cl, CF\Sub{3}	&	1	\\
HCFC-141b	&	H\Sub{3}C\Sub{2}FCl\Sub{2}	&	H\Sub{2}C\Sub{2}Cl\Sub{2}, H\Sub{3}C\Sub{2}FCl	&	2 (1: C\Sub{2}H\Sub{4}FCl)	\\
CHCl\Sub{3}	&	CHCl\Sub{3}	&	yes	&	1	\\
CFC-12	&	CF\Sub{2}Cl\Sub{2}	&	yes	&	1	\\
C\Sub{2}HCl\Sub{3}	&	C\Sub{2}HCl\Sub{3}	&	yes	&	1	\\
CFC-11	&	CFCl\Sub{3}	&	yes	&	1	\\
HCFC-124	&	HC\Sub{2}F\Sub{4}Cl	&	yes	&	1	\\
PFC-116	&	C\Sub{2}F\Sub{6}	&	C\Sub{2}F\Sub{5}	&	1	\\
CH\Sub{3}I	&	CH\Sub{3}I	&	yes	&	1	\\
SF\Sub{6}	&	SF\Sub{6}	&	SF\Sub{5}	&	1	\\
Halon-1301	&	CF\Sub{3}Br	&	yes	&	1	\\
CCl\Sub{4}	&	CCl\Sub{4}	&	CCl\Sub{3}	&	2 (1: CHCl\Sub{3})	\\
CFC-115	&	C\Sub{2}F\Sub{5}Cl	&	C\Sub{2}F\Sub{4}Cl, C\Sub{2}F\Sub{5}	&	1	\\
CCl\Sub{2}=CCl\Sub{2}	&	C\Sub{2}Cl\Sub{4}	&	yes	&	1	\\
Halon-1211	&	CF\Sub{2}ClBr	&	CFClBr, CF\Sub{2}Br, CF\Sub{2}Cl	&	1	\\
CFC-114	&	C\Sub{2}F\Sub{4}Cl\Sub{2}	&	C\Sub{2}F\Sub{3}Cl\Sub{2}, C\Sub{2}F\Sub{4}Cl	&	1	\\
CH\Sub{2}Br\Sub{2}	&	CH\Sub{2}Br\Sub{2}	&	yes	&	1	\\
CFC-113	&	C\Sub{2}F\Sub{3}Cl\Sub{3}	&	yes	&	1	\\
PFC-218	&	C\Sub{3}F\Sub{8}	&	C\Sub{3}F\Sub{7}	&	1	\\
SF\Sub{5}CF\Sub{3}	&	SF\Sub{5}CF\Sub{3}	&	SF\Sub{5}, CF\Sub{3}	&	SF\Sub{5}, CF\Sub{4}	\\
PFC-c318	&	C\Sub{4}F\Sub{8}	&	C\Sub{3}F\Sub{5}	&	C\Sub{3}F\Sub{6}	\\
Halon-2402	&	C\Sub{2}F\Sub{4}Br\Sub{2}	&	yes	&	1	\\
C\Sub{6}F\Sub{14}	&	C\Sub{6}F\Sub{14}	&	C\Sub{5}F\Sub{9}	&	C\Sub{4}F\Sub{10}	\\

\hline
  \end{tabular}
\end{table}

\begin{table}[ht]
  \caption{Known compounds used as validation set: presence of the molecular ion. If the molecular ion is absent, we give the detected maximal fragments instead. Note that several maximal fragments may be detected for one substance. The last column indicates if the correct molecular ion is reconstructed by our method, with its ranking in parenthesis, or which molecular ion(s) is reconstructed (if any - for brevity only the two first ranked wrong molecular ions are reported).}
    
    \label{tab:validation-set-mol-ion}
  \begin{tabular}{l l l l}
\hline
Compound	& \begin{tabular}{@{}l@{}}Chemical \\formula \end{tabular}	 &	\begin{tabular}{@{}l@{}}Molecular \\ion present \end{tabular}	& \begin{tabular}{@{}l@{}}Reconstructed \\mol. ion \end{tabular} \\
\hline
\multicolumn{4}{c}{Kigali Amendment to the Montreal Protocol}\\
\hline
HFC-41	&	CH\Sub{3}F	&	yes	&	1	\\
HFC-32	&	CH\Sub{2}F\Sub{2}	&	yes	&	1	\\
HFC-152	&	C\Sub{2}H\Sub{4}F\Sub{2}	&	yes	&	1	\\
HFC-152a	&	C\Sub{2}H\Sub{4}F\Sub{2}	&	yes	&	1	\\
HFC-23	&	CHF\Sub{3}	&	CF\Sub{3}, HCF\Sub{2}	&	1	\\
HFC-143	&	C\Sub{2}H\Sub{3}F\Sub{3}	&	yes	&	1	\\
HFC-143a	&	C\Sub{2}H\Sub{3}F\Sub{3}	&	yes	&	1	\\
HFC-134	&	C\Sub{2}H\Sub{2}F\Sub{4}	&	yes	&	1	\\
HFC-134a	&	C\Sub{2}H\Sub{2}F\Sub{4}	&	yes	&	1	\\
HFC-125	&	C\Sub{2}HF\Sub{5}	&	C\Sub{2}F\Sub{5}, HC\Sub{2}F\Sub{4}	&	1	\\
HFC-245ca	&	C\Sub{3}H\Sub{3}F\Sub{5}	&	C\Sub{3}H\Sub{2}F\Sub{3}, C\Sub{2}H\Sub{3}F\Sub{2}, C\Sub{3}HF\Sub{4}	&	C\Sub{3}H\Sub{2}F\Sub{4} (1), C\Sub{3}H\Sub{3}F\Sub{3} (2)	\\
HFC-245fa	&	C\Sub{3}H\Sub{3}F\Sub{5}	&	yes	&	1	\\
HFC-365mfc	&	C\Sub{4}H\Sub{5}F\Sub{5}	&	C\Sub{4}H\Sub{5}F\Sub{4}, C\Sub{3}H\Sub{2}F\Sub{5}	&	C\Sub{4}H\Sub{6}F\Sub{4} (1), C\Sub{3}H\Sub{3}F\Sub{5} (2)	\\
HFC-236cb	&	C\Sub{3}H\Sub{2}F\Sub{6}	&	C\Sub{3}H\Sub{2}F\Sub{5}, C\Sub{3}HF\Sub{6}	&	1	\\
HFC-236ea	&	C\Sub{3}H\Sub{2}F\Sub{6}	&	C\Sub{3}H\Sub{2}F\Sub{5}	&	1	\\
HFC-236fa	&	C\Sub{3}H\Sub{2}F\Sub{6}	&	C\Sub{3}H\Sub{2}F\Sub{5}	&	1	\\
HFC-227ea	&	C\Sub{3}HF\Sub{7}	&	C\Sub{3}HF\Sub{6}	&	1	\\
HFC-43-10mee	&	C\Sub{5}H\Sub{2}F\Sub{10}	&	C\Sub{4}H\Sub{2}F\Sub{7}, C\Sub{5}HF8, C\Sub{5}H\Sub{2}	&	C\Sub{4}H\Sub{2}F\Sub{8} (1), C\Sub{4}H\Sub{3}F\Sub{7} (2) 	\\

\hline
\multicolumn{4}{c}{HFOs}					\\
\hline
HFO-1234yf	&	C\Sub{3}H\Sub{2}F\Sub{4}	&	yes	&	1	\\
HFO-1234ze(E)	&	C\Sub{3}H\Sub{2}F\Sub{4}	&	yes	&	1	\\
HCFO-1233zd(E)	&	C\Sub{3}H\Sub{2}ClF\Sub{3}	&	yes	&	1	\\

\hline
\multicolumn{4}{c}{Halogenated compounds with high boiling point} \\
\hline
HCBD	&	C\Sub{4}Cl\Sub{6}	&	yes	&	1	\\
TCHFB	&	C\Sub{4}Cl\Sub{4}F\Sub{6}	&	C\Sub{4}ClF\Sub{6}, C\Sub{3}Cl\Sub{3}F\Sub{4}	&	C\Sub{3}Cl\Sub{3}F\Sub{5} (1), C\Sub{4}Cl\Sub{2}F\Sub{6} (2)	\\

\hline
  \end{tabular}
\end{table}

\begin{table}[ht]
  \caption{Behaviour of the likelihood estimator: ten first knapsack fragments for CFC-11 and CCl\Sub{4}, set out in order according to their likelihood value calculated at the first iteration of Step 5. Some fragments may be deleted at subsequent iterations. 1\Sup{st} column: chemical formula of the fragment, containing only abundant isotopes. * indicates that the fragment is not part of the molecular ion. 2\Sup{nd} col.: calculated exact mass of ionised fragment. 3\Sup{rd} col.: percent signal of the fragment and its isotopocules relative to the total measured signal. 4\Sup{th} col.: percent signal of the fragment, all its sub-fragments and all associated isotopocules relative to the total measured signal, computed from Eq.~\eqref{eq:bad-estimator}. 5\Sup{th} col.: likelihood value computed from Eq.~\eqref{eq:likelihood-estimator-2}. 6\Sup{th} col.: ranking according to decreasing likelihood value. For CFC-11, CFCl\Sub{3} ranked first is the molecular ion.}
  \label{tab:CFC11-results}

  \begin{tabular}{crrrrcc}
\hline
Fragment	&	Exact mass	&	 \begin{tabular}{@{}r@{}} \% Assigned \\ signal of the \\ fragment\end{tabular} & 
\begin{tabular}{@{}r@{}} \% Assigned \\ signal of all \\ (sub)fragments\end{tabular} & Likelihood & Ranking & \begin{tabular}{@{}r@{}} Maximal \\ fragment?\end{tabular} 	\\

\hline
\multicolumn{7}{c}{CFC-11 (CFCl\Sub{3})}\\
\hline

CFCl\Sub{3}\Sup{+}	&	135.90496	&	0.0	&	98.4	&	88.6	&	1	&	True  \\
CFCl\Sub{2}\Sup{+}	&	100.93611	&	81.9	&	96.3	&	84.3	&	2	&	False  \\
CFCl\Sup{+}	&	65.96726	&	5.6	&	11.9	&	11.9	&	3	&	False  \\
*CHFCl\Sup{+}	&	66.97508	&	0.0	&	12.0	&	9.3	&	4	&	True  \\
CCl\Sub{3}\Sup{+}	&	116.90656	&	2.1	&	10.0	&	8.0	&	5	&	False  \\
*CHCl\Sub{3}\Sup{+}	&	117.91438	&	0.0	&	10.2	&	7.9	&	6	&	True  \\
CCl\Sub{2}\Sup{+}	&	81.93771	&	2.6	&	8.0	&	6.0	&	7	&	False  \\
*CHCl\Sub{2}\Sup{+}	&	82.94553	&	0.0	&	8.1	&	5.8	&	8	&	False  \\
CCl\Sup{+}	&	46.96885	&	3.1	&	5.4	&	5.4	&	9	&	False  \\
*C\Sub{3}Cl\Sub{2}\Sup{+}	&	105.93771	&	0.0	&	8.0	&	3.2	&	10	&	False  \\
\hline

\multicolumn{7}{c}{CCl\Sub{4}}\\

\hline
CCl\Sub{3}\Sup{+}	&	116.90656	&	71.2	&	92.4	&	73.9	&	1	&	False   \\
*CHCl\Sub{3}\Sup{+}	&	117.91438	&	0.4	&	93.9	&	73.0	&	2	&	True   \\
*CHCl\Sub{2}\Sup{+}	&	82.94553	&	0.5	&	22.3	&	16.0	&	3	&	False   \\
CCl\Sub{2}\Sup{+}	&	81.93771	&	11.6	&	21.2	&	15.9	&	4	&	False   \\
*COCl\Sub{2}\Sup{+}	&	97.93262	&	0.3	&	21.8	&	12.1	&	5	&	True   \\
CCl\Sup{+}	&	46.96885	&	5.5	&	9.5	&	9.5	&	6	&	False   \\
*HCl\Sup{+}	&	35.97668	&	0.7	&	4.7	&	4.7	&	7	&	False   \\
Cl\Sup{+}	&	34.96885	&	4.0	&	4.0	&	4.0	&	8	&	False   \\
*CF\Sub{2}Cl\Sup{+}	&	84.96566	&	0.2	&	9.7	&	3.6	&	9	&	True   \\
*OCl\Sub{2}\Sup{+}	&	85.93262	&	0.3	&	4.3	&	2.2	&	10	&	False   \\

\hline
  \end{tabular}
\end{table}

\begin{table}[ht]
  \caption{Numerical values for the obtained runtime on the training set, in seconds. These values are displayed on Fig.~\ref{fig:method-runtime}. Step 1: knapsack enumeration of fragment formulae. Step 2: graph construction. Step 3: isotopocule enumeration. Step 7: optimisation of multiple isotopocule sets together using \texttt{lmfit}.}
  \label{tab:runtimes}

  \begin{tabular}{lrrrrrr}
\hline

Compound	&	\begin{tabular}{@{}l@{}} No. knapsack\\ solutions\end{tabular} & \begin{tabular}{@{}l@{}} Step 1: \\ Knapsack\end{tabular}	&	\begin{tabular}{@{}l@{}} Step 2: \\ Graph \end{tabular}	&	\begin{tabular}{@{}l@{}} Step 3: \\ Iso. Enum. \end{tabular}	&	\begin{tabular}{@{}l@{}} Step 7: \\ Optimisation \end{tabular}	&	Total \\	
				
\hline
C\Sub{2}H\Sub{6}	&	4	&	0	&	0	&	0.00035	&	0.58114	&	0.61831  \\
C\Sub{3}H\Sub{8}	&	12	&	0.001	&	0.00103	&	0.00277	&	0.93511	&	1.02196  \\
CH\Sub{3}Cl	&	6	&	0.001	&	0	&	0.001	&	0.22414	&	0.28026  \\
COS	&	5	&	0.00175	&	0	&	0.001	&	0.05579	&	0.09874  \\
NF\Sub{3}	&	11	&	0.001	&	0	&	0.00089	&	0.09805	&	0.11600  \\
Benzene	&	47	&	0.00399	&	0.00798	&	0.01695	&	1.36925	&	1.54867  \\
CH\Sub{2}Cl\Sub{2}	&	15	&	0.00739	&	0.001	&	0.00499	&	0.29121	&	0.39700  \\
HCFC-22	&	17	&	0.0039	&	0.001	&	0.0032	&	0.23608	&	0.31227  \\
CF\Sub{4}	&	3	&	0.00068	&	0	&	0.001	&	0.04567	&	0.06479  \\
Toluene	&	106	&	0.00894	&	0.03889	&	0.05063	&	1.23265	&	1.67224  \\
CH\Sub{3}Br	&	15	&	0.00484	&	0.001	&	0.00345	&	0.06066	&	0.09972  \\
HCFC-142b	&	127	&	0.00697	&	0.0389	&	0.03164	&	3.55514	&	3.90335  \\
SO\Sub{2}F\Sub{2}	&	16	&	0.00299	&	0.00099	&	0.00399	&	0.19949	&	0.23456  \\
CFC-13	&	9	&	0.00298	&	0	&	0.00099	&	1.99773	&	2.02956  \\
HCFC-141b	&	46	&	0.00499	&	0.00603	&	0.01056	&	1.52992	&	1.65464  \\
CHCl\Sub{3}	&	17	&	0.00939	&	0.00105	&	0.00499	&	0.08471	&	0.16614  \\
CFC-12	&	35	&	0.0139	&	0.00794	&	0.01178	&	0.28523	&	0.42641  \\
C\Sub{2}HCl\Sub{3}	&	66	&	0.01396	&	0.01615	&	0.02093	&	0.09776	&	0.30321  \\
CFC-11	&	44	&	0.01503	&	0.00757	&	0.01496	&	0.14162	&	0.30119  \\
HCFC-124	&	110	&	0.01198	&	0.02792	&	0.02887	&	0.10436	&	0.34243  \\
PFC-116	&	78	&	0.00891	&	0.01794	&	0.02439	&	0.08703	&	0.34109  \\
CH\Sub{3}I	&	45	&	0.01097	&	0.00499	&	0.01476	&	1.05608	&	1.17037  \\
SF\Sub{6}	&	99	&	0.00901	&	0.0299	&	0.03683	&	0.05829	&	0.35405  \\
Halon-1301	&	39	&	0.01396	&	0.00895	&	0.01894	&	0.12531	&	0.26130  \\
CCl\Sub{4}	&	23	&	0.01103	&	0.002	&	0.0047	&	0.78681	&	0.86528  \\
CFC-115	&	183	&	0.01696	&	0.11013	&	0.07779	&	0.42488	&	1.15757  \\
C\Sub{2}Cl\Sub{4}	&	92	&	0.06682	&	0.02865	&	0.05261	&	0.09374	&	0.54953  \\
Halon-1211	&	78	&	0.01994	&	0.01795	&	0.02094	&	0.50948	&	0.73878  \\
CFC-114	&	358	&	0.02792	&	0.36901	&	0.17453	&	0.47408	&	2.15301  \\
CH\Sub{2}Br\Sub{2}	&	11	&	0.02094	&	0.00047	&	0.00233	&	0.06566	&	0.11770  \\
CFC-113	&	698	&	0.06004	&	1.08788	&	0.42879	&	0.61436	&	4.44405  \\
PFC-218	&	317	&	0.01794	&	0.27526	&	0.13997	&	0.06965	&	2.87218  \\
SF\Sub{5}CF\Sub{3}	&	66	&	0.00698	&	0.00997	&	0.01695	&	0.08213	&	0.23013  \\
PFC-c318	&	149	&	0.01241	&	0.04537	&	0.06582	&	0.09975	&	0.58743  \\
Halon-2402	&	1362	&	0.12367	&	4.2448	&	0.7849	&	0.14161	&	11.59514  \\
C\Sub{6}F\Sub{14}	&	3096	&	0.09473	&	23.70908	&	1.79919	&	0.21742	&	76.37336  \\

\hline
  \end{tabular}
\end{table}


\begin{table}[ht]
  \caption{Numerical values for the obtained runtime on the validation set, in seconds. These values are displayed on Fig.~\ref{fig:method-runtime}. Step 1: knapsack enumeration of fragment formulae. Step 2: graph construction. Step 3: isotopocule enumeration. Step 7: optimisation of multiple isotopocule sets together using \texttt{lmfit}.}
  \label{tab:runtimes-validation}

  \begin{tabular}{lrrrrrr}
\hline

Compound	&	\begin{tabular}{@{}l@{}} No. knapsack\\ solutions\end{tabular} & \begin{tabular}{@{}l@{}} Step 1: \\ Knapsack\end{tabular}	&	\begin{tabular}{@{}l@{}} Step 2: \\ Graph \end{tabular}	&	\begin{tabular}{@{}l@{}} Step 3: \\ Iso. Enum. \end{tabular}	&	\begin{tabular}{@{}l@{}} Step 7: \\ Optimisation \end{tabular}	&	Total \\

\hline
HFC-41	&	3	&	0.00098	&	0	&	0.00199	&	0.0748	&	0.09376	\\
HFC-32	&	5	&	0.001	&	0	&	0.001	&	0.14962	&	0.17655	\\
HFC-152	&	20	&	0.00199	&	0.00198	&	0.00499	&	0.48074	&	0.53358	\\
HFC-152a	&	22	&	0.00199	&	0.00199	&	0.00598	&	0.63579	&	0.70561	\\
HFC-23	&	5	&	0.00099	&	0	&	0.00299	&	0.10472	&	0.13266	\\
HFC-143	&	33	&	0.00299	&	0.00299	&	0.00698	&	0.56412	&	0.64393	\\
HFC-143a	&	39	&	0.00499	&	0.00798	&	0.00898	&	1.28358	&	1.38034	\\
HFC-134	&	51	&	0.00499	&	0.00997	&	0.01396	&	0.75513	&	0.8778	\\
HFC-134a	&	69	&	0.00598	&	0.01396	&	0.01795	&	1.15894	&	1.34445	\\
HFC-125	&	62	&	0.00899	&	0.01297	&	0.01795	&	0.31715	&	0.51916	\\
HFC-245ca	&	119	&	0.00897	&	0.04189	&	0.05086	&	1.04231	&	1.44332	\\
HFC-245fa	&	204	&	0.01695	&	0.10971	&	0.06982	&	0.73306	&	1.49405	\\
HFC-365mfc	&	356	&	0.02094	&	0.40094	&	0.13765	&	3.05672	&	5.10083	\\
HFC-236cb	&	404	&	0.02693	&	0.43587	&	0.16956	&	0.91956	&	3.0124	\\
HFC-236ea	&	352	&	0.01995	&	0.35804	&	0.13165	&	0.58447	&	2.39165	\\
HFC-236fa	&	157	&	0.01297	&	0.06383	&	0.06148	&	0.17143	&	0.85424	\\
HFC-227ea	&	369	&	0.02592	&	0.34808	&	0.15559	&	0.49173	&	2.80308	\\
HFC-43-10mee	&	3192	&	0.1516	&	27.8139	&	1.71247	&	1.455	&	83.46999	\\
HFO-1234yf	&	184	&	0.00798	&	0.08976	&	0.06084	&	0.15259	&	0.72908	\\
HFO-1234ze(E)	&	153	&	0.00997	&	0.04887	&	0.04488	&	0.44783	&	0.87914	\\
HCFO-1233zd(E)	&	189	&	0.01296	&	0.10572	&	0.07579	&	0.08976	&	0.74725	\\
HCBD	&	790	&	0.13464	&	1.53374	&	0.77896	&	0.14859	&	6.21535	\\
TCHFB	&	3452	&	0.15757	&	28.7482	&	2.01913	&	0.85859	&	52.94723	\\

\hline
  \end{tabular}
\end{table}

\clearpage


\end{backmatter}
\end{document}


\begin{frontmatter}

\begin{fmbox}
\dochead{Supplement}


\title{Supplement for: Automated fragment formula annotation for electron ionisation, high resolution mass spectrometry: application to atmospheric measurements of halocarbons}


\author[
   addressref={aff1},                   
   email={myriam.guillevic@empa.ch}   
]{\inits{MG}\fnm{Myriam} \snm{Guillevic}}
\author[
   addressref={aff2},
   email={aurore.guillevic@inria.fr}
]{\inits{AG}\fnm{Aurore} \snm{Guillevic}}
\author[
   addressref={aff1},                   
   email={martin.vollmer@empa.ch}   
]{\inits{MKV}\fnm{Martin K.} \snm{Vollmer}}
\author[
   addressref={aff1},                   
   email={paul.schlauri@empa.ch}   
]{\inits{PS}\fnm{Paul} \snm{Schlauri}}
\author[
   addressref={aff1},                   
   email={matthias.hill@empa.ch}   
]{\inits{MH}\fnm{Matthias} \snm{Hill}}
\author[
   addressref={aff1},                   
   email={lukas.emmenegger@empa.ch}   
]{\inits{LE}\fnm{Lukas} \snm{Emmenegger}}
\author[
   addressref={aff1},                   
   email={stefan.reimann@empa.ch}   
]{\inits{SR}\fnm{Stefan} \snm{Reimann}}

\address[id=aff1]{
  \orgname{Laboratory for Air Pollution /Environmental Technology, Empa, Swiss Federal Laboratories for Materials Science and Technology}, 
  \street{Ueberlandstrasse 129},                     %
  \postcode{8600}                                
  \city{D\"ubendorf},                              
  \cny{Switzerland}                                    
}
\address[id=aff2]{%
  \orgname{Universit\'e de Lorraine, CNRS, Inria, LORIA},
  \street{}
  \postcode{54000}
  \city{Nancy},
  \cny{France}
}


\begin{artnotes}
\end{artnotes}

\end{fmbox}

\end{frontmatter}



\newpage

\section{Training set and validation set: SMILES codes}

We provide here the SMILES codes for each of the substances in the training and validation set and if a mass spectrum is documented in the NIST spectral database \citep{NISTwebsite}.

\begin{table}[ht]
  \caption{Known compounds used as training set: CAS numbers, SMILES codes and presence in the NIST chemistry webbook \citep{NISTwebsite}. All values are taken from the ChemSpider website \citep{ChemSpiderwebsite}, the PubChem website \citep{PubChem_website} and the NIST website \citep{NISTwebsite}.}
  \label{tab:training-set-SMILES}
  \begin{tabular}{l l l l l}
\hline
Compound	&	\begin{tabular}{@{}l@{}}Chemical \\formula \end{tabular}	&	CAS number & SMILES code	& \begin{tabular}{@{}l@{}}NIST \\spectrum \end{tabular} \\
\hline

C\Sub{2}H\Sub{6}	&	C\Sub{2}H\Sub{6}	&	74-84-0	&	CC	&	yes	\\
C\Sub{3}H\Sub{8}	&	C\Sub{3}H\Sub{8}	&	74-98-6	&	CCC	&	yes	\\
CH\Sub{3}Cl	&	CH\Sub{3}Cl	&	74-87-3	&	CCl	&	yes	\\
COS	&	COS	&	463-58-1	&	C(=O)=S	&	yes	\\
NF\Sub{3}	&	NF\Sub{3}	&	7783-54-2	&	N(F)(F)F	&	yes	\\
Benzene	&	C\Sub{6}H\Sub{6}	&	71-43-2	&	C1=CC=CC=C1	&	yes	\\
CH\Sub{2}Cl\Sub{2}	&	CH\Sub{2}Cl\Sub{2}	&	75-09-2	&	C(Cl)Cl	&	yes	\\
HCFC-22	&	HCF\Sub{2}Cl	&	75-45-6	&	C(F)(F)Cl	&	yes	\\
CF\Sub{4}	&	CF\Sub{4}	&	75-73-0	&	C(F)(F)(F)F	&	yes	\\
Toluene	&	C\Sub{7}H\Sub{8}	&	108-88-3	&	CC1=CC=CC=C1	&	yes	\\
CH\Sub{3}Br	&	CH\Sub{3}Br	&	74-83-9	&	CBr	&	yes	\\
HCFC-142b	&	H\Sub{3}C\Sub{2}F\Sub{2}Cl	&	75-68-3	&	CC(F)(F)Cl	&	yes	\\
SO\Sub{2}F\Sub{2}	&	SO\Sub{2}F\Sub{2}	&	2699-79-8	&	O=S(=O)(F)F	&	yes	\\
CFC-13	&	CF\Sub{3}Cl	&	75-72-9	&	C(F)(F)(F)Cl	&	yes	\\
HCFC-141b	&	H\Sub{3}C\Sub{2}FCl\Sub{2}	&	1717-00-6	&	CC(F)(Cl)Cl	&	yes	\\
CHCl\Sub{3}	&	CHCl\Sub{3}	&	67-66-3	&	C(Cl)(Cl)Cl	&	yes	\\
CFC-12	&	CF\Sub{2}Cl\Sub{2}	&	75-71-8	&	C(F)(F)(Cl)Cl	&	yes	\\
C\Sub{2}HCl\Sub{3}	&	C\Sub{2}HCl\Sub{3}	&	79-01-6	&	C(=C(Cl)Cl)Cl	&	yes	\\
CFC-11	&	CFCl\Sub{3}	&	75-69-4	&	C(F)(Cl)(Cl)Cl	&	yes	\\
HCFC-124	&	HC\Sub{2}F\Sub{4}Cl	&	2837-89-0	&	C(C(F)(F)F)(F)Cl	&	yes	\\
PFC-116	&	C\Sub{2}F\Sub{6}	&	76-16-4	&	C(C(F)(F)F)(F)(F)F	&	yes	\\
CH\Sub{3}I	&	CH\Sub{3}I	&	74-88-4	&	CI	&	yes	\\
SF\Sub{6}	&	SF\Sub{6}	&	2551-62-4	&	FS(F)(F)(F)(F)F	&	no	\\
Halon-1301	&	CF\Sub{3}Br	&	75-63-8	&	C(F)(F)(F)Br	&	no	\\
CCl\Sub{4}	&	CCl\Sub{4}	&	56-23-5	&	C(Cl)(Cl)(Cl)Cl	&	yes	\\
CFC-115	&	C\Sub{2}F\Sub{5}Cl	&	76-15-3	&	C(C(F)(F)Cl)(F)(F)F	&	yes	\\
C\Sub{2}Cl\Sub{4}	&	C\Sub{2}Cl\Sub{4}	&	127-18-4	&	C(=C(Cl)Cl)(Cl)Cl	&	yes	\\
Halon-1211	&	CF\Sub{2}ClBr	&	353-59-3	&	C(F)(F)(Cl)Br	&	yes	\\
CFC-114	&	C\Sub{2}F\Sub{4}Cl\Sub{2}	&	76-14-2	&	C(C(F)(F)Cl)(F)(F)Cl	&	yes	\\
CH\Sub{2}Br\Sub{2}	&	CH\Sub{2}Br\Sub{2}	&	74-95-3	&	C(Br)Br	&	yes	\\
CFC-113	&	C\Sub{2}F\Sub{3}Cl\Sub{3}	&	76-13-1	&	C(C(F)(Cl)Cl)(F)(F)Cl	&	yes	\\
PFC-218	&	C\Sub{3}F\Sub{8}	&	76-19-7	&	C(C(F)(F)F)(C(F)(F)F)(F)F	&	yes	\\
SF\Sub{5}CF\Sub{3}	&	SF\Sub{5}CF\Sub{3}	&	373-80-8	&	C(F)(F)(F)S(F)(F)(F)(F)F	&	yes	\\
PFC-c318	&	C\Sub{4}F\Sub{8}	&	115-25-3	&	C1(C(C(C1(F)F)(F)F)(F)F)(F)F	&	yes	\\
Halon-2402	&	C\Sub{2}F\Sub{4}Br\Sub{2}	&	124-73-2	&	C(C(F)(F)Br)(F)(F)Br	&	yes	\\
C\Sub{6}F\Sub{14}	&	C\Sub{6}F\Sub{14}	&	355-42-0	&	C(C(C(C(F)(F)F)(F)F)(F)F)(C(C(F)(F)F)(F)F)(F)F	&	yes	\\

\hline
  \end{tabular}
\end{table}

\begin{table}[ht]
  \caption{Known compounds used as validation set: CAS numbers, SMILES codes and presence in the NIST chemistry webbook \citep{NISTwebsite}. All values are taken from the ChemSpider website \citep{ChemSpiderwebsite}, the PubChem website \citep{PubChem_website} and the NIST website \citep{NISTwebsite}.}
  \label{tab:validation-set-SMILES}
  \begin{tabular}{l l l l l}
\hline
Compound	&	\begin{tabular}{@{}l@{}}Chemical \\formula \end{tabular}	&	CAS number & SMILES code	& \begin{tabular}{@{}l@{}}NIST \\spectrum \end{tabular} \\
\hline

\multicolumn{5}{c}{Kigali Amendment to the Montreal Protocol}\\
\hline
HFC-41	&	H\Sub{3}CF	&	593-53-3	&	CF	&	yes	\\
HFC-32	&	H\Sub{2}CF\Sub{2}	&	75-10-5	&	C(F)F	&	yes	\\
HFC-152	&	H\Sub{4}C\Sub{2}F\Sub{2}	&	624-72-6	&	C(CF)F	&	yes	\\
HFC-152a	&	H\Sub{4}C\Sub{2}F\Sub{2}	&	75-37-6	&	CC(F)F	&	yes	\\
HFC-23	&	HCF\Sub{3}	&	75-46-7	&	C(F)(F)F	&	yes	\\
HFC-143	&	H\Sub{3}C\Sub{2}F\Sub{3}	&	430-66-0	&	C(C(F)F)F	&	yes	\\
HFC-143a	&	H\Sub{3}C\Sub{2}F\Sub{3}	&	420-46-2	&	CC(F)(F)F	&	yes	\\
HFC-134	&	H\Sub{2}C\Sub{2}F\Sub{4}	&	359-35-3	&	C(C(F)F)(F)F	&	yes	\\
HFC-134a	&	H\Sub{2}C\Sub{2}F\Sub{4}	&	811-97-2	&	C(C(F)(F)F)F	&	yes	\\
HFC-125	&	HC\Sub{2}F\Sub{5}	&	354-33-6	&	C(C(F)(F)F)(F)F	&	yes	\\
HFC-245ca	&	H\Sub{3}C\Sub{3}F\Sub{5}	&	679-86-7	&	C(C(C(F)F)(F)F)F	&	yes	\\
HFC-245fa	&	H\Sub{3}C\Sub{3}F\Sub{5}	&	460-73-1	&	C(C(F)F)C(F)(F)F	&	no	\\
HFC-365mfc	&	H\Sub{5}C\Sub{4}F\Sub{5}	&	406-58-6	&	CC(CC(F)(F)F)(F)F	&	no	\\
HFC-236cb	&	H\Sub{2}C\Sub{3}F\Sub{6}	&	677-56-5	&	C(C(C(F)(F)F)(F)F)F	&	no	\\
HFC-236ea	&	H\Sub{2}C\Sub{3}F\Sub{6}	&	431-63-0	&	C(C(F)F)(C(F)(F)F)F	&	yes	\\
HFC-236fa	&	H\Sub{2}C\Sub{3}F\Sub{6}	&	690-39-1	&	C(C(F)(F)F)C(F)(F)F	&	yes	\\
HFC-227ea	&	HC\Sub{3}F\Sub{7}	&	431-89-0	&	C(C(F)(F)F)(C(F)(F)F)F	&	no	\\
HFC-43-10mee	&	H\Sub{2}C\Sub{5}F\Sub{10}	&	138495-42-8	&	C(C(C(F)(F)F)F)(C(C(F)(F)F)(F)F)F	&	no	\\

\hline
\multicolumn{5}{c}{HFOs}					\\
\hline

HFO-1234yf	&	H\Sub{2}C\Sub{3}F\Sub{4}	&	754-12-1	&	C=C(C(F)(F)F)F	&	no	\\
HFO-1234ze(E)	&	H\Sub{2}C\Sub{3}F\Sub{4}	&	29118-24-9	&	C(=CF)C(F)(F)F	&	no	\\
HCFO-1233zd(E)	&	H\Sub{2}C\Sub{3}F\Sub{3}Cl	&	102687-65-0	&	C(=CCl)C(F)(F)F	&	no	\\

\hline
\multicolumn{5}{c}{{Halogenated compounds with high boiling point}} \\
\hline
									
{HCBD}	&	C\Sub{4}Cl\Sub{6}	&	87-68-3 	&	C(=C(Cl)Cl)(C(=C(Cl)Cl)Cl)Cl	&	yes	\\
{TCHFB}	&	C\Sub{4}Cl\Sub{4}F\Sub{6}	&	375-45-1	&	C(C(C(F)(F)Cl)(F)Cl)(C(F)(F)Cl)(F)Cl	&	no	\\

\hline
  \end{tabular}
\end{table}

\clearpage


\section{Mass calibration procedure}

A time-of-flight (ToF) instrument measures a time, elapsed between two events: the
extraction and the detection, when the ions hit the detector plate. To convert
this time measurement into a mass measurement in the most possible accurate
manner, internal mass calibration proves to be a good strategy. Known masses detected during a measurement are used to establish the calibration function between ToF and mass. In our cases, we use known masses produced by fragmentation of a mass calibration substance, perfluoroperhydrophenanthrene (PFPHP), of chemical formula C$_{14}$F$_{24}$. Only two types of atoms are present in this molecule, carbon (mass 12.000000) and fluorine (mass 18.99840316). The most abundant peaks can therefore be associated to a unique molecular formula. For example, the peak at integer mass 69 can be associated to CF$_3^+$ only, of exact mass 68.99466108.

Based on the NIST mass spectrum and our measured mass spectrum for PFPHP, we have chosen a list of {13} masses present with a sufficient abundance to be used for mass calibration. In addition, we use the masses of N\Sub{2}, O\Sub{2}, Ar, which are the most abundant air components and may slightly leak in our system, as well as Cl, also observed to be always present in our detector. The masses are chosen to be evenly distributed between the minimum and maximum masses, covering a range from m/z~= 28.0055994348 (N$_2^+$) to m/z~= 292.98188618 (C$_7$F$_{11}^+$).

We have observed that our ToF detector is subject to mass drift of up to 100~ppm during a run. This drift is possibly due to temperature variation of our preceding GC. To correct for this drift, we perform a mass calibration every {two} minutes, using average data of the preceding and following two minutes. For each set of four-minutes-averaged data, all peaks in the mass domain near the exact masses of the selected list are detected. Note that several mass peaks can be detected where only one exact mass from the calibrant is expected. Each detected peak is fitted using a pseudo-Voigt function, which is a combination of a Gaussian and a Lorentzian function:
\begin{equation}\label{eq:pseudoVoigt}
f(x; A, \mu, \sigma, \alpha, b) = (1 - \alpha) \frac{A}{\sigma \sqrt{2 \pi}} \exp\left( - \frac{(x - \mu)^2}{2 \sigma ^2} \right) +  \alpha \frac{A}{\pi} \left( \frac{\sigma_{L}}{ (x - \mu)^2 + {\sigma_L}^2}   \right)  + b
\end{equation}

where
\begin{equation}
\sigma_L = \sigma \sqrt{ 2 \ln(2)}
\end{equation}
and the full-with-at-half-maximum (FWHM)
\begin{equation}
\FWHM = 2 \sigma \sqrt{2 \ln(2)} = 2 \sigma_{L}~.
\end{equation}
The parameter FWHM is used later on ({Main article, Section 2.5.1}) to generate candidate mass peaks with the appropriate peak broadness.

Then, all obtained centres of ToF are associated to the closed expected exact mass. The entire set of pairs (ToF; expected exact mass) is used to fit a calibration function of the form:
\begin{equation}
i_{\ToF} = p_1 m^{p_3} + p_2
\end{equation}
with $i_{\ToF}$ the time of flight index and $m$ the exact mass.
The parameters $p_1$, $p_2$ and $p_3$ are optimised using the Python lmfit package. {According to theory, $p_3$ should be equal to 0.5 \footnote{\url{https://en.wikipedia.org/wiki/Time-of-flight_mass_spectrometry}}. However a better accuracy is obtained when $p_3$ is allowed to slightly vary.} If one or several pairs are further away from the fit than a set maximum value (20~ppm in our case), the furthest away pair is eliminated and the optimisation routine is repeated. This one-by-one pair elimination improves the robustness of the algorithm. This was proven necessary as when measuring real air from the industrial area where Empa is located, once in while a prominent pollution event occurs, producing outstanding mass peaks that may be in the vicinity of the expected peak, even masking it, therefore disturbing the mass calibration function.

Once all residuals are below the set value, the mass calibration is complete. Any ToF value can then be converted to a m/z value using:
\begin{equation}
m =  \left( \frac{i_{\ToF} - p_2}{p_1} \right)^{\frac{1}{p_3}}
\end{equation}
where $p_1$, $p_2$ and $p_3$ are calculated at any given specific time as linear interpolation using their time-bracketing optimised values.

\section{Uncertainty of the mass calibration}

\begin{figure}[htb]
  \centering
\includegraphics[width=10cm]{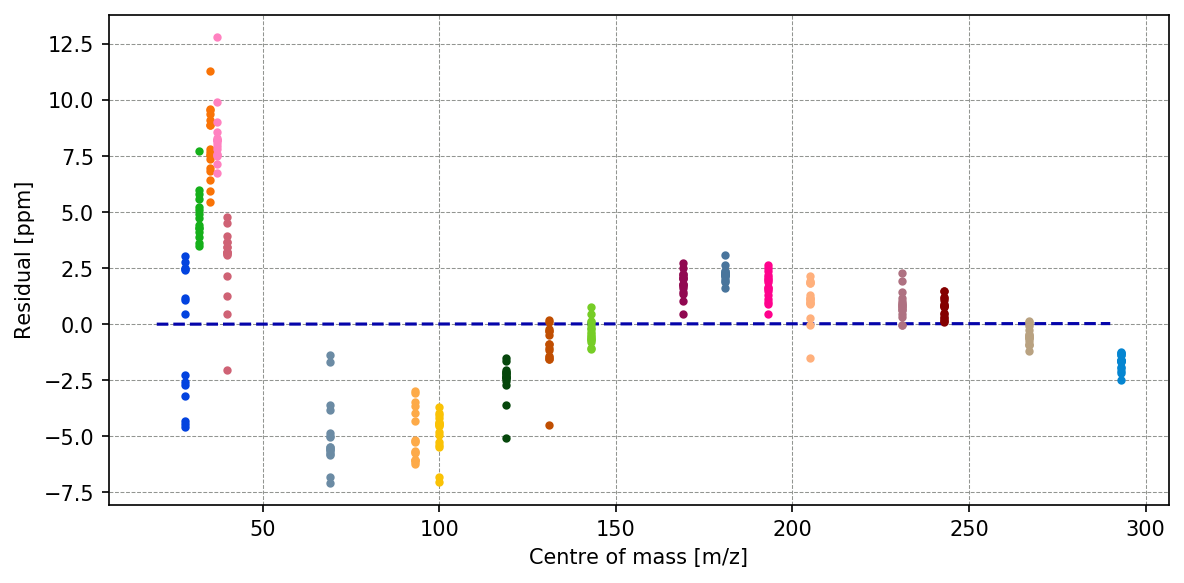}
\caption{Residuals of the mass calibration vs mass. Dots of the same colour represent the same mass peak, but at a different time slices.}
\label{fig:mass-cal-residual-vs-mass}
\end{figure}

After the optimisation, the obtained fit parameters are used to calculate the reconstructed m/z values; these values are then compared to the expected exact m/z, for each time slice of four minutes. The obtained offsets, expressed in ppm over the mass domain, are displayed in Fig.~\ref{fig:mass-cal-residual-vs-mass}. With our instrument, the observed mass accuracy is better for larger masses, with residuals usually below 5~ppm for masses higher than 100~m/z, while the accuracy can deteriorate to 20~ppm below 50~m/z. This is potentially due to the mass resolution of our instrument that is around 3000 for masses smaller than 50~m/z but around 4000 for larger masses.

To reflect this varying mass accuracy over the mass domain, for each used exact mass, we use as uncertainty the maximum observed offset at this mass. Then for any measured mass, its uncertainty is calculated as a linear interpolation between uncertainty at bracketing masses. This constitutes the mass calibration uncertainty.

\section{Validation set: preparation of qualitative standards for compounds newly regulated by the Kigali amendment to the Montreal Protocol}

Eighteen substances listed under the Kigali Amendment to the Montreal Protocol were part of the validation set. 

First, substances were separated in two groups, with the aim that each group should not contain isomers, to make sure each substance could be identified by its mass spectrum only. 
Group A contained: HFC-41, HFC-143a, HFC-134a, HFC-227ea, HFC-236ea, HFC-245fa, HFC-43-10-mee, HFC-152 and HFC-236cb.
Group B contained: HFC-32, HFC-23, HFC-125, HFC-152a, HFC-365mfc, HFC-143, HFC-236fa, HFC-245ca and HFC-134.

The pure substances were bought from Synquest Laboratories (Florida, USA). For each group, the pure substances were spiked one after the other into synthetic air, and the mixture was pressurised into a flask. The two obtained mixtures were prepared at approximately 6.5 nmol$\cdot$mol$^{-1}$.

Then, each mixture was measured by our preconcentration, gas chromatography, time-of-flight mass spectrometry instrumentation.
Data analysis then followed the same procedure as explained in the main article.

\section{Algorithmic improvements}
We describe our algorithmic improvements to speed-up the running-time of some
critical steps.

\subsection{Organisation of the chemical formulae in a directed acyclic graph (DAG)}
\label{sup-DAG}
After running the knapsack algorithm, many candidate chemical formulae are obtained, {hereafter simply referred to as 'formulae'}.
We organise them in a graph. In this graph, a node $n_j$ is a descendant
of a node $n_i$ if the node's fragment $s_j$ is a sub-fragment of the fragment
$s_i$ of the node $n_i$. For
example, CCl is a descendant of CCl\Sub{3}.
Conversely, a node $n_i$ is an ancestor of a node $n_j$ if its fragment $s_i$ is a
sup-fragment of the fragment $s_j$.
This define a \emph{partial order} on the formulae that we formally
define in Definition~\ref{def:partial-order}.

\begin{definition}[Partial order] \label{def:partial-order}
  We define the following partial order on the formulae.
  Let $s_i$ and $s_j$ be two formulae encoded as vectors of non-negative integers.
\begin{itemize}
\item The formula $s_j$ is \emph{smaller than} the formula $s_i$, denoted
  $s_j \leq s_i$, if $s_j$ is a sub-fragment of $s_i$;
\item The  formula $s_j$ is \emph{greater than} the formula $s_i$, denoted
  $s_j \geq s_i$, if $s_j$ is a sup-fragment of $s_i$;
\item otherwise $s_i$ and $s_j$ are \emph{incomparable}.
\end{itemize}
\end{definition}

Organising a set $\mathcal{S}$ of $n$ items (here
fragments) in a DAG (directed acyclic graph)
can takes as many as $n^2$ comparisons of items, but since $\mathcal{S}$ can
be quite large (e.g., $n=10000$), it makes sense to reduce the number of
comparisons.
Moreover, the graph should have as less edges as possible, that is, two
formulae $s_i \geq s_j$ are binded with an edge if and only if there is no other
formula $s_{i'}$ that could be inserted between them like $s_i \geq s_{i'}
\geq s_j$. For example with CCl, CCl\Sub{2} and CCl\Sub{3}, we will define the
graph with minimal edges on the left, not the one of the right (Fig.~\ref{fig:dag}).

\begin{figure}[hbt]
\begin{center}
\begin{tikzpicture}
  \node[circle, draw=black, inner sep=0, minimum width=2.25em] at (0,4\baselineskip) (n3) {CCl\Sub{3}};
  \node[circle, draw=black, inner sep=0, minimum width=2.25em] at (0,2\baselineskip) (n2) {CCl\Sub{2}};
  \node[circle, draw=black, inner sep=0, minimum width=2.25em] at (0,0) (n1) {CCl};

  \draw[->] (n3) -- (n2) ;
  \draw[->] (n2) -- (n1) ;
\end{tikzpicture}
\quad \quad
\begin{tikzpicture}
  \node[circle, draw=black, inner sep=0, minimum width=2.25em] at (0,4\baselineskip) (n3) {CCl\Sub{3}};
  \node[circle, draw=black, inner sep=0, minimum width=2.25em] at (0,2\baselineskip) (n2) {CCl\Sub{2}};
  \node[circle, draw=black, inner sep=0, minimum width=2.25em] at (0,0) (n1) {CCl};

  \draw[->] (n3) -- (n2) ;
  \draw[->] (n2) -- (n1) ;
  \draw[->, gray, densely dashed] (n3.south west) to[out=225,in=135, edge label=no, swap] (n1.north west) ;
\end{tikzpicture}
\end{center}
\caption{Directed acyclic graph.}
\label{fig:dag}
\end{figure}

We now explain how we reduced the number of comparisons between fragments to set
the edges, thus improving the complexity of building the graph.
First the target masses are sorted in decreasing order before the knapsack
step. Hence the output of the knapsack is made of batches of chemical formulae, each
one for a given target mass, in decreasing order.
\emph{Because the mass uncertainty is far thinner than $0.5 m/z$, all
  chemical formulae for one target mass are \emph{incomparable}: it is not possible
  to find that a formula is a sub-fragment of another, for the same target
  mass, otherwise the mass difference between the two would be at least $1 m/z$.}

Therefore we have a list of formulae $\{s_i\}_{1\leq i \leq \#\mathcal{S}}$
such that for any formula $s_i$, the formulae in the preceeding batches
weight more and are either incomparable or contain $s_i$ as a sub-fragment, and
the formulae in the forthcoming batches are lighter and either incomparable
or sub-fragments of $s_i$.
The maximal fragments will be the nodes at the ``roots'' of the DAG. They are made of the
formulae of the first (heaviest) batch, and some other formulae from other batches.

We maintain a list of the root nodes (maximal fragments) of the graph. They have no ancestor,
and they are incomparable to each other. To add a new formula in the graph,
because of the ordering of the formulae, we know that it is either
incomparable, or a subfragment of any node of the graph. \emph{It cannot be a
sup-fragment (a parent) of any node of the graph}. We compare the new
formula to each of the maximal fragments. If it is incomparable to any of
them, we add it as as new maximal fragment. Otherwise, for each maximal
fragment that has the new formula as sub-fragment, we compare
the new one to its children. If it is incomparable to any of the children, we
add it as a new child of the maximal fragment (we add an edge toward it). Otherwise, we recursively
explore the children of the children that have this new formula as sub-fragment.
In this way, we avoid many useless comparisons: all children of incomparable
nodes are omitted.

Thanks to the list of maximal fragments, the singletons are identified right
away: they are the maximal fragments without any child.
Figure~\refmainfigdag in the main article shows the graph obtained for CCl\Sub{4}.

\subsection{Removing a node and updating the edges}
\begin{figure}[htb]
\begin{center}
  \begin{tikzpicture}
  \node[circle, draw=black, inner sep=0, minimum width=2em, fill=blue!30!white] at (0,0) (n) {n};
  
  \node[circle, draw=black, inner sep=0, minimum width=2em] at (-1,1) (p1) {p\Sub{1}};
  \node[circle, draw=black, inner sep=0, minimum width=2em] at ( 0,1) (p2) {p\Sub{2}};
  \node[circle, draw=black, inner sep=0, minimum width=2em] at ( 1,1) (p3) {p\Sub{3}};

  \draw[->, blue] (p3) -- (n) ;
  \draw[->, blue] (p2) -- (n) ;
  \draw[->, blue] (p1) -- (n) ;

  \node[circle, draw=black, inner sep=0, minimum width=2em] at (-1,-1) (c1) {c\Sub{1}};
  \node[circle, draw=black, inner sep=0, minimum width=2em] at ( 0,-1) (c2) {c\Sub{2}};
  \node[circle, draw=black, inner sep=0, minimum width=2em] at ( 1,-1) (c3) {c\Sub{3}};

  \draw[->, blue] (n) -- (c1) ;
  \draw[->, blue] (n) -- (c2) ;
  \draw[->, blue] (n) -- (c3) ;

  \node[circle, draw=black, inner sep=0, minimum width=2em] at (2,0) (n2) {n$'$};

  \draw[->] (p3) -- (n2) ;
  \draw[->] (n2) -- (c3) ;

  \node[circle, draw=black, inner sep=0, minimum width=2em] at (-2,0) (n3) {n$''$};
  \draw[->] (n3) -- (c1) ;
       
\end{tikzpicture}
\quad \quad
\begin{tikzpicture}  
  \node[circle, draw=black, inner sep=0, minimum width=2em] at (-1,1) (p1) {p\Sub{1}};
  \node[circle, draw=black, inner sep=0, minimum width=2em] at ( 0,1) (p2) {p\Sub{2}};
  \node[circle, draw=black, inner sep=0, minimum width=2em] at ( 1,1) (p3) {p\Sub{3}};

  \node[circle, draw=black, inner sep=0, minimum width=2em] at (-1,-1) (c1) {c\Sub{1}};
  \node[circle, draw=black, inner sep=0, minimum width=2em] at ( 0,-1) (c2) {c\Sub{2}};
  \node[circle, draw=black, inner sep=0, minimum width=2em] at ( 1,-1) (c3) {c\Sub{3}};

  \draw[->, blue] (p3) -- (c1) ;
  \draw[->, blue] (p3) -- (c2) ;

  \draw[->, blue] (p2) -- (c1) ;
  \draw[->, blue] (p2) -- (c2) ;
  \draw[->, blue] (p2) -- (c3) ;
  
  \draw[->, blue] (p1) --node[anchor=east]{yes} (c1) ;
  \draw[->, blue] (p1) -- (c2) ;
  \draw[->, blue] (p1) -- (c3) ;

  \node[circle, draw=black, inner sep=0, minimum width=2em] at (2,0) (n2) {n$'$};

  \draw[->] (p3) -- (n2) ;
  \draw[->] (n2) -- (c3) ;
  \draw[->, red, densely dashed] (p3) to[edge label=no] (c3) ;

  \node[circle, draw=black, inner sep=0, minimum width=2em] at (-2,0) (n3) {n$''$};
  \draw[->] (n3) -- (c1) ;
\end{tikzpicture}
\end{center}
\caption{Example for Algorithm~\ref{alg:remove-node-update-edge}: removing a
  node and updating the edges.}
\label{fig:rm-node-up-edges}
\end{figure}
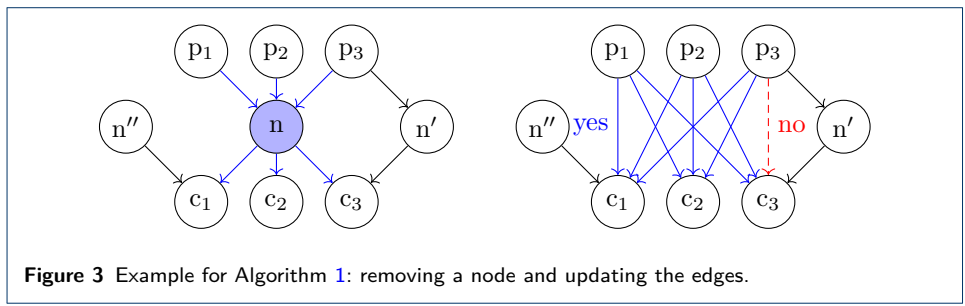

A node n to be removed has parents (closest sup-fragments) connected with one
edge, and children (closest sub-fragments) connected with one edge.
We wrote this procedure to remove a node and update the edges and list of
maximal fragments (Alg.~\ref{alg:remove-node-update-edge}, example in Fig.~\ref{fig:rm-node-up-edges}).

\begin{algorithm}[htb]
  \caption{remove the node n and update the edges}
  \label{alg:remove-node-update-edge}
\For{each parent p\Sub{i} of n}{
  remove the edge from p\Sub{i} to n \;
  \For{each child c\Sub{i} of n}{
    lists its own parents q\Sub{j} (without n) \;
    \If{none of q\Sub{j} is a sub-fragment of p\Sub{i}}{
       add an edge from p\Sub{i} to c\Sub{i}
      \tcp{(if there is one q\Sub{j} which is a sub-fragment of p\Sub{i}, do nothing: the edges
        are already fine)}
      }
  }
}
\For{each child c\Sub{i} of n}{
  remove the edge from n to c\Sub{i} \;
  \If{c\Sub{i} has no more parent after removing n}{
    add it in the list of maximal fragments
  }
}
\If{n is a maximal fragment (it has no parent)}{
  remove it from the list of maximal fragments
}
Remove n
\end{algorithm}

\subsection{Enumeration of isotopocules}
\label{sup-enum-isotopocule}

We now recall the computation of the relative intensities of the minor isotopocules
(see~\citep{Yergey1983} for a detailed computation).
The abundant formula has proportion (of the set of all isotopocules)
$pr = \prod_{\{\text{element}~e\}} (a_{e,0})^{n_e}$.
The product is over all the distinct atoms (denoted $e$), $a_{e,0}$ is the abundance
of the most abundant isotope of an atom, and $n_e$ is the number of occurrences
of that atom in the chemical formula. For example, with CCl\Sub{4} one computes
$pr = a_{\text{C}}a_{\text{Cl}}^4=0.326$.
An isotopocule with only one element $e$ and $i$ minor isotopes of abundance $a_{e,i}$ has proportion
\begin{align*}
  pr_{e} &= a_{e,0}^{n_{e,0}} {\genfrac(){0pt}{2}{n_e}{n_{e,0}}}
  a_{e,1}^{n_{e,1}} {\genfrac(){0pt}{2}{n_e-n_{e,0}}{n_{e,1}}}
  a_{e,2}^{n_{e,2}} {\genfrac(){0pt}{2}{n_e-n_{e,0}-n_{e,1}}{n_{e,2}}}
  \cdots
  a_{e,i}^{n_{e,i}} {\genfrac(){0pt}{2}{n_e-n_{e,0}-n_{e,1}-\ldots -n_{e,i-1}}{n_{e,i}}} \\
 &= a_{e,0}^{n_{e,0}} a_{e,1}^{n_{e,1}} a_{e,2}^{n_{e,2}} \cdots a_{e,i}^{n_{e,i}}
   \tfrac{n_{e}!} {n_{e,0}! n_{e,1}! n_{e,2}! \cdots n_{e,i}!}
\end{align*}
where the terms ${\genfrac(){0pt}{2}{n}{m}} = \frac{n!}{m!(n-m)!}$ are binomal coefficients
with $n \geq m$, denoting the number
of ways to choose $m$ items in a set of size $n$.
Their product simplifies in $n_{e}! / (n_{e,0}! n_{e,1}! n_{e,2}! \cdots n_{e,i}!)$.
An isotopocule has proportion
\begin{equation}
 pr 
 = \prod_e n_e! \prod_i (a_{e,i})^{n_{e,i}}/(n_{e,i}!)
\end{equation}
where $e$ ranges over the elements, $i$ ranges over the isotopes of an
element, $n_e$ is the total number of an element (with all isotopes), $n_{e,i}$ is
the number of occurences of one isotope.
The relative intensity of a minor-isotope formula is the ratio (see~\citep{Yergey1983})
\begin{align}
p = \frac{  \prod_e n_e! \prod_i (a_{e,i})^{n_{e,i}}/(n_{e,i}!)}
{\prod_{e} (a_{e,0})^{n_e}}
  \label{eq:relative-proba}
= \prod_e n_e! \prod_{i > 0} \left(\frac{a_{e,i}}{a_{e,0}}\right)^{n_{e,i}} \frac{1}{n_{e,i}!}
~.
\end{align}
One needs to enumerate all the possible combinations of isotopes of a given
element. This is a classical problem in combinatorics. Denote by $i$ the number
of isotopes (the abundant one included). Enumerate all the ways to
sum at most $i$ positive integers to obtain $n_e$.
It can be seen as a
knapsack-like problem: given all isotopes each of
``weight'' 1, finds how to sum to $n_e$. In particular, each isotope is allowed
at most $n_e$ times.

All ratios are relative to the abundant chemical formula, whose maximum possible
intensity is known: this is the intensity of the corresponding measured mass.
To improve the running-time of the enumeration, we do not list the minor
isotopocules whose intensity would be below the detection threshold of the
ToF-MS.
For this purpose, we consider the elements one after the other.
We maintain a list of partial isotopocules with their partial relative
intensity, made of the elements processed so far. The list is ordered by
decreasing relative intensity.
Given a new element $e$ and its occurence $n_e$, we generate its isotopic
patterns, the abundant one included, and sort them in decreasing order of
relative intensity.
Reading the two lists in decreasing order of relative intensity, we combine the
new isotopes to the partial solutions and multiply together the
intensities. A loop over a list stops as soon as the product of
intensities is below the partial threshold. Once the new list of partial solutions is
computed, it is sorted in decreasing order of relative intensity. Then, the next element is
processed in the same manner, until all elements are done.
The first item of the resulting list is the isotopocule of highest relative
intensity (it can be greater than one). For implementation purpose, we scale the
list and divide all numbers by this highest relative intensity, so that
everything is in the interval $[0,1]$.

\section{Full Numerical Example with carbon tetrachloride}
\label{sup:CCl4-full-example}

For CCl\Sub{4} found at a retention time of 1708.27~s, nineteen masses are observed, listed in
Table~\ref{tab:CCl4-measured-masses} with uncertainty and intensity.

\subsection{Knapsack algorithm with two lists but without considering separately multi-valent and mono-valent atoms}
\label{subsubsec:knapsack-two-naive-sets}
In this paragraph we present an example of a knapsack algorithm with two sets of
atoms, and two lists of intermediate masses.
The candidate atoms are H, C, N, O, F, S, Cl, Br, I.
We arbitrarily define two subsets: \{C, N, O, S, Br\} and \{H, F, Cl, I\}.
The knapsack algorithm is run with input the masses of the atoms of each set, and the minimal
mass is set to 1. One obtains two lists: A and B given in Table~\ref{tab:knapsack-A-B-naive}.
The lists are sorted in increasing order of mass. One obtains
A = [(12.0, C), (14.0030740074, N), (15.9949146223, O), (24.0, C\Sub{2}),
  (26.003074007400002, CN), (27.9949146223, CO), (28.0061480148, N\Sub{2}),
  (29.9979886297, NO), (31.97207073, S), (31.9898292446, O\Sub{2})]
and
B = [(1.0078250319, H), (2.0156500638, H\Sub{2}), $\ldots$(all H\Sub{3$\ldots$17}), (18.140850574199998,
  H\Sub{18}), (18.99840316, F), (19.1486756061, H\Sub{19}), (20.0062281919, HF),
  $\ldots$(all H\Sub{20$\ldots$33} and H\Sub{2$\ldots$14}F),
  (34.1157786385, H\Sub{15}F), (34.2660510846, H\Sub{34}), (34.96885271, Cl)].
Then pairing the masses of
the two lists, one obtains masses within the target interval (if any): there is
one solution (34.96885271, Cl). Finally,
the DBE is computed and solutions with a negative DBE are discared.
This first approach has a major drawback: many impossible sub-fragments are
enumerated, im particular because of the Hydrogen which as a very small mass,
and is mono-valent.

Here is a detailed example to compute the lists A and B for the first target mass
$m_{\max}$ = 34.97006625322406.
First one computes multiples of each mass up to $m_{\max}$.
One obtains Tables~\ref{tab:init-knapsack-A} and~\ref{tab:init-knapsack-B}.
Then one combines at most one mass per column, starting the enumeration with the
lightest mass of each column (the first row value). One obtains the fragments
of the first row of Table~\ref{tab:knapsack-A-B-naive}.

\begin{table}[h]
\caption{Initial partial list of masses before computing the list A made of
  fragments of atoms \{C, N, O, S, Br\} and masses up to $m_{\max}$ = 34.97006625322406.}
\begin{tabular}{c|c|c|c|c|c}
\# & C & N & O & S & Br \\
\hline
1 & 12 & 14.0030740074    & 15.9949146223    & 31.97207073      &                  \\
2 & 24 & 28.0061480148    & 31.9898292446    &                  &                  \\
\end{tabular}
\label{tab:init-knapsack-A}
\end{table}

\begin{table}[h]
\caption{Initial partial list of masses before computing the list B made of
  fragments of atoms \{H, F, Cl, I\} and masses up to
  $m_{\max}$ = 34.97006625322406. There are 34 multiples of Hydrogen: H,
  H\Sub{2}, up to H\Sub{34}.
}
\begin{tabular}{c|c|c|c|c}
 & H & F & Cl & I \\
\hline
 1 & 1.0078250319     & 18.99840316      & 34.96885271      &                  \\
 2 & 2.0156500638     &                  &                  &                  \\
$\vdots$ & $\vdots$ & & & \\
 34 & 34.2660510846    &                  &                  &                  \\
\end{tabular}
\label{tab:init-knapsack-B}
\end{table}

\begin{table}[h]
\caption{Intermediate lists in the knapsack algorithm with two sets of atoms
  \{C, N, O, S, Br\} for the list A and \{H, F, Cl, I\} for the list B.
  The notation H\Sub{1-34} means all the 34 fragments made of one to 34 atoms of
  Hydrogen. The notation C\Sub{1-2}O means CO and C\Sub{2}O.}
\begin{tabular}{p{2.5cm}|r|p{2.6cm}|r|p{1.7cm}|r|l|r|l}
  target mass interval & \#A & A & \#B & B & \multicolumn{2}{l|}{all sol.} &
  \multicolumn{2}{l}{DBE $\geq 0$ } \\
\hline
34.96751070677594, 34.97006625322406 & 10 & S, O\Sub{1-2}, NO, CO, N\Sub{1-2}, CN, C\Sub{1-2} & 51 & Cl, H\Sub{1-15}F, F, H\Sub{1-34} & 1 & Cl & 1 & Cl \\
35.974137795964566, 35.97778836403543 & 10 & S, O\Sub{1-2}, NO, CO, N\Sub{1-2}, CN, C\Sub{1-2} & 54 & HCl, Cl, H\Sub{1-16}F, F, H\Sub{1-35} & 1 & HCl & 1 & HCl \\
36.96406557296814, 36.967505987031856 & 11 & S, O\Sub{1-2}, NO, CO, N\Sub{1-2}, CN, C\Sub{1-3} & 56 & HCl, Cl, H\Sub{1-17}F, F, H\Sub{1-36} & 0 &  & 0 &  \\
46.96648952357635, 46.970287436423654 & 21 & NS, CS, S, NO\Sub{2}, CO\Sub{2}, O\Sub{1-2}, N\Sub{2}O, NO, CNO, C\Sub{1-2}O, N\Sub{1-3}, CN\Sub{2}, C\Sub{1-2}N, C\Sub{1-3} & 95 & H\Sub{1-11}Cl, Cl, H\Sub{1-8}F\Sub{2}, F\Sub{1-2}, H\Sub{1-27}F, H\Sub{1-46} & 1 & CCl & 1 & CCl \\
48.962558174089345, 48.96840118591066 & 24 & OS, NS, CS, S, O\Sub{1-3}, NO\Sub{2}, CO\Sub{2}, N\Sub{2}O, NO, CNO, C\Sub{1-2}O, N\Sub{1-3}, CN\Sub{2}, C\Sub{1-2}N, C\Sub{1-4} & 103 & H\Sub{1-13}Cl, Cl, H\Sub{1-10}F\Sub{2}, F\Sub{1-2}, H\Sub{1-29}F, H\Sub{1-48} & 0 &  & 0 &  \\
\end{tabular}
\label{tab:knapsack-A-B-naive}
\end{table}

\subsection{Faster knapsack: avoiding enumerating fragments of negative DBE value}
\label{subsubsec:knapsack-DBE-opt}

\begin{table}[p]
\caption{Intermediate lists in the knapsack algorithm with two sets of atoms
  \{C, N, O, S, Br\} for the list A and \{H, F, Cl, I\} for the list B,
  then with a set of multi-valent atoms \{C, N, O, S\} giving the list M, and a
  set of mono-valent atoms \{H, F, Cl, Br, I\} giving the list m.
After enumerating the list M, the maximum possible valence is computed and used
as an upper bound on the number of mono-valent atoms to generate the list
m. One observes that this technique allows to divide by more than two the
length of the second list.}
\begin{tabular}{r@{.}l@{}r@{.}l |r|r|c||r|c|r|c}
\multicolumn{4}{c|}{target mass interval} & \#A & \#B & \begin{tabular}{@{}c@{}}\#\\sol.\end{tabular} & \#M & \begin{tabular}{@{}c@{}}2 max\\DBE\end{tabular} & \#m & \begin{tabular}{@{}c@{}}DBE\\$\geq0$\end{tabular} \\
\hline
 34&96751070677594, & 34&97006625322406   & 10 & 51 & 1 & 10 & 6 & 13 & 1 \\
 35&974137795964566,& 35&97778836403543   & 10 & 54 & 1 & 10 & 6 & 14 & 1 \\
 36&96406557296814, & 36&967505987031856  & 11 & 56 & 0 & 11 & 8 & 18 & 0 \\
 46&96648952357635, & 46&970287436423654  & 21 & 95 & 1 & 21 & 8 & 31 & 1 \\
 48&962558174089345,& 48&96840118591066   & 24 & 103 & 0 & 24 & 10 & 39 & 0 \\
 59&960220186945,   & 59&971315773055004  & 37 & 156 & 1 & 37 & 10 & 48 & 1 \\
 81&9330864132061,  & 81&9395331467939    & 90 & 315 & 1 & 89 & 14 & 110 & 1 \\
 82&93831758560036, & 82&95111397439963   & 94 & 324 & 3 & 93 & 14 & 115 & 2 \\
 83&93024931474109, & 83&9372426452589    & 94 & 333 & 0 & 93 & 14 & 117 & 0 \\
 84&92634272134954, & 84&97112963865045   & 101 & 342 & 7 & 100 & 16 & 135 & 4 \\
 85&92419323227152, & 85&93924312772849   & 101 & 351 & 1 & 100 & 16 & 137 & 1 \\
 97&9236485334006,  & 97&9389656265994    & 154 & 477 & 3 & 150 & 18 & 189 & 2 \\
 99&90729357057015, & 99&94127718942985   & 161 & 501 & 4 & 157 & 18 & 197 & 3 \\
116&90200574284233, &116&90847941715768   & 274 & 735 & 2 & 262 & 20 & 293 & 1 \\
117&89455759263474, &117&92205636736527   & 275 & 751 & 5 & 262 & 20 & 300 & 3 \\
118&89897942315969, &118&9056775368403    & 287 & 766 & 0 & 274 & 20 & 305 & 0 \\
119&89648859190275, &119&91785116809724   & 290 & 782 & 2 & 275 & 20 & 311 & 1 \\
120&89523104367791, &120&90302931632209   & 306 & 798 & 1 & 291 & 22 & 345 & 0 \\
122&8875535007597,  &122&90537265924031   & 323 & 830 & 1 & 305 & 22 & 357 & 1 \\
\end{tabular}
\label{tab:knapsack-A-B-M-m}
\end{table}

\newcommand{\knap}[1]{\textcolor{blue}{#1}}
\newcommand{\isot}[1]{\textcolor{orange}{#1}}
\begin{table}[p]
  \caption{Measured masses at RT = 1708.27s. In blue, the correct guess made by
    knapsack. In orange, the identified isotopocules.}
  \centering
  \begin{tabular}{r r@{.}l@{}r@{.}l r@{.}l p{2cm} l}
  \begin{tabular}{@{}c@{}}measured \\mass (m/z)
  \end{tabular}
    & \multicolumn{4}{c}{mass range with uncertainty} &
                                                                          \multicolumn{2}{c}{intensity}
    & knapsack & identified \\
 34.96878848 & [ 34&96751070677594, &  34&97006625322406]  &  2722&2042 & \knap{Cl}  & \knap{Cl} \\
 35.97596308 & [ 35&974137795964566,&  35&97778836403543]  &  1051&6898 & \knap{HCl} & \knap{HCl} \\
 36.96578578 & [ 36&96406557296814, &  36&967505987031856] &   914&6638 & --  & \isot{[\Sup{37}Cl]} \\
 46.96838848 & [ 46&96648952357635, &  46&970287436423654] &  3784&4981 & \knap{CCl} & \knap{CCl}\\
 48.96547968 & [ 48&962558174089345,&  48&96840118591066]  &  1192&8077 & --  & \isot{C[\Sup{37}Cl]}\\
 59.96576798 & [ 59&960220186945,   &  59&971315773055004] &   120&657  & COS & --\\
 81.93630978 & [ 81&9330864132061,  &  81&9395331467939]   &  6160&9695 & \knap{CCl\Sub{2}} &\knap{CCl\Sub{2}} \\
 82.94471578 & [ 82&93831758560036, &  82&95111397439963]  &   319&247  & S\Sub{2}F, \knap{HCCl\Sub{2}} &\knap{HCCl\Sub{2}}\\
 83.93374598 & [ 83&93024931474109, &  83&9372426452589]   &  3956&1947 & -- & \isot{CCl[\Sup{37}Cl]}\\
 84.94873618 & [ 84&92634272134954, &  84&97112963865045]  &   140&2337 & H\Sub{2}S\Sub{2}F, H\Sub{2}OSCl, HNCl\Sub{2}, CF\Sub{2}Cl&\isot{HCCl[\Sup{37}Cl]}\\
 85.93171818 & [ 85&92419323227152, &  85&93924312772849]  &   564&31   & OCl\Sub{2}&\isot{C[\Sup{37}Cl]\Sub{2}}\\
 97.93130708 & [ 97&9236485334006,  &  97&9389656265994]   &   134&1543 & H\Sub{2}S\Sub{3}, \knap{COCl\Sub{2}}&\knap{COCl\Sub{2}}\\
 99.92428538 & [ 99&90729357057015, &  99&94127718942985]  &   106&7792 & HS\Sub{2}Cl, HO\Sub{2}SCl, NOCl\Sub{2}&\isot{COCl[\Sup{37}Cl]}\\
116.90524258 & [116&90200574284233, & 116&90847941715768]  & 28974&7117 & \knap{CCl\Sub{3}}&\knap{CCl\Sub{3}}\\
117.90830698 & [117&89455759263474, & 117&92205636736527]  &   189&527  & S\Sub{2}FCl, OSCl\Sub{2}, HCCl\Sub{3} &\isot{[\Sup{13}C]Cl\Sub{3}}\\
118.90232848 & [118&89897942315969, & 118&9056775368403]   & 29078&7276 & -- & \isot{CCl\Sub{2}[\Sup{37}Cl]}\\
119.90716988 & [119&89648859190275, & 119&91785116809724]  &   182&0316 & C\Sub{2}S\Sub{3} &\isot{[13C]Cl\Sub{2}[\Sup{37}Cl]}\\
120.89913018 & [120&89523104367791, & 120&90302931632209]  &  9220&1959 & -- &\isot{CCl[\Sup{37}Cl]\Sub{2}}\\
122.89646308 & [122&8875535007597,  & 122&90537265924031]  &   886&6747 & CSBr &\isot{C[\Sup{37}Cl]\Sub{3}}\\
  \end{tabular}
  \label{tab:CCl4-measured-masses}
\end{table}

With two arbitrary lists, many impossible partial fragments are enumerated, in
particular with too many hydrogens. It would speed-up the process to know an
upper bound on the number of mono-valent atoms. To be able to compute such
value, the first list, denoted M, is now made of the multi-valent atoms and the
second list, denoted m,
made of mono-valent atoms only. In this way, after computing the list M, one can
compute the DBE value of each partial fragment made of multi-valent atoms only.
An upper bound on the number of mono-valent atoms is two times the maximum DBE
value obtained over the list M. This upper bound allows to constraint the
enumeration of the list m, hence reducing the running-time and the length of the
second list. Table~\ref{tab:knapsack-A-B-M-m} presents a comparison of the
lengths of the lists A, B, M and m for the target masses of CCl\Sub{4}.
We observed that the list m is at least two times smaller than the
list B.

The chosen possible atoms are H, C, N, O, F, S, Cl, Br, I. At start, only the
abundant atoms are considered. The multi-valent atoms are C, N, O, S,
the mono-valent are H, F, Cl, Br, I.

Consider now the target mass $m$ = 116.90524258 m/z, with uncertainty range
$m_{\min}$ = 116.90200574284233, $m_{\max}$ = 116.90847941715768.
Our knapsack algorithm first lists all possible
chemical formulae made of any number of the multi-valent atoms and so that the mass of the
fragment is at most $m_{\max}$.
There are 263 combinations whose DBE ranges from to 2 to 20.
For each fragment, its DBE value $n$ is computed and a second knapsack algorithm is run to find a complement
fragment made of at most $n$ mono-valent atoms so that the total mass fits
within the bounds $m_{\min}, m_{\max}$ and the total DBE is positive or zero.
For the mass $m$ = 116.90524258 m/z, there is one solution: CCl3.

To account for chemical formulae made of mono-valent atoms only, a final knapsack
algorithm is run to find the chemical formulae with only one or two mono-valent atoms
whose mass fits in the uncertainty range (this gives the solution Cl for the
mass 34.96878848).

\begin{table}[h]
  \caption{isotopocules of CCl\Sub{4} and relative intensity. See also Fig.~\refmainfigisotopocules in the main article.}
  \label{tab:isotopocules-CCl4}
  \begin{tabular}{|l|r|c|c|}
\hline
isotopocule   & mass (m/z)  &proportion& relative intensity \\
\hline
{C           } & 12.00000000 & 0.988922 & 1.000000 \\
{[\Sup{13}C] } & 13.00335484 & 0.011078 & 0.011202 \\
\hline
{Cl          } & 34.96885271 & 0.757647 & 1.000000 \\
{[\Sup{37}Cl]} & 36.96590260 & 0.242353 & 0.319876 \\
\hline
{CCl\Sub{4}                             } & 151.87541084 & 0.325859 & 1.000000 \\
{CCl\Sub{3}[\Sup{37}Cl]                 } & 153.87246073 & 0.416938 & 1.279504 \\
{CCl\Sub{2}[\Sup{37}Cl]\Sub{2}          } & 155.86951062 & 0.200052 & 0.613923 \\
{CCl[\Sup{37}Cl]\Sub{3}                 } & 157.86656051 & 0.042661 & 0.130920 \\
{C[\Sup{37}Cl]\Sub{4}                   } & 159.86361040 & 0.003412 & 0.010470 \\
{[\Sup{13}C]Cl\Sub{4}                   } & 152.87876568 & 0.003650 & 0.011202 \\
{[\Sup{13}C]Cl\Sub{3}[\Sup{37}Cl]       } & 154.87581557 & 0.004671 & 0.014333 \\
{[\Sup{13}C]Cl\Sub{2}[\Sup{37}Cl]\Sub{2}} & 156.87286546 & 0.002241 & 0.006877 \\
{[\Sup{13}C]Cl[\Sup{37}Cl]\Sub{3}       } & 158.86991535 & 0.000478 & 0.001467 \\
{[\Sup{13}C][\Sup{37}Cl]\Sub{4}         } & 160.86696524 & 0.000038 & 0.000117 \\
\hline
  \end{tabular}
\end{table}


\section{Results for the validation set}
\begin{figure}[htbp]
  \centering
\includegraphics[width=10cm]{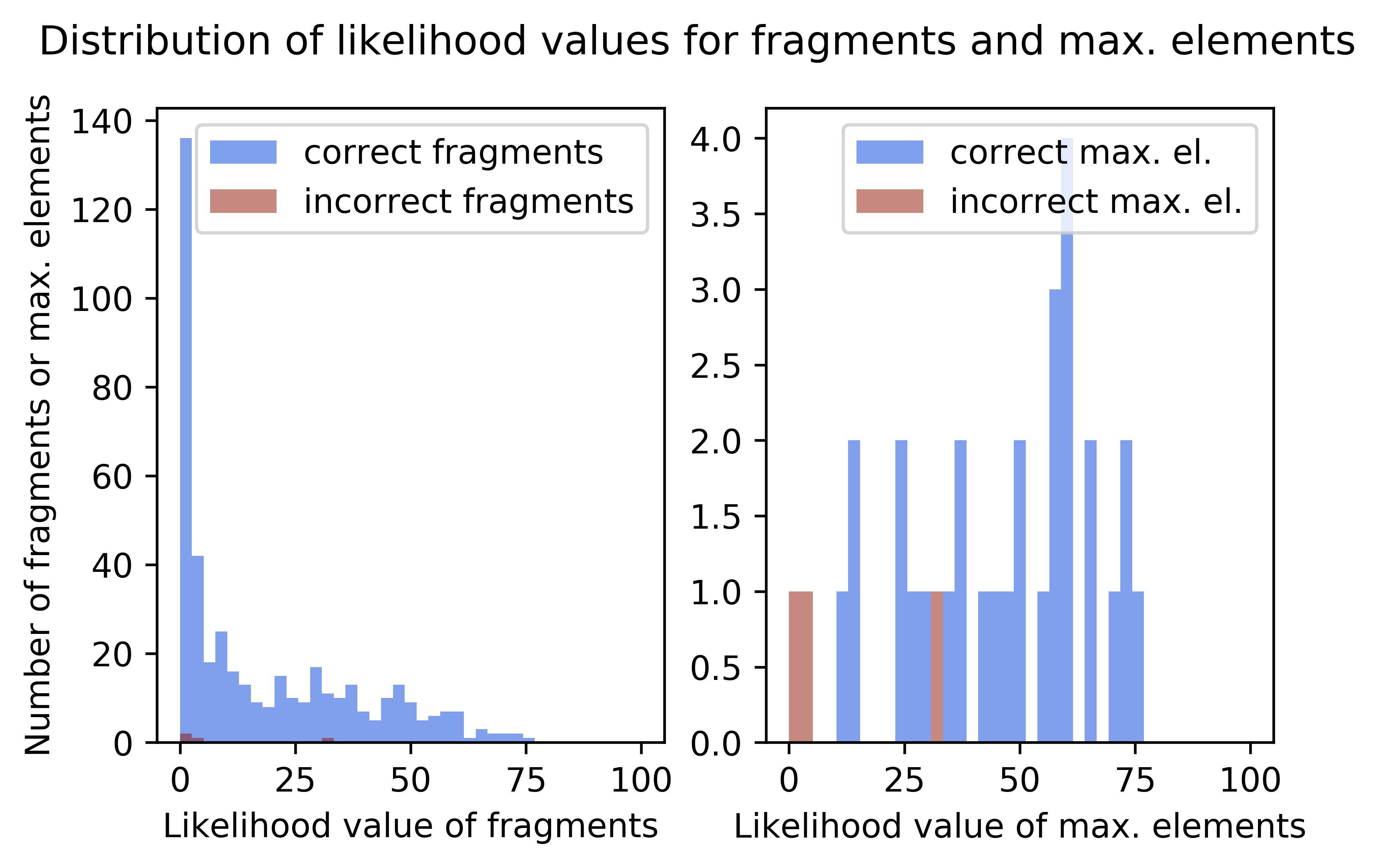}
\caption{Distribution of likelihood values of fragments (left) and maximal fragments (right)  assigned to a measured mass, for the validation set (23 compounds). A likelihood value of 100 indicates that the chemical formula of the fragment or maximal fragment is highly likely. Blue: distribution for correctly identified fragments/maximal fragments  assigned to a measured mass. Red: distribution for wrongly identified fragments/maximal fragments. }
\label{fig:val-hist-likelihood-frag-att}
\end{figure}

\begin{figure}[hbtp]
  \centering
\includegraphics[width=10cm]{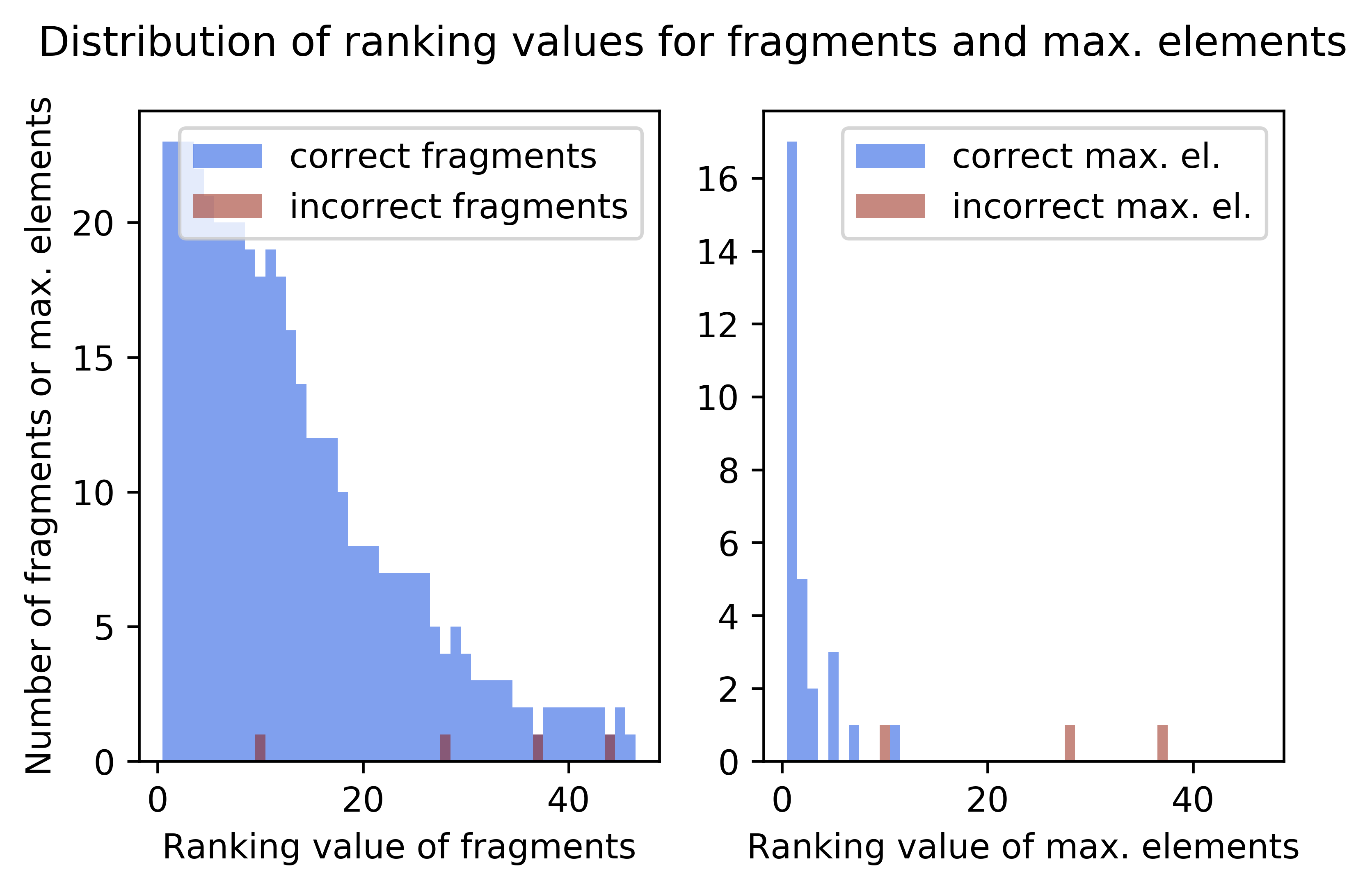}
\caption{Distribution of ranking values for fragments and maximal fragments assigned to a measured mass, for the validation set (23 compounds). A ranking value of 1 means that the fragment/maximal fragments was ranked as most likely (maximum likelihood value within the set of fragments/maximal fragments). Blue: distribution of ranking for correctly identified fragments/maximal fragments  assigned to a measured mass. Red: distribution of ranking for wrongly identified fragments/maximal fragments.  }
\label{fig:val-hist-ranking-frag-att}
\end{figure}

Results for the validation set are given in
Fig.~\ref{fig:val-hist-likelihood-frag-att} and Fig.~\ref{fig:val-hist-ranking-frag-att}.

\clearpage

\section{Testing the MOLGEN-MS commercial software}

MOLGEN-MS is a commercial software aiming at reconstucting the correct chemical formula and structure of an unknown substance for which the EI spectrum is provided, with unit mass resolution. The MOLGEN-MS workflow is made of the following steps \citep{molgenms_manual}:
\begin{itemize}
	\item MSclass: module to identify which chemical classes are likely present or not in the EI mass spectrum. This module uses a pre-defined list;
	\item {ElCoCo}: module to guess the most likely weight(s) of the molecular ion, with unit mass resolution. It can (or not) uses inputs form MSclass. Using the guessed molecular weight(s), all matching chemical formulae are generated, using a list of chemical elements that can be user-defined.
	\item MOLGEN: using all generated chemical formulae, generates all possible chemical strucutures. These structures are then fragmented and the obtained fragments are compared to the measured ones. All candidate structures are ranked, the most likely first.
\end{itemize}

We used a 90-days free version kindly provided by Markus Meringer, version 1.0.1.5.
We first converted all our data into a compatible format (.TRA). All masses were converted to unit mass resolution.
We used the default settings in MOLGEN-MS and keeping user intervention at the minimum. 
We used the same chemical elements as in our software: H, C, N, O, F, S, Cl, Br, I. We do not use Si and P that occur very seldom in our type of samples.
We tested MOLGEN-MS using two different settings: with and without the MSclass module.
When using the MSclass module, the output was then directly used in the {ElCoCo} module, without any user manipulation of the results.
For the {ElCoCo} module, we used the default parameter for the confidence interval for guessing the molecular weight (98). We did not use a minimum match value to eliminate candidate formula with a match value lower than a threshold. We then checked if the correct molecular weight was listed within the solutions.

We do not include the following substances in the discussion below because their true molecular wieght is outside our detection mass range: C\Sub{6}F\Sub{14} in the training set and TCHFB in the validation set.

In MOLGEN-MS, when using MSclass followed by {ElCoCo}, on the training set, only three substances have the correct molecular formula listed (COS, benzene and dichlorodifluoromethane). We actually obtain better results when not using the submodule MS-class.
When using {ElCoCo} alone, with the default precision value for the molecular weight (98), and no minimum precision for the molecular formula, the correct molecular formula is listed in 15 cases of the training set (out of 35, this is 43\%), with rankings from 1 to 45. If a user manually puts in the correct molecular weight, then the correct molecular formula is listed in 28 cases (out of 35, this is 80\%). Two cases with no correct solution require the Sulphur atom to have a set valence of 6, which may not be the default setting in MOLGEN-MS. In 6 cases, the correct molecular formula was listed only if the correct number of each atom was given as input. We do not have an explanation for this.

MOLGEN-MS provides better results on the validation set: for 17 compounds out of 22 (77\%), the correct molecular formula is listed (without using MSclass), with ranking from 1 to 46. 
{On} average, the correct molecular formula is listed using MOLGEN-MS-{ElCoCo} in 56\% of the cases, in the worse conditions where no user intervention is provided at all.

We provide all input data file in a MOLGEN-MS compatible format as zip file.

We believe that to identify the structure of an unknown compound, a promising setup may be to first run our ALPINAC software on the high resolution mass spectrum, and use the suggested molecular ion(s) together with the unit mass resolution spectrum as input to MOLGEN-MS, where the MOLGEN module would be used.


\bibliographystyle{bmc-mathphys} 
\bibliography{ALPINAC_20210319}      